# A General Perspective of Fe-Mn-Al-C Steels

O.A. Zambrano[a,b]

[a]Research Group of Fatigue and Surfaces (GIFS), Mechanical Engineering School, Universidad del Valle, Cali, Colombia

[b]Research Group of Tribology, Polymers, Powder Metallurgy and Processing of Solid Waste (TPMR), Materials Engineering School, Universidad del Valle, Cali, Colombia

∗Corresponding author. Tel: +57 (2) 3966134-3212133.

E-mail address: oscar.zambrano@correounivalle.edu.co

**Abstract**: During the last years, the scientific and industrial community have focused on the astonishing properties of Fe-Mn-Al-C steels. These high advanced steels allow high-density reductions about ~18% lighter than conventional steels, high corrosion resistance, high strength (ultimate tensile strength (UTS) ~1 Gpa) and at the same time ductility above 60%. The increase of the tensile or yield strength and the ductility at the same time is almost a special feature of this kind of new steels, which makes them so interesting for many applications such as in the automotive, armor and mining industry. The control of these properties depends on a complex relationship between the chemical composition of the steel, the test temperature, the external loads and the processing parameters of the steel. This review has been conceived to tried to elucidate these complex relations and gather the most important aspects of Fe-Mn-Al-C steels developed so far.

**Keywords:** Fe-Mn-Al-C steels; κ-carbide; lightweight steel; TRIP steel; TWIP steel; stacking fault energy; deformation mechanisms; corrosion; mechanical properties

# Contents

# 1   Introduction

During the last 50 years, a large body of work has been published about different aspects of Fe-Mn-Al-C alloys. Some very important compilations and viewpoint sets from an automotive, mechanical, and microstructural perspective have been published [1-7]. Nevertheless, this work is oriented to the process of understanding the different aspects of Fe-Mn-Al-C steels, and at the same time including other topics and points of view that are not in the existing literature. Recently, the author pre-published a very early version of this paper in arXiv [8]. However, in this manuscript, new sections, improvements and discussions are included, as well as recent literature.

The main goal of this review is to provide a critical overview of the current state of the art in Fe-Mn-Al-C steels in an easy reading way, bringing to the reader the last advances and developments in Fe-Mn-Al-C steels as well as the most important topics which have not been covered so far. This manuscript is divided into 7 sections, and each section of this assessment is intended to be self-consistent, that is, each section can be read independently. Section 1 is an introduction, which aims to give a general idea of Fe-Mn-Al-C alloys, and its historical development in a very abbreviated form. For this reason, all the contributions made by different researchers are not included here, and those studies will be discussed in other sections. Section 2 discusses the phase constitution in equilibrium and non-equilibrium conditions, and due to the importance also the kappa carbide. The mechanical properties and its microstructural relation are described in detail in Section 3, where fundamental concepts such as the stacking fault, stacking fault energy and its relation to deformation mechanisms are covered. The advances in the wear behavior are shown in Section 4, some processing techniques of Fe-Mn-Al-C steels are describing in Section 5, and the corrosion behavior is reviewed and analyzed in Section 6. This review closes suggesting future investigation directions in the light of the current development in these steels in Section 7.

## 1.1 Historical sketch

From a historical point of view, the development of Fe-Mn-Al-C steels dates back since Robert Hadfield in 1882 [9, 10] published his investigations about the high toughness and wear resistance of a Fe-13Mn-1.2C steel. Then he patented in 1890 some earliest Fe-Mn-Al-C steels [11]. Since then, it has been a meaningful advance in the study of these alloys. Some early works during the 50's, 60's, and 70's were focused on the development of the physical metallurgy of the alloys [12-18]. However, it was not until the 1980s that these steels took an important role, when they were conceived a possible substitutes for conventional stainless steels (Fe-Cr-Ni alloy) [19-22], changing the nickel for manganese and chromium for aluminum, which are less expensive alloy elements and have very similar effects in the physical metallurgy of these steels i.e. promote the growth of a protective oxide layer ($Al_2O_3$), and stabilize the fcc structure, respectively. For these reasons, in those days these steels were called nickel-free [21], nickel-chromium free [23] or even, the "the poor man stainless steel" [24]. However, the results to that date did not show that it was possible to satisfactorily replace the stainless steels. In the same decade, there was also a solid research effort about the mechanical properties at cryogenic temperatures [21, 25-27], obtaining promising results in comparison with conventional steels. The phenomenon responsible for this improvement was attributed to activation of specific deformation mechanisms. It was at that moment that elucidating and understanding the relationship between deformation mechanisms and mechanical properties substantially attracted the attention of the academic community. Indeed, for this reason, in nowadays these steels are referred depending on its deformation modes.

Following the chronological development of these steels, at the end of 90's and beginning of the 2000s decade, the Prof. Dr. Frommeyer [28], former head of the Department of Materials Technology at the Max Planck Institute for Iron Research (Düsseldorf, Germany), was pioneer in the quick progress of these steels, providing a strong background and comprehension about the deformation mechanisms and its relation with mechanical properties [29-35]. This opened the door to use this alloy system in the automotive industry. Particularly, oriented in the conformability of automotive body frames, where Frommeyer patented several alloys [36-38]. The main reason behind this success is that these special deformation mechanisms extend the plastic deformation beyond the usual ductility values, without any special treatment or expensive alloy elements, which permitted obtaining more complex geometries i.e. it was possible a greater degree of plastic deformation during the processing, until then unknown. Besides, with the same reasoning, it can be inferred that a high amount of energy can be absorbed by this kind of steels during a car-crash, adding an extra-safety to the passengers. In the same decade, the Prof. Wei-Chun Cheng [39-50] at the Department of Mechanical Engineering of the National Taiwan University of Science and Technology (Taipei, Taiwan) made huge contributions in the field of the physical metallurgy of these steels, such as several discoveries about its phase transformations.

In nowadays, the research group of Prof. Dr. Dierk Raabe at the Max Planck Institute for Iron Research (Düsseldorf, Germany), in conjunction with other important research groups around the world have taken the lead in researching these steels. Among his innumerable contributions to the science and the physical metallurgy of these steels, the most important topics could be summarized as follows:: the knowledge of dislocation structures as deformation progresses, the role of κ-carbide on the strain hardening behavior, the effect of grain size on the TWIP response, the effect of Al content on mechanical properties, the report of new mechanisms of plasticity, the impact of nanodifusion on the stacking fault energy, the use of *ab-initio* methodologies to guide the alloy design, the processing of these alloys, a crystal plasticity model for twinning and transformation-induced plasticity, among many others [51-68]. Obviously, at the present, different research groups are currently making great efforts to study different aspects of Fe-Mn-Al-C steels, which are not yet understood or investigated. On the other hand, there are also efforts in massifying the use of these steels at the industrial level and to be applied in different fields.

To end this section, it is presented in Figure 1, the chronological results obtained from Scopus®, where it can be seen the growing interest from different research groups in these last years.

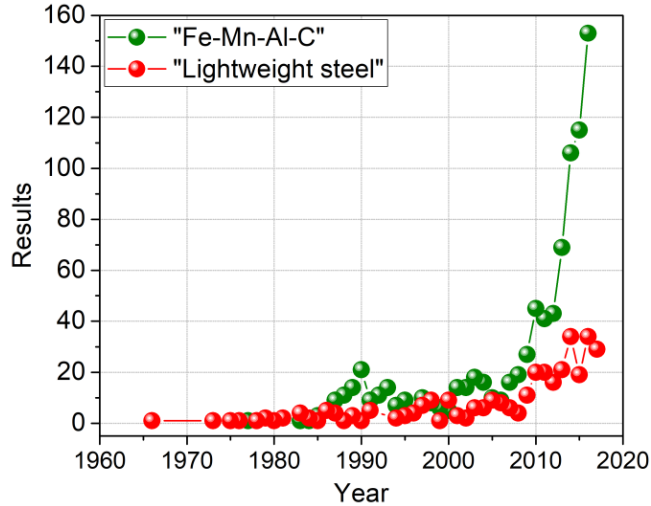

Figure 1. Chronological results in Scopus® during the last 50 years for Fe-Mn-Al-C steels. The results were obtained using the words: "Fe-Mn-Al-C" and "Lightweight steel" filtered to the materials engineering subject. The results obtained can be different depending on the words used as well as the subject filtered. However, this graphic represents the increasing tendency of research in Fe-Mn-Al-C alloys during the last decade.

## *1.2 General effects of the alloying elements*

The main alloy elements used in the Fe-Mn-Al-C system are; carbon (C), manganese (Mn), aluminum (Al) and silicon (Si), and their main effect are shown in Table 1. This table can be used as an initial tool to know quickly the main effect of some alloying elements.

| Element | Main effect |
|---|---|
| Mn | Stabilize the γ phase [26, 69], it can promote the β-Mn phase (which is brittle. possible the first report of this phases was done by [15]). The $\gamma \rightarrow \varepsilon \rightarrow \alpha'$ is favored for manganese contents ranging between 5-12%Mn and $\gamma \rightarrow \varepsilon$ is favored for manganese contents between 15-30%Mn [65, 70], improve the oxidation resistance through a (Fe,Mn)O scale [71, 72] |
| C | Stabilize the γ phase [69, 73, 74], decrease the fatigue strength in LCF (low cycle fatigue) and ELCF (extremely low cycle fatigue) [75] regimes. |
| Al | Stabilizes the α phase [69, 73, 74], promotes a protective layer of alumina ($Al_2O_3$), reduce the carbon diffusivity, increases the stacking fault energy (SFE) of the steel thereby suppressing the austenite γ to ε phase transformation [21, 26], reduce drastically the density of the steel [30]. |
| Si | Improves formation of an alumina (Al$_2$O$_3$) layer, solid solution strengthening austenite [76], favored the precipitation of (Fe,Mn)$_5$(Si,Al)C in ferrite based steels which reduce the ductility [77], increases the fluidity and decreases the melting point by 30° C)/wt pct Si [78], increase the kinetics of zone formation previous to κ-carbide precipitation [78], prevent the β-Mn phase reaction [79], favored the oxidation resistance by forming a protective passive film of SiO$_2$ |
| High-C and High-Al | Stabilizes the κ-phase [47, 80-87] |
| Mo | Increase the strain energy between the κ-phase and γ, and also delay the precipitation of κ-carbides during aging [88] |

Table 1. Summary of the alloy elements effects in the Fe-Mn-Al-C system

## *1.3 General mechanical properties and its low density*

The Fe-Mn-Al-C steels are attractive mainly because if possible achieve higher density reduction around ~18% (6.5 g/cm$^3$) than conventional steels [16, 30, 61, 89], high strength (ultimate tensile strength (UTS) ~1 GPa [55, 90], and recently Sohn *et al.* (2017) [91] reported 1.5 GPa) with an excellent ductility (>80% [90, 92]), which makes them promising for many applications. A comparison of these steels with other alloys systems from the mechanical point of view as well as the density reduction perspective is shown in Figure 2. It can be easily identified the astonishing combination of the tensile strength, elongation and density reduction achieve by Fe-Mn-Al-C steels. The other conventional alloy systems do not show this extraordinary combination of mechanical properties. The exceptional mechanical behavior relies on their deformation mechanisms; transformation induced plasticity (TRIP), twinning induced plasticity (TWIP), micro-band induced plasticity (MBIP) and dynamic slip band refinement (DSBR) [64, 93, 94], respectively. The control of these deformation mechanisms, allow having wide-ranging mechanical properties. However, the understanding and control of the deformation mechanisms in Fe-Mn-Al-C steels have been challenging during the last decades, mainly due to the complex relationship between alloy elements, phase constitution, mechanical properties and the external conditions imposed.

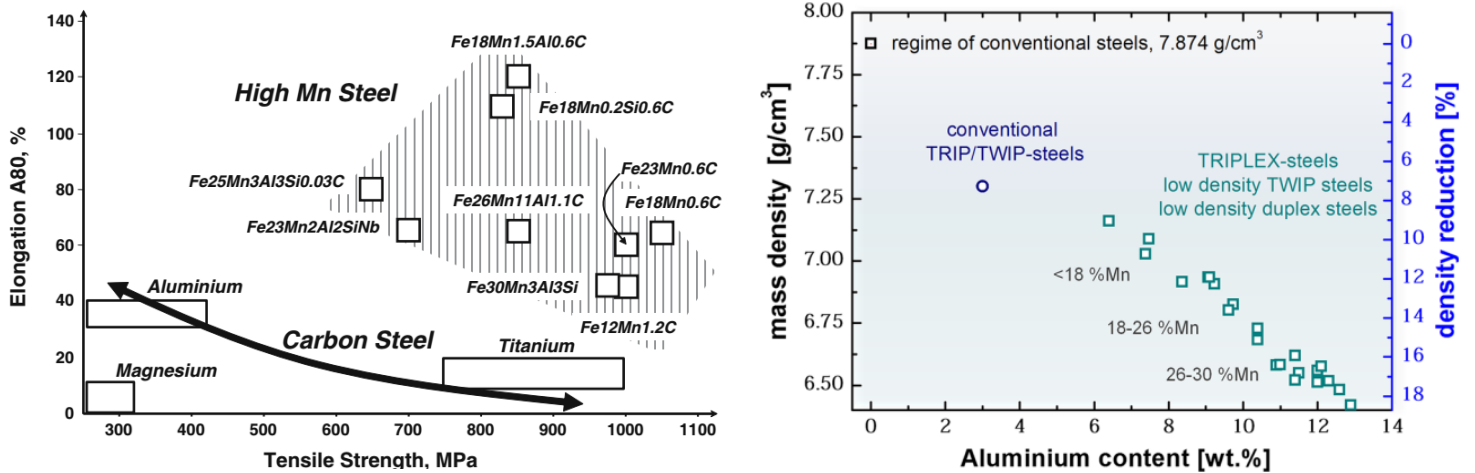

Figure 2. (left) tensile strength and elongation of different alloys and Fe-Mn-Al-C steels, reproduced from [95] with permission of Springer®, and (right) density reduction in Fe-Mn-Al-C steels as a function of Al content, reproduced from [61] with permission of Springer®

## 2 Phase constitution

Describing the phase diagrams of Fe-Mn-Al-C is not an easy task. Only in the last 8 years, new phases and phase transformations have been identified [41, 44, 48, 50, 62, 96] in Fe-Mn-Al-C alloys. Thus, a full description of these phases is still challenging and maybe in some degree premature yet. For this reason, in this section, a general description of the quaternary Fe-Mn-Al-C system and its main equilibrium phases (α-ferrite, kappa carbide (κ-carbide), γ-austenite, $M_3C$ carbide (θ), and β-Mn) will be presented, as well as some non-equilibrium phases. In this section, it will be using the weight percent (wt. %) unless otherwise specified.

### 2.1 Equilibrium phases

The first description of the Fe-Mn-Al ternary system was present by Köster *et al.* (1933) [97], who presented the α/γ phase equilibrium in the isothermal section at 1200°C. This study served as a base for the investigation performed by Schmatz (1959) [13] who discovered the β-manganese phase in Fe-Mn-Al system with high manganese contents. In this sense, Chakrabarti (1977) [98] corroborate the presence of β-manganese and expanded the Fe-Mn-Al system based on metallography examination and the X-ray identification. However, in his work, there was difficult with assessment relationship at the Mn-Al side, which was corrected with a more accurate phase equilibria for Fe-Mn-Al ternary system by Liu *et al.* (1993) [99] and Liu *et al.* (1996) [100]. On the other hand, the establishment of the Fe-Al-C system by Kumar *et al.* (1991) [101], Raghavan (1993) [102] and Palm *et al.* (1995) [103] were the first steps to understand the formation of the κ-phase (which is an intermetallic compound of $(Fe,Mn)_3AlC$ [14], with an $L1_2$ ordered fcc. structure) in the global Fe-Mn-Al-C system. The former authors determined some important properties of the κ-phase, like the lattice constant as a function of chemical composition, the microhardness and thermal expansion coefficient. Additionally, some updates in the Fe-Al-C phase diagrams were also performed by Raghavan (2002) [104] and Raghavan (2007) [105].

The first phase diagram of the Fe-Mn-Al-C system, were determined experimentally by Goretskii *et al.* (1990) [74] and Ishida *et al.* (1990) [73], holding the samples isothermally up to 250 hours between the temperatures of 627° to 1127°C and from 14 h to 210 h in a temperature range from 900ºC to 1200ºC. The first one established the phase diagram of a Fe-XMn-10Al-C system from 20% to 35% of Mn and 0.4 to 1.4% of C and the Ishida et al investigation increased the range of carbon and aluminum content up to 5% and 18%, respectively for the Fe-20Mn-XAl-XC and Fe-30Mn-XAl-XC system. The work performed by Goretskii *et al* showed that the β-Mn phase is stable at Mn contents greater than 25% and is present in general up to 800°C depending on the carbon content. In contrast, the work of Ishida *et al* shows that the β-Mn phase does not appear to form above 900°C for the range of compositions studied. In this sense, both works agree and are complementary to establish the initial Fe-Mn-Al-C phase diagram.

In recent years, Chin *et al.* (2010) [106] constructed a Fe-20Mn-Al-C phase diagram from the thermodynamical point of view based on the CALPHAD approach [107]. In this new effort, they take it into account better quaternary parameters, enhanced the κ-phase stability $((Fe,Mn)_3Al_1(C,Va)_1)$ and γ/α phase equilibria, comparing their results with experimental data obtained by Ishida *et al.* (1990) [73], finding a very good agreement. However, Kim and Kang (2015) [108, 109] and Phan *et al.* (2014) [110] developed a more accurate thermodynamic database for modeling the Fe-XMn-XAl-XC system from 900°C to 1200°C (with XMn from 0 to 40%, XAl from 0% to 20% and XC from 0% to 6%). In this new approach they take into account the thermodynamic properties of the liquid (they improving and updating the ternary systems (Fe-Mn-Al, Fe-Mn-C, Fe-Al-C, and Mn-Al-C)), enhanced the thermodynamic function of κ phase stability $((Fe,Mn)_3(Fe,Mn,Al)_1(Va,C)_1,)$, improving the accuracy experimentation with quaternary Fe-Mn-Al-C steels and comparison their results with the previous ones of Chin *et al.* and Ishida *et al.* to valid their thermodynamic assessment. In summary, this is one of the most completes phase diagrams in the literature for Fe-Mn-Al-C steels, and only some isothermals of the Fe–10Mn–Al–C system are reproduced here in Figure 3.

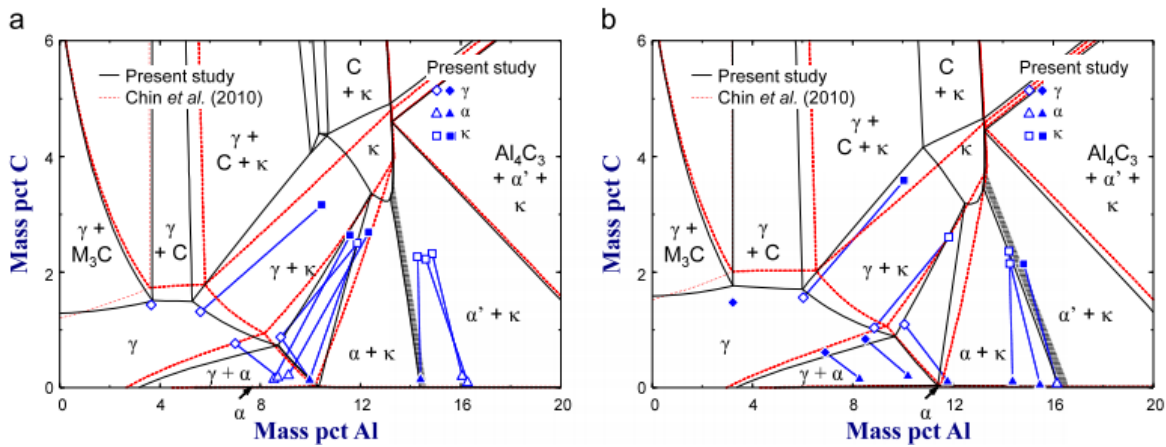

Figure 3. Isothermal sections of Fe-10Mn-Al-C at (a) 900 °C, and (b) 1000 °C. Experimental data are shown in symbols: solid symbol: deviation of Mn content within ±2%, open symbol: deviation of Mn content larger than ±2%, reproduced from [108] with permission of Elsevier®

To complete the outlook of the phase diagrams in the Fe-Mn-Al-C steel, Hansoo *et al.* (2013) [2] constructed low manganese phase diagram for the Fe–5Mn–XAl–XC system (with XAl varying from 0% to 9% and XC varying from 0% to 2%) from 500 to 1200°C using the thermodynamic date from [106] and is shown in abbreviated form in Figure 4.

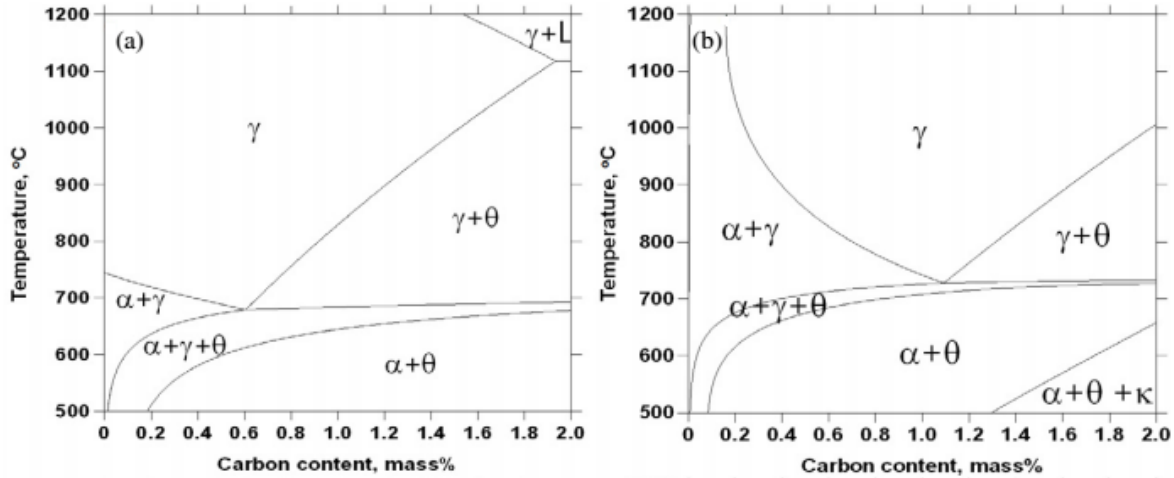

Figure 4. Phase diagrams of Fe–5Mn–XAl–XC alloys as a function of carbon content at Al concentrations of (a) 0%, and (b) 3% [2]

Recent discoveries in the Fe-Mn-Al-C equilibrium constituents have been reported. For instance, the eutectoid reactions; $\gamma \rightarrow \alpha + M_{23}C_6$ [96], $\gamma \rightarrow \alpha + kappa\ carbide + M_{23}C_6$ [46] and $\gamma \rightarrow \gamma + \alpha + kappa\ carbide + M_{23}C_6$ [49] in the Fe-13.4Mn-3.0Al-0.63C, Fe-13.5-Mn-6.3-Al-0.78-C and Fe-17.9-Mn-7.1-Al-0.85-C steels, respectively. The reader is referred to these references to analyses these phase diagrams that have not been included in the present review. To conclude, I would like to remark that the comprehension of Fe-Mn-Al-C phase diagrams are not completed yet. Conversely, much effort has to be made in this direction. For instance, this year Hallstedt *et al.* (2017) [111] published a reassessed database (taking into account the new description of the Fe-Mn-C system [112]) for construct some Fe-Mn-Al-C phase diagrams. Their contributions include an improved description of the Fe–Mn–Al–C system as well as the descriptions of cubic carbides and nitrides precipitation.

## 2.2 *Non-equilibrium phases*

In the last decades, it has been reported new phases outside of equilibrium conditions. For instance, the phase resulted after an homogenization at 1300°C and followed by air cooling to room temperature, which is different from the 18R martensite (which is formed in the bcc grains of Fe-Mn-Al alloys during water quenching from 1300°C [69]) or the austenite Widmanstätten side plate (because the austenite Widmanstätten side plates have a larger plate than this new phase) [113]. This new phase was identified by Cheng (2005) [50] as simple cubic (SC) Bravais lattice with a Widmanstätten morphology distributed uniformly in bcc grains. The main features of each microstructure are shown in Figure 5.

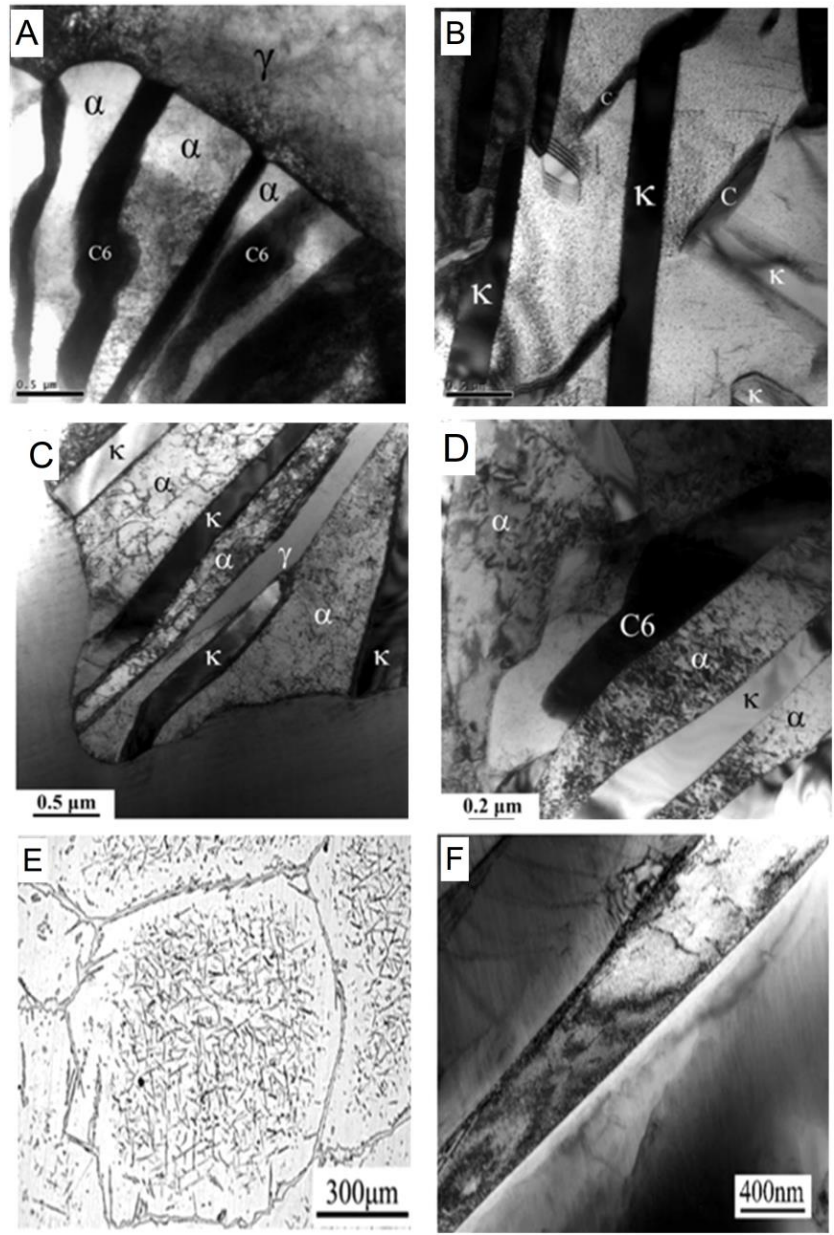

Figure 5. TEM for the (a) Fe-13.4Mn-3.0Al-0.63C steel after isothermal holding at 600°C for 100 h (γ: austenite; α: ferrite; and C6: $M_{23}C_6$), reproduced from [96] with permission of Springer®, (b) Fe-13.5-Mn-6.3-Al-0.78-C steel after isothermal holding at 650°C for 100 h (κ: κ-carbide and C: $M_{23}C_6$), reproduced from [46] with permission of Springer®, (c, d) Fe-17.9-Mn-7.1-Al-0.85-C steel after isothermal holding at 600°C for 100 h (γ: austenite; α: ferrite; κ: κ-carbide and C6: $M_{23}C_6$), reproduced from [49] with permission of Springer® and (e) Fe-Mn-Al alloy heated at 1573 K with argon protection for 30 min and air cooled, and (f) a TEM micrograph of (e), reproduced from [50] with permission of Springer®

## 2.3 Kappa-carbide (κ) in Fe-Mn-Al-C steels

Due to the importance of κ-carbide in the Fe-Mn-Al-C system, the main literature and features related to the κ-phase structure is discussed here. The influence of the κ-phase on the mechanical and other properties are covered in Section 3.3. One of the first reports about κ-carbides was conducted by James (1969) [14], who observed with the increase of carbon and aluminum content in the Fe-Al system, the age hardening effect, which produced the highest hardness in the range of 500°C and 550°C of aging temperature. However, the first mention of the precipitation hardening in Fe-Mn-Al-C steel was done by Kayak (1969) [15] who attribute the increase in mechanical properties after quenching from 1150°C and aging 16h at 550°C due to precipitation of iron-aluminum carbides. Later, Alekseenko *et al.* (1972) [16], manifested difficulties using X-ray and metallographic analysis, the proposed a magnetic method (saturation magnetization) to analyze quantitatively the $(Fe,Mn)_3AlC_x$ (due to $(Fe, Mn)_3AlC_x$ is ferromagnetic and $\gamma$ is not) precipitation evolution for aging temperatures from 300°C to 600°C at 8 h. They found the maximum increase of yield strength at 550°C wherein is the highest amount of hardening phase, however, at this aging temperature, there was a decrease in ductility from 60% to 40% and toughness from 20 to 10 kg m/cm$^2$. Then, Inoue *et al.* (1981) [114] studied the structure of this carbide in Fe-Mn-Al-C steels in a chemical composition range of 7 to 65 Mn, 3 to 9 Al and 0.8 to 2.4 C. They suggested that this precipitate has an L1$_2$ structure (order) with a formula of $(Fe,Mn)_3AlC$. However, at that time there was not a clear evidence if the κ-carbide had an L1$_2$ or L'1$_2$ order. In this regard, Han *et al.* (1983) [115] and Yang *et al.* (1990) [116] showed that κ-carbide has a fcc-based phase with L'1$_2$ ordered crystal structure (also termed as E2$_1$) which is similar to the perovskite oxides (like $BaTiO_3$), where the aluminum atoms occupy each corner, iron, and manganese atoms are located on face centers and the carbon atom is placed at the center of the unit cell (octahedral site) as is shown in Figure 6. They also supported its results from the study performed by Yang *et al.* (1990) [116], performing structure factor calculations, showing that in L'1$_2$ structure the {100} superlattice spots in TEM are more intense than the {110} superlattice spots, and this difference was more pronounced when the interstitial sites were fully occupied (carbon atom in the $\frac{1}{2},\frac{1}{2},\frac{1}{2}$ position), which supports the L'1$_2$ ordered crystal structure of the κ-carbide (Figure 6).

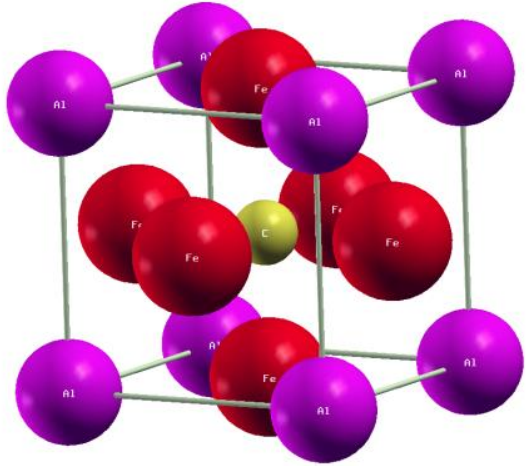

Figure 6. Unit cell of κ-carbide: aluminum occupies each corner, iron and manganese are located on face centers and the carbon atom is placed at the center of the unit cell which is also an octahedral site made by iron and manganese [117]

From the precipitation mechanism point of view, the former authors showed also that the Inoue et al's suggestion was not technically correct, they used X-ray diffraction and analyze the line profile around the (200) austenite peak for rapidly solidified alloys and also with aging treatments, showed the appearance of sidebands (Figure 7). These results suggested that a supersaturated austenitic phase has a decomposition accompanied by the sideband phenomenon even during the rapid cooling from the melting.

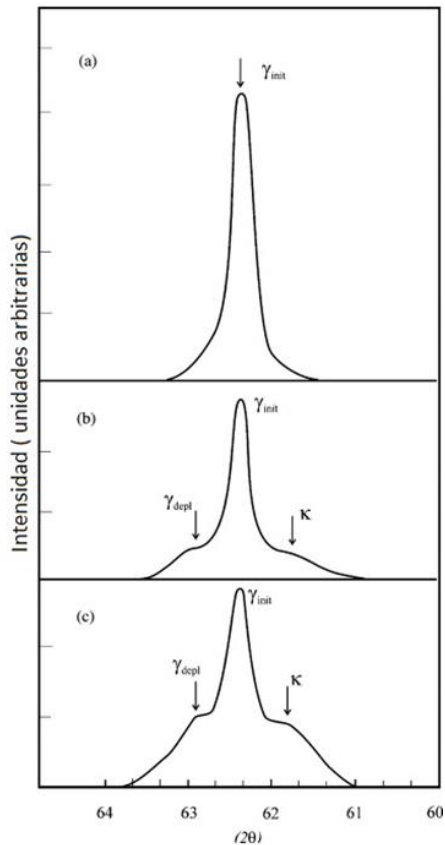

Figure 7. X-ray diffraction around the peak (200) in: (a) as-quenched, (b) after conventional aging at 550 8C for 16 h and (c) after second-step aging at 520 8C for 10 h, reproduced from [118] with permission of Elsevier®

In this sense, Bentley (1986) [119] has shown, using a TEM (Figure 8), that the carbide precipitation is not a result of nucleation and growth mechanism caused by the aging treatment. Instead, he showed evidence phase decomposition starts via solute composition fluctuation that is small in amplitude but is extended over the entire grain, found that the mechanism is a spinodal decomposition and rapid ordering during quenching. Similar results were found by Chu *et al.* (1992) [120], who studied the growth kinetics of κ-carbide for a Fe-30Mn-10Al-1C-1Si steel during aging. They proposed a relationship between of κ-carbide size and time ($r = r_0 + kt^{1/3}$), founding that even at zero time, there were presence of these carbides. Another important parameter that they determined was the activation energy for the κ-carbide coarsening during aging, which is around 196 kJ/mol. Yang *et al.* (2016) [121] studied the κ-carbide formation from a thermodynamic point of view, they revealed that the (FeMn)$_3$AlC carbide precipitation change from a spinodal decomposition to a classical nucleation-growth manner gradually, and the Al content has the strongest effect during the growth stage of κ-carbide.

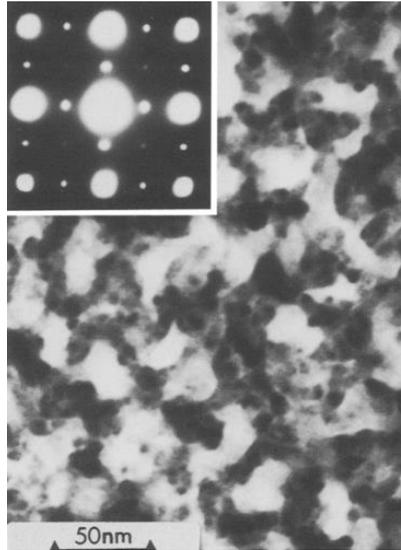

Figure 8. Transmission electron micrograph of 11% Al-32% Mn-0.8%C alloy after oil quenching from 1000°C. The diffraction pattern shows the LI$_2$ ordering, reproduced from [119] with permission of Springer®

Bartlett *et al.* (2014) [78] used APT, XRD, TEM and *ab-initio* methods to study the influence of Si on the κ-carbide composition, morphology (Figure 9), thermodynamics, and precipitation kinetics. They determined the lattice parameter of κ-carbide to be is practically independent of the silicon content and has an almost a constant value of 0.3712 nm. Additionally, they found a stoichiometric formula for the κ-carbide of (Fe, Mn)$_3$AlC$_x$ with x=0.3 after 60 h of aging time at 530 °C. On the other hand, the stability and kinetics of the κ-carbide precipitation increase with silicon content due to the increase of zone formation and some partitioning of manganese into the κ-carbide (how has been previously reported by Chao *et al.* (1991) [122]). They showed the partitioning of the elements after the aging treatment in Figure 10. These results were also recently confirmed by Kim *et al.* (2016) [123], who varied in a Fe–30Mn–9Al–0.9C–0.5Mo steel the silicon content from 0 to 1%. In a similar way, recently Yao *et al.* (2016) [66] used APT and DFT techniques to study experimentally and theoretically the chemical composition and site-occupancy of κ-carbides, founding the cuboidal κ-carbides particles are arranged in three dimensions in the matrix as shown in Figure 10 (similar findings for κ-carbides partitioning in a ferrite base Fe-Mn-Al-C steel, was performed by Seol *et al.* (2013) [62]). Furthermore, the partitioning of carbon and aluminum into κ-carbides and depletion of manganese are confirmed through the values of the partitioning coefficients; 0.92, 1.23 and 3.07 for Mn, Al, and C, respectively. They also determined the κ-carbides deviates from the ideal stoichiometry composition ((Fe, Mn)$_3$AlC, 60% (Fe+Mn), 20%(Al) and 20%(C)) to be ((Fe$_{2.00}$Mn$_1$)(Mn$_{0.09}$Al$_{0.91}$)(C$_{0.61}$Vac$_{0.39}$)). This is interesting because implies that the carbon atoms are not completely occupied in the octahedral sites, instead, they found the existence of C vacancies in the octahedral sites. Similar analysis shows that Mn atoms can also occupy the corners sites. In fact, recently Dey *et al.* (2017) [124] confirmed the deviation from the perfect stoichiometry of the κ-carbide using ab-initio methods. They even showed that depending on the κ-carbide location i.e. at the grain interior or located at the grain boundaries, the difference from the ideal stoichiometry was greater or lesser. These results are important for future works, specifically in the thermodynamic modeling of κ-carbide stability and modeling of plasticity mechanism such MBIP.

In summary, the initial stage during the aging in Fe-Mn-Al-C alloys is a compositional modulation and ordering produced by spinodal decomposition in carbon-rich or carbon and aluminum-rich zones and depleted zones. This produces the general kinetic path $\gamma^{ht} \to \gamma^{\uparrow C} + \gamma^{\downarrow C} \to \gamma^{\downarrow C} + L1_2 \to \gamma^{\downarrow C} + \kappa$ (however, the recent work of Emo *et al.* (2017) [125] proposed an alternative path using an atomic mean-field model for the κ-carbide kinetic). In summary, homogeneous and coherent precipitation of nano-sized (Fe,Mn)$_3$AlCx carbide has a E2$_1$ cubic perovskite crystal structure where the aluminum atoms occupies corner positions, iron and manganese atoms occupy face-centered positions, and carbon is located at the body center interstitial octahedral site.

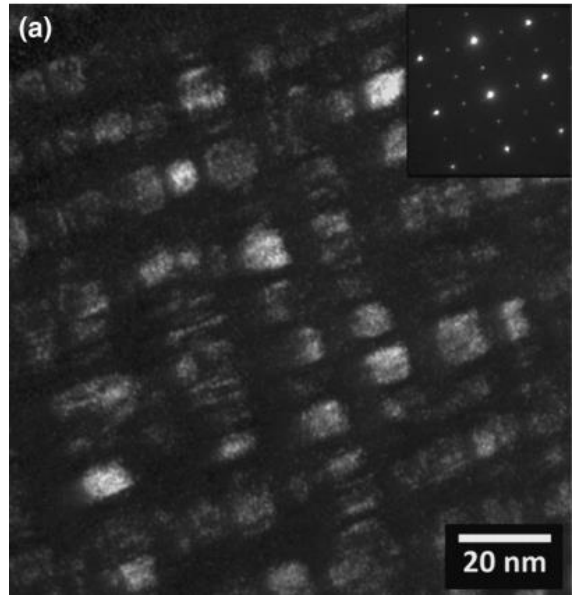

Figure 9. Dark-field images of the (a) 1.07 pct Si and aged for 100 h at 530 °C show κ-carbide as cuboidal particles with an average size of 10 nm cube length with an average center-to-center particle spacing along cube directions of 16 nm in an austenitic matrix, reproduced from [78] with permission of Springer®

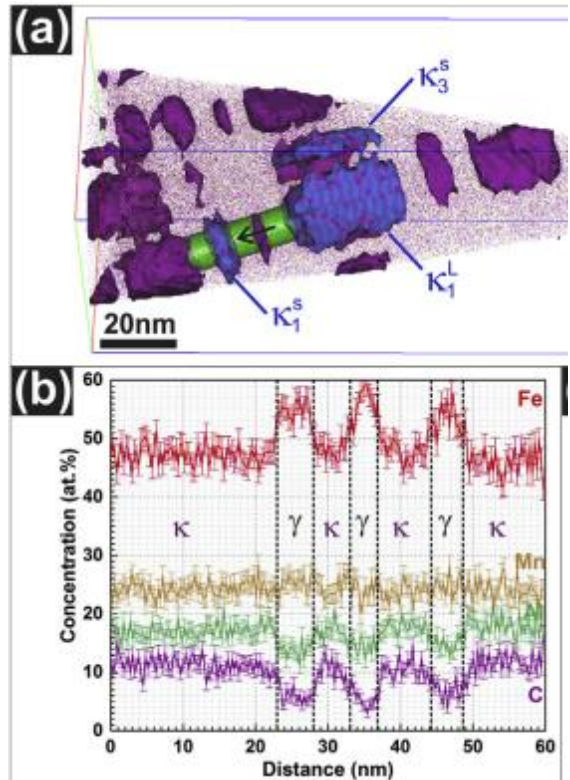

Figure 10. (a) APT maps of C (purple), Fe (red), Al (green) and Mn (yellow) atoms. κ-carbide precipitates are visualized by a 9 at.%C iso-concentration surface; (b) 1D concentration profile of elements along the green cylinder (F ¼ 10 nm) in (a) with a bin size of 0.3 nm. Steel solution treated at 1100 °C for 2 h followed by water quenching and aged at 600 °C for 24 h, reproduced from [66] with permission of Elsevier®

Regarding the formation of κ-carbides during aging treatments in austenitic Fe-Mn-Al-C steels, Song *et al.* (2015) [126] found, using DFT and synchrotron x-ray diffraction, that the κ-carbide is more thermodynamically stable in the ordered state than in disordered state. Another important contribution was to establish that the aluminum content has a more impact in ordering the κ-carbide than the carbon content. This latter aspect can also help to improve thermodynamics works about the κ-carbide stabilization. On the other hand, recently, Yao *et al.* (2016) [66] showed, using DFT, TEM, APT and nanoindentation, that the molybdenum delays the precipitation kinetics of κ-carbide precipitation. Basically, because the Mo atoms do not partition to the κ-carbide because the formation energy of κ-carbide increases with the substitution of Fe or Mn by Mo, and the strain energy between the κ phase and γ increased due to the Mo additions as is shown in Figure 11.

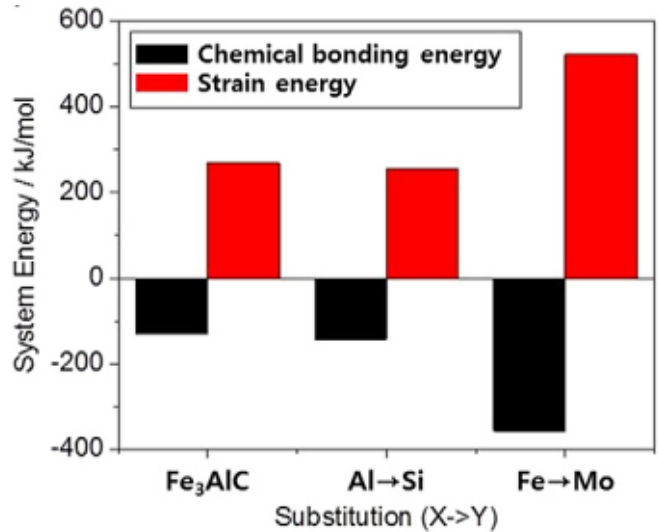

Figure 11. The calculated chemical bonding and strain contribution to the interfacial energy. The bars at Fe₃AlC indicate the values for κ-carbide with Fe3AlC formula unit. The bars at Al → Si and Fe → Mo indicates the values for the structures in which Si or Fe atoms are replaced by Al or Mo, respectively. Reproduced from [88] with permission of Elsevier®

# 3 Mechanical properties

The mechanical properties of the Fe-Mn-Al-C steel, such as yield strength, ductility, hardness and tensile strength, among others, can be modified either through solid solution hardening, precipitation hardening (carbides-κ) or by the control of the deformation mechanisms; transformation plasticity (TRIP), twinning induced plasticity (TWIP), slip dislocations and micro-bands induced plasticity (MBIP) induced. At present, both the effect of κ-carbides in the mechanical properties as well as the understanding of the deformation mechanisms of these alloys is still under study [53]. The main variables directly related to the control of mechanical properties and/or deformation mechanisms in these alloys is reviewed.

## 3.1 Stacking faults (SF), stacking fault energy (SFE) and its relationship with the deformation mechanisms

The activation of different deformation mechanisms in high manganese steels is intimately related to the stacking fault energy (SFE) of austenite. To initiate the discussion of the deformation mechanisms is important to define first the SFE. Regarding this, it is possible to conceive the ideal or perfect fcc crystal structure as a stacking sequence of {111} close-packed planes in the arrangement of ABCABCABC… and for the ideal hcp crystal structure a stacking of {0001} planes in a sequence of ABABAB... As is shown in Figure 12.

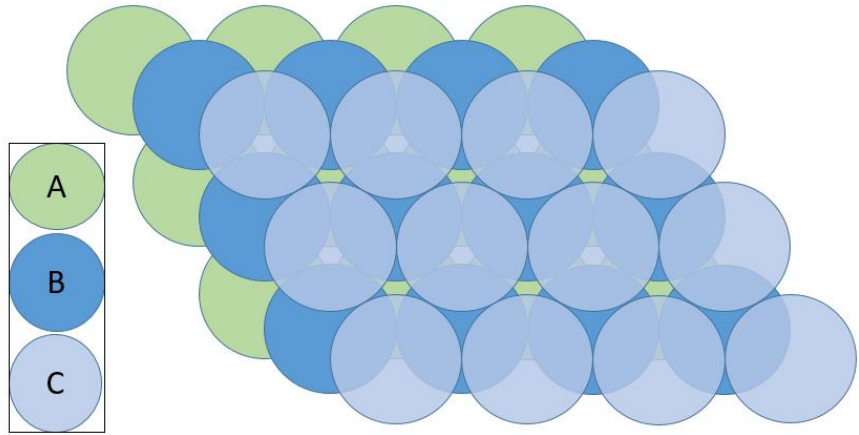

Figure 12. Schematic representation of the stacking sequence of a perfect (fcc) structure

However, in the reality, the perfect sequence is not preserved and one or two-layer interruption may occur due to the plastic deformation or due to the solidification process, which changes the packing sequence, giving rise faults. This is explained in detail as follows; In fcc crystals, the slip phenomenon occurs in the close-packed planes and directions. The slip of dislocations in these slip systems will leave behind a perfect fcc lattice, so this kind of dislocations are called perfect dislocations. However, in general, a perfect dislocation is dissociated in two partial dislocations (named Shockley partial dislocation) in the slip system (111)[$\bar{1}$10] through the dissociate reaction: $(\frac{a}{2}[\bar{1}10]) \rightarrow (\frac{a}{6}[\bar{2}11]) + (\frac{a}{6}[\bar{1}2\bar{1}])$ which is schematically represented in Figure 13. Since the energy per dislocation length is proportional to the square of the Burgers vector, it can be easily check that the dissociation of the perfect dislocation is energetically favored. This implies that it is easier for the (B) atoms to move along a zigzag path in the valleys between (A) atoms instead of "climbing over" them. The first partial dislocation with vector will move the B atoms into C position, which will produce a defect crystal containing a SF with the stacking sequence ABCACABC… It appears as if the (B) layer has been removed from the stacking sequence. The correct stacking of the fcc lattice will not be restored until the second partial dislocation passes.

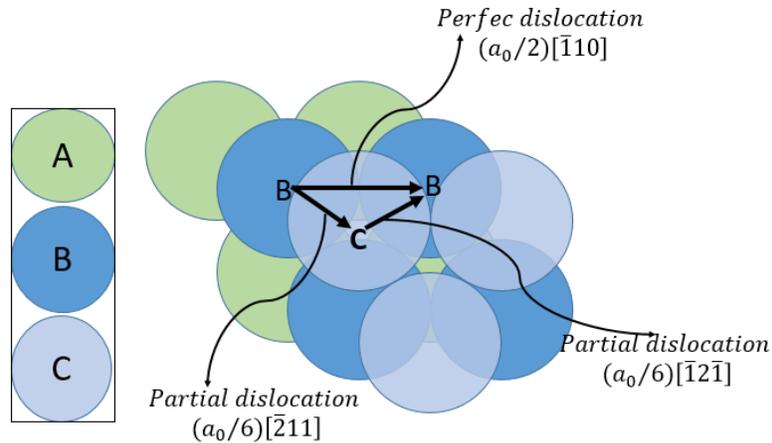

Figure 13. Illustration of the fcc lattice where the slip of a perfect dislocation and a partial dislocation is represented by arrows.

So, it can be concluded that when a single partial dislocation passes through the crystal, it leaves behind a region with abnormal stacking fault sequence as is shown in Figure 14 (a) and Figure 14(b).

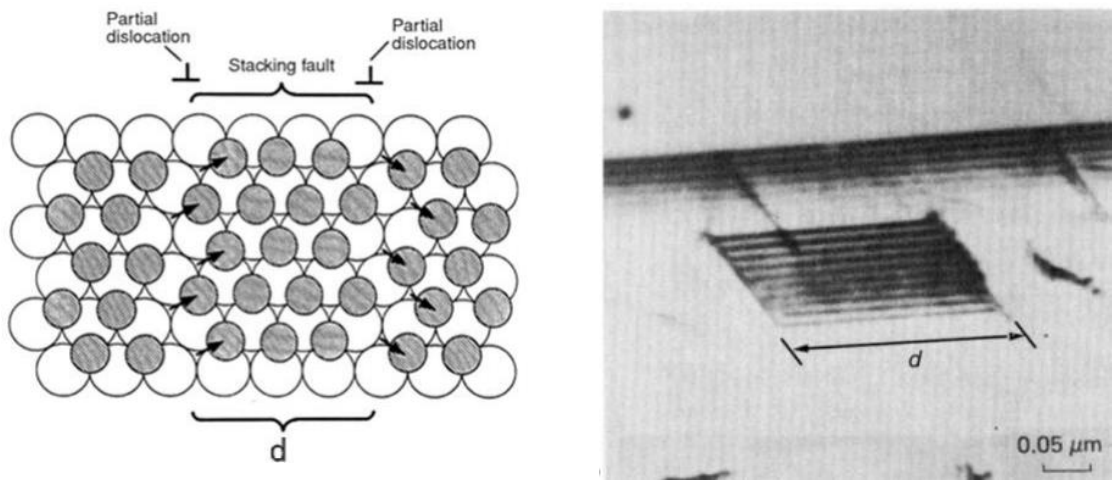

Figure 14. Equilibrium distance between two partial dislocations and stacking fault (a) schematically [127] and (b) experimentally obtained by TEM [128]

The fault described above, i.e. with the stacking sequence ABC<u>AC</u>ABC..., that is created by the passage of a single Shockley partial is called intrinsic. An intrinsic SF is considered to consist of two planes of the hexagonal close-packed (hcp) structure. Another kind of stacking fault could be defined, the extrinsic SF or twin SF, which has a stacking sequence of ABC<u>ACB</u>CAB… The extrinsic SF is formed if an additional C layer has been inserted in the ABCABC... ideal fcc stacking, which produces a twin with the stacking sequence ACB (both changes in the stacking fault are shown schematically in Figure 15). Thus, from the crystallographic point of view, the strain-induced ε-martensite or twinning induced plasticity effect could be explained and considered to originate from stacking faults. Experimental evidence of the role of the intrinsic and extrinsic SF on the nucleation of these mechanisms was addressed by Idrissi *et al.* (2009) [129].

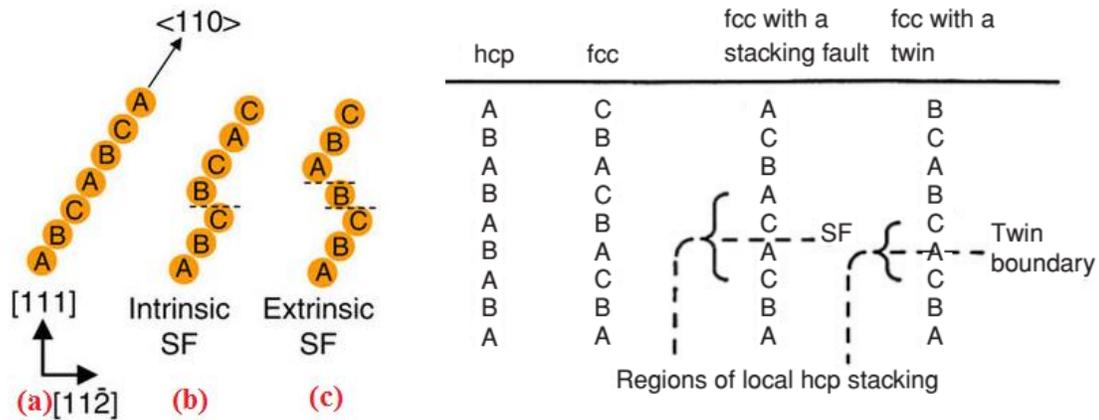

Figure 15. Stacking sequence for an (a) perfect fcc lattice, intrinsic SF and extrinsic SF, reproduced from [130] with permission of Nature Publishing Group®, and (b) how the character of the SF cause a local hcp sequence or twin boundary [127]

The stacking fault (SF) and the energy related to this stacking fault (since this is not the equilibrium structure, the SF increases the total energy of the lattice) is known as stacking fault energy. The width (d) of an SF (Figure 14) resulted from the split of a perfect dislocation in two Shockley partial dislocations is an indirect measure of the SFE. This represents the equilibrium distance between partial dislocations caused by the repulsive and attractive forces (surface tension). Thus, low SFE implies or is associated with high separation distance whereas high SFE implies that the partial dislocations separations are narrow. According to the aforementioned concepts, it can be inferred that low SFE materials show a wider stacking fault which difficult the cross-slip and climb of dislocations. In high SFE materials, the opposite is observed. These facts strongly alter the mobility of dislocations during plastic deformation in Fe-Mn-Al-C steels, that is changing the deformation mechanisms.

The SFE has been found to depend mainly on the chemical composition, microsegregation of alloying elements (Suzuki effect [131]), temperature and grain size [132-135]. However, why is so important this parameter in high manganese steels? Basically, the high ductility, strength and high strain-hardening rate of these steels depend upon austenite stability which is controlled by the SFE. In this context, the estimation of the SFE has been done so far with mainly four techniques; transmission electron microscopy (TEM) through the dislocation-node radius method [136, 137], which is adequate for low SFE steels [138], X-ray diffraction methods [139, 140], *ab-initio* calculations [141, 142] and thermodynamics-based methods [134, 143, 144]. This last method is extensively used to measure the SFE in manganese steels [134, 144-146] (a critical assessment of this kind of modeling was addressed by Geissler *et al.* (2014) [147]). As a final point, the reader is referred to the work performed by Das (2015) [148], who did an extensive review during the last 60 years about the SFE estimation in steels and propose a model based on a Bayesian neural network perspective to analyze the parameters which influence the SFE. Lately, a novel machine learning approach was also used for predicting the SFE of austenitic steels, particularly to classify strictly the main deformation mechanisms, and this approach showed promising results [149].

The stability of the austenite (governed by the SFE), controls the facility of the deformation-induced martensitic transformation of austenite (γ) into ε-martensite, α`-martensite, mechanical twinning or permit the dislocation gliding. In general, the stress-induced $\gamma \rightarrow \varepsilon$ transformation (TRIP) occurs in steels with SFEs ≤ 20 mJ/m$^2$ [33, 150], the deformation twinning (TWIP) has been observed in steels with SFEs between 20 mJ/m$^2$ to 40 mJ/m$^2$ [150], partial and/or perfect dislocation gliding above 40 mJ/m$^2$ and predominant microband-induced plasticity (MBIP) or the recently discover of the dynamic slip band refinement (DSBR) effect has been reported in steels with SFEs above 60 mJ/m$^2$ [51, 64, 92, 151].

The summary of these values is shown in Table 2 and its microstructural schematic representation is show in Figure 16. It can be seen that the austenite during deformation can produce martensite needles or twin, or dislocations bands, which reduce dynamically the dislocations movement.

| *SFE (mJ/m$^2$)* | < 20 | 20 - 40 | 40 - 60 | >60 |
|---|---|---|---|---|
| Main deformation mechanism | ε-martensite | Twinning | Cross slip and activation of slip systems | Micro-bands or Dynamic slip band refinement |

Table 2. Deformation mechanisms in austenitic Fe-Mn-Al-C steels as a function of SFE (these values represent general tendencies)

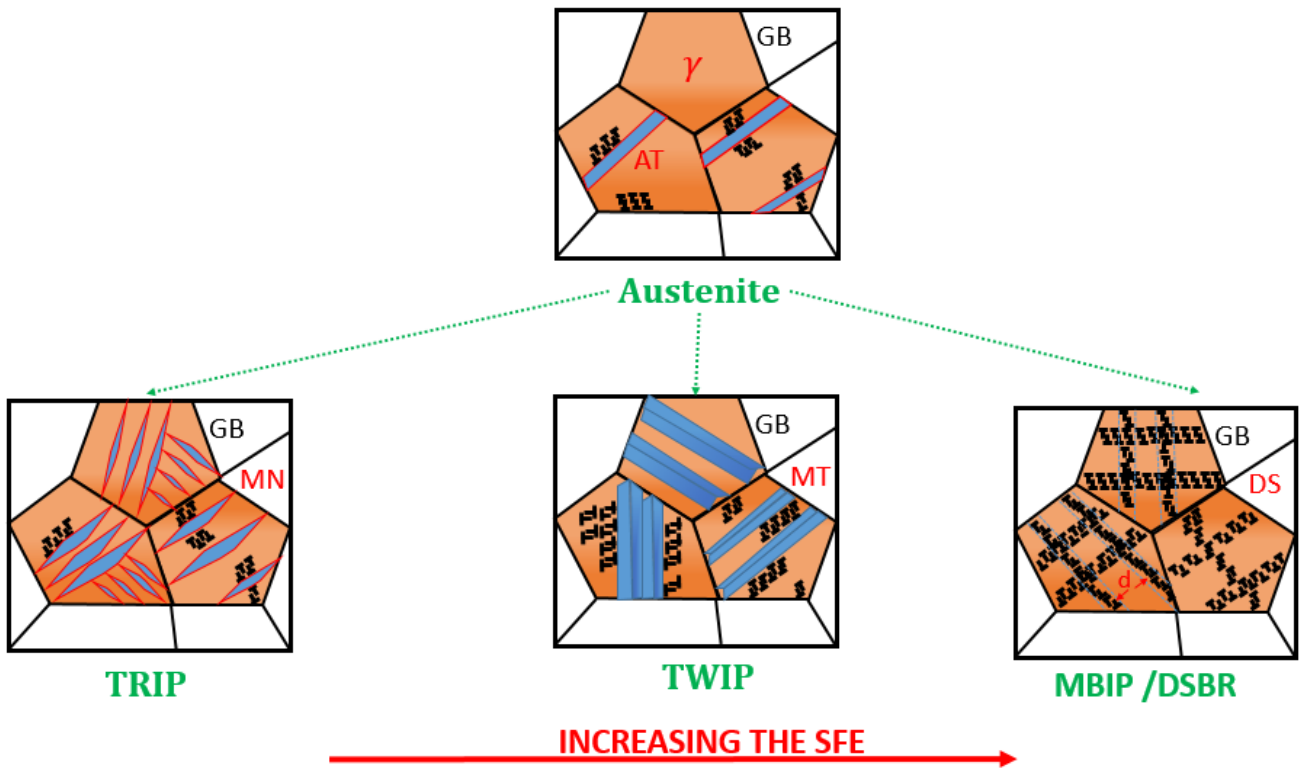

Figure 16. A schematic representation of the main deformation mechanisms in austenitic Fe-Mn-Al-C steels. GB: Grain boundary, AT: Annealing twin, MN: Martensite needle, MT: Mechanical twin, DS: Dislocations, d: spacing.

It is important to point out that the activation of these deformation mechanisms is not an immovable boundary. For example, Koyama *et al.* (2015) [152] showed that the operation of the twinning deformation mechanism was activated even in SFE values as high as 55 $mJ/m^2$. They showed that this behavior cannot be explained only by the strict dependence of twinning and the SFE. Instead, they proposed that the DSA mechanism assists the deformation twinning. A schematic representation of the DSA phenomenon in high manganese steels is shown in Figure 17.

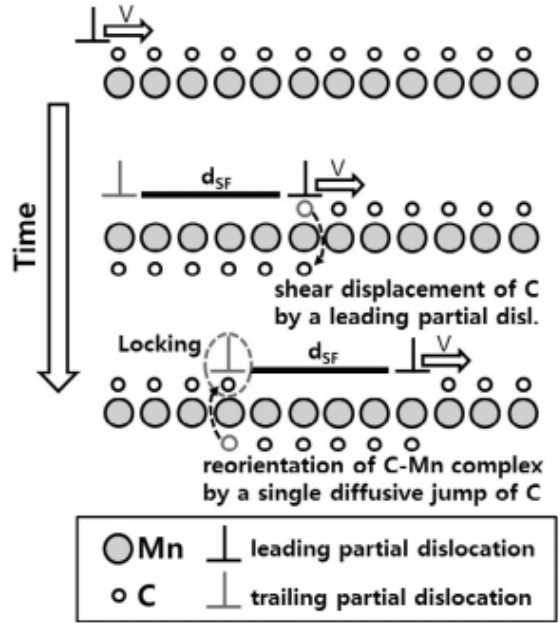

Figure 17. Schematic illustrating of the DSA phenomenon in high-Mn TWIP steel, explaining the C-Mn interaction with the SF, reproduced from [153] with permission of Elsevier®

From the technology point of view, the SFE can be changed by adjusting the chemical composition of the alloy. Mainly because the alloying elements are one of most influential variables on the SFE and so the deformation modes. Lastly, I suggest an updated review of the TRIP effect, which includes its definition as well as its historical development and the applicability of this phenomenon in commercial steels, which can be found in [154].

### 3.1.1 Factors affecting the SFE in Fe-Mn-Al-C steels

The SFE alloy varies depending on factors such as chemical composition, the grain size of austenite [133, 155] and deformation temperature [132-134, 156-158]. Each of this factors will be discussing shortly to give a general idea of its effects.

*3.1.1.1 Temperature effect*

In general, with the increase of the temperature the dislocations nodes decrease, i.e. the SFE increases with temperature. This fact was reported originally by Remy (1977) [159] and Rémy *et al.* (1978) [160] through TEM observations, in which the dislocations nodes decreases as temperature increase. Recently, Hickel *et al.* (2014) [58] through *ab-initio* calculations and TEM observations found the opposite trend using as a start temperature a cryogenic temperature to study the nano-diffusion effect (Suzuki segregation) of carbon in the stacking faults. They found with increasing the test temperature the partial dislocation width increases which produce a decrease of the SFE. Recently, the variation of the SFE with temperature was plotted in Figure 18 [161].

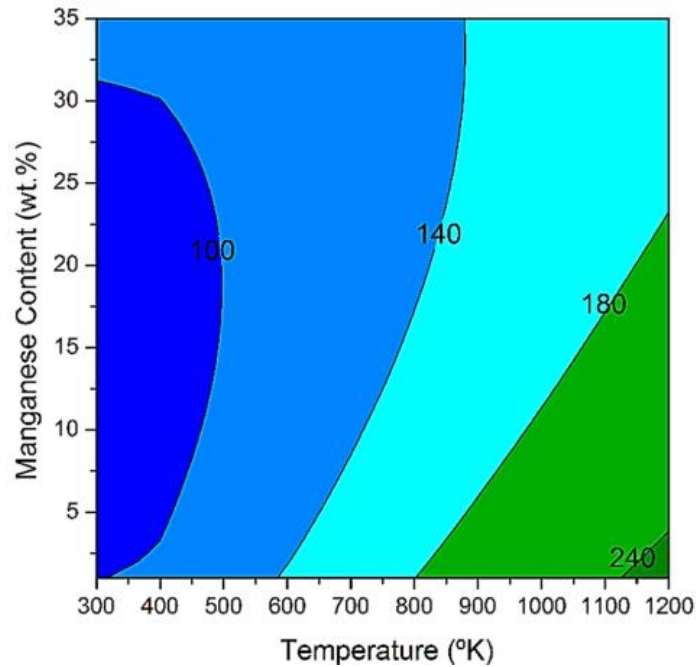

Figure 18. Two-dimensional SFE (unit: mJ/m$^2$) map for Fe–XMn–10Al–0.5 C alloy system with different temperatures [161]

### *3.1.1.2 Chemical composition effect*

The chemical composition of the alloy affects drastically the SFE. For instance, the **manganese** content increases the SFE how have been reported by Lee *et al.* (2000) [155] in Fe-Mn binary alloys and Dai *et al.* (2002) [162] for austenitic steels. The **silicon** has been reported to decrease the SFE in Fe-Mn-Si-C steels using X-ray diffraction analysis [163] and thermodynamic calculations [164] and also have been reported in Fe-Mn-C and Fe-Mn-Al-C steels using thermodynamic models [132, 165]. A brief summary of effect of some alloy elements is show in Figure 19.

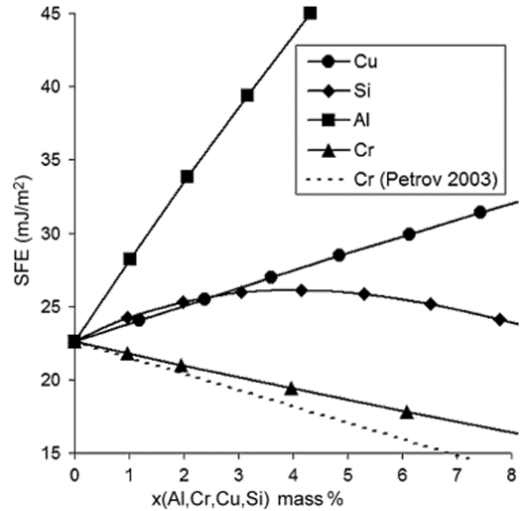

Figure 19. Influence of alloy elements on the SFE in the Fe-22Mn-0.6C system, reproduced from [132] with permission of Elsevier®

There are many reports in the literature about the drastic increase of the SFE by **aluminum** additions. For instance, Chen *et al.* (1993) [166] in Fe-20Mn-Al-C steels, observed a decrease of strain-induced martensite and deformation twinning by increases the aluminum additions i.e. an increase on SFE. Even an addition of 1 wt.%Al inhibited the TRIP effect. In this sense, Tian *et al.* (2013) [167] also reported the suppression of the strain induced $\gamma \rightarrow \varepsilon$ transformation and deformation twinning by the increase of SFE by aluminum additions in Fe-25Mn-Al-C steels. Previous studies about the **carbon** effect on the SFE shown a remarkably different trends; a strong increase on the SFE with the increase of the carbon content [139, 168] and negligible effect on the SFE by the carbon content [169]. Recently it was concluded, using *ab-initio* calculations, that the carbon increase the SFE in Fe-C [170] and FCC for pure iron [141, 171]. Additionally, the variation of the SFE with the carbon and aluminum content for the Fe-Mn-Al-C system [161] is shows in Figure 20.

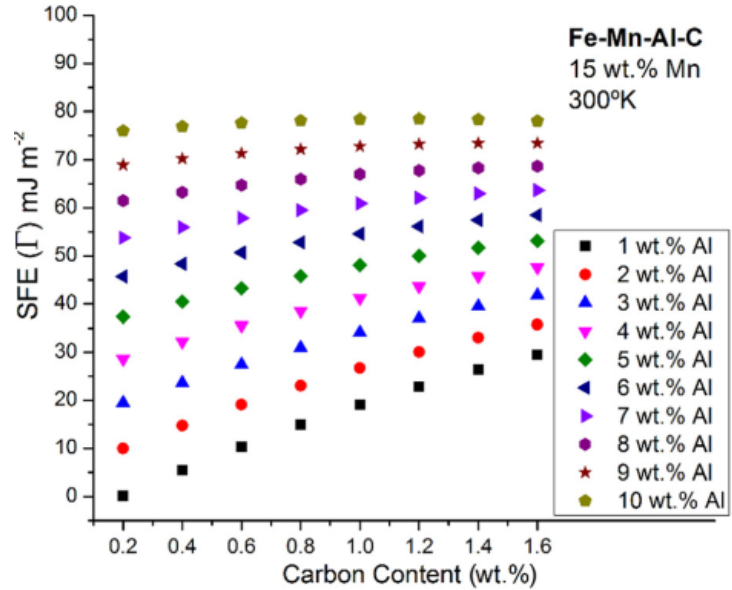

Figure 20. Two-dimensional SFE (unit: mJ/m$^2$) map for Fe–15Mn–XAl–0.5 C alloy system with carbon contents [161]

*3.1.1.3 Grain size effect*

Is not clear yet how the grain size affects the SFE in austenitic steels and there are only a few reports in the literature about this relation. One of these few investigations were performed by Jun *et al.* (1998) [172] who studied in Fe-Mn alloys the relation between grain size and SFE. They found that decreasing the grain size below 35 μm the SFE increases exponentially. Conversely, if the grain size increases above 35 μm, the SFE decrease. The work of Lee *et al.* (2000) [155] in Fe-Mn alloys in a range of austenitic grain sizes from 5 to 150 μm, confirms that increasing the grain size, the SFE drastically decreases for all manganese contents. Moreover, even in low SFE steels, the effect of grain size is relevant for both kind of martensite (ε, and $\alpha'$) and the suppression of this effect [173]. In a larger range of SFEs values, it is concluded that at higher grain sizes, the deformation modes of TRIP and TWIP (that require low SFEs) are more likely to occur than in the case of fine grain sizes. Some experimental studies in high manganese steels show this phenomenon i.e. the TRIP and TWIP effects are reduced or suppressed in fine grain sizes [133, 174, 175] or promoted in coarse grain sizes [172]. The variation of the SFE with the austenitic grain size in the Fe-Mn-Al-C system [161] is shows in Figure 21.

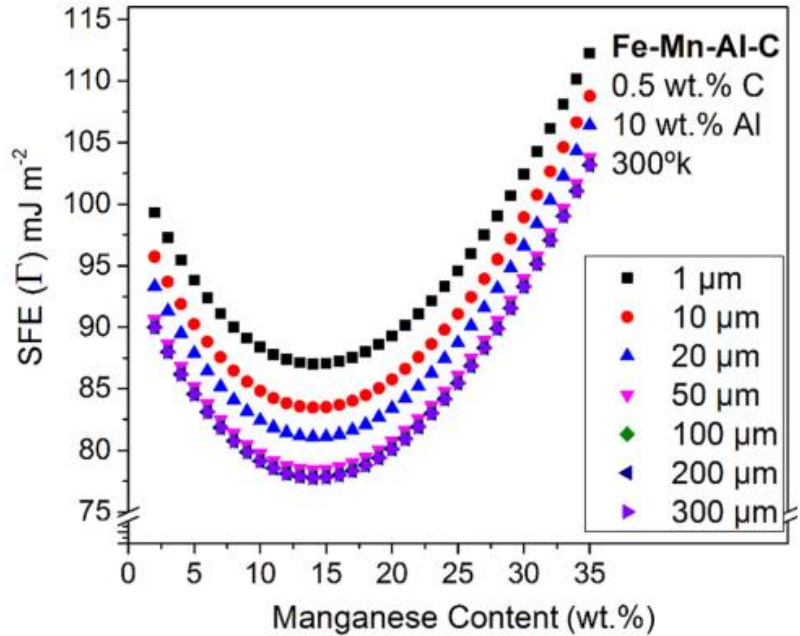

Figure 21. Variations in SFE map by increasing grain size and manganese and aluminum contents: for 10 wt.% Al [161]

Recently, it was reported that Nb additions in Fe-Mn-Al-C steels [176], produced a drastic decrease of the twin activity, mainly caused by the grain size refinement in the microstructure. This could confirm the previous investigations about the role of the grain size on the deformation mechanisms.

### *3.1.1.4 Solid solution hardening*

Jung *et al.* (2013) [153] determined the effect of aluminum on mechanical properties at different temperatures for a Fe-18Mn-xAl-0.6C system. Figure 22 shows that the increase in aluminum at each temperature produces an increase in the yield stress for all test temperatures. This behavior is due to the solid solution hardening gives the aluminum, producing a high distortion of the crystal lattice.

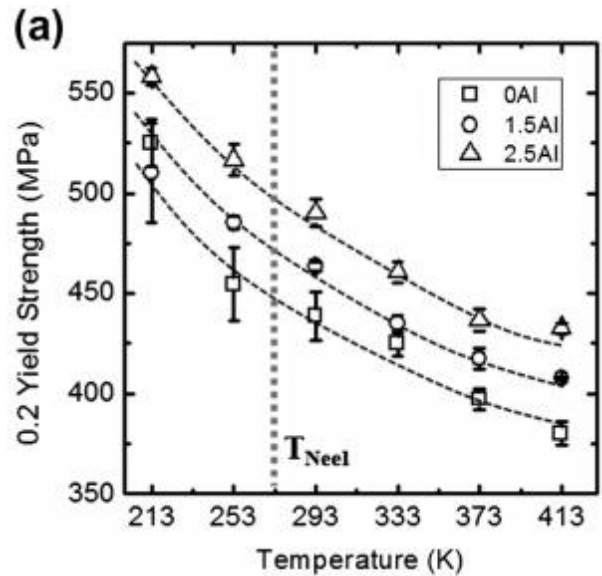

Figure 22. Effect of the 0Al, 1.5Al and 2.5Al content on the 0.2% yield strength of a TWIP steel, reproduced from [153] with permission of Elsevier®

The effect of aluminum on the increase of the yield strength has also been reported in these alloys by Raabe *et al.* (2014) [61]. However, Chen *et al.* (1993) [166] found that with increasing aluminum there is a decrease in yield strength and the ultimate yield strength which is due to the decrease in the SFE.

As for the effect of the manganese in solid solution hardening, it has been determined that this element does not necessarily increase the mechanical properties, but it can occur either an increase or a decrease in the mechanical properties depending on the percentage of manganese employed in the steel [76, 177, 178]. On the other hand, carbon as expected increases both the yield stress and the tensile strength due to interstitial solid solution hardening [89]. Finally, silicon promotes the increase in mechanical properties due to solid solution hardening, as recently reported by Li *et al.* (2015) [179] who determined the effect of additions of silicon in the Fe-Mn-Al system, concluded that this element produces a significant increase in yield stress, tensile strength, and hardness caused by solid solution in the austenite.

## 3.2 Strain hardening rate ($d\sigma/d\varepsilon$)

The work hardening behavior in Fe-Mn-Al-C steels is a vital feature from the scientific and technological point of view, mainly because of influence the adequate conditions on the material processing as well as the selection for specific structural applications. This topic is so important, that is termed strain-hardening engineering [180], this means that we can only treat in this review some important aspects of the strain hardening behavior in Fe-Mn-Al-C steels. The work hardening is described generally by the semi-empirical constitutive Hollomon equation ($\sigma = k\varepsilon^n$). However, the Hollomon formulation is valid always as the relationship between log $\sigma$ and log $\varepsilon$ in the whole plastic flow range maintain linear. It is also assumed that microstructure does not vary or change during the plastic deformation. In Fe-Mn-Al-C steels, this last assumption is not strictly valid, because several metallurgical factors could occur during deformation which produce different work hardening stages [51, 64, 181-184], which create a deviation from the Hollomon prediction. In this sense, different authors have proposed models in manganese steels to include effect of martensite transformation [185, 186], twinning [187] or recently, both phenomena [65] on the flow stress curves. The interested reader in a more general strain hardening models and its discussions is referred to reference [188].

The high work hardening phenomenon in Fe-Mn-Al-C steels is due microstructural changes during deformation. Four distinct stages can be described the work hardening behavior as is shown in Figure 23. The stage I is easily identified for the drastic decrease in the work-hardening rate and is commonly presented to almost any steel independent of its SFE value. This can be attributed to the easy glide of dislocations in a single system (it does not occur when the deformation takes place by multiple slips from the beginning [189]), because if each mobile dislocation could pass through the crystal without any barrier, the work hardening rate would be exactly zero [190].

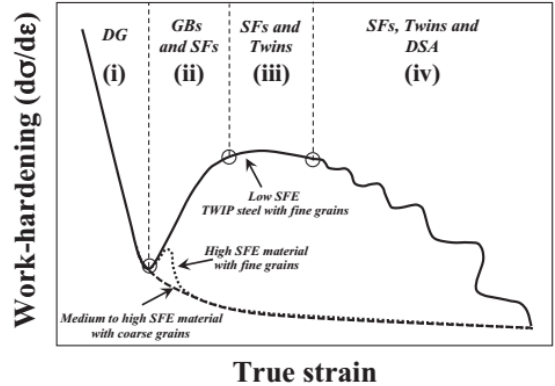

Figure 23. Schematic representation of the work-hardening behavior of fcc materials with different stacking fault energies (SFE) under tensile testing. Dash line curve represents the material with medium to high SFE having coarse grain size. Dot line the curve shows the material with medium to high SFE with small grain size. Solid line curve represents the low SFE TWIP steel with fine grain. DG, GB, SF, and DSA are dislocation glide, grain boundaries, stacking faults and dynamic strain aging, respectively. Reproduced from [184] with permission of Elsevier®

Stage II is characterized by the emission of dislocations by the stacking faults which interact with other glide dislocations [184], which increase the work-hardening. However, this is not the only metallurgical feature involved in the increase of the work-hardening. For instance, in this stage could occur the activation of the cross-slip, which allow that primary and secondary slip systems interact with themselves, which increase the stress required to further dislocation movement [191]. On the other hand, dynamic transformation during deformation from austenite to martensite or twinning austenite (dislocation–mechanical twin boundary interactions [192]) (TRIP, TWIP, MBIP or DSBR) also produce an increase of the work hardening during the stage II due to the creation of effective obstacles inside the grain for the dislocations movement (Zhou *et al.* (2015) [193] have shown that deformation twins not only act as a barrier for dislocation glide but also as a site for dislocation accumulation and as a source of partial dislocations), producing a similar effect to the Hall-Petch effect, but in this case, the microstructure refinement is produced during deformation and hence, is termed *dynamic* Hall-Petch effect [52]. The static strain aging (SSA) and dynamic strain aging (DSA), which is mainly due to the dynamic interaction between solid solution and interstitial atoms with the dislocation cores (C-Mn pairs require small thermal energy [194] and are a strong obstacle for the dislocations movement) during deformation, is also responsible for the increase of the work-hardening during the stage II. In this sense, Shun *et al.* (1992) [195] have shown in two Fe-Mn-Al-C steels, that there was not a relation between twinning and the work hardening, rather, it was the DSA effect. Usually, the DSA is limited due to the exhaustion of the carbon atoms in the alloy which can potentially lock or pin mobile dislocations. However, they argued that if there is enough carbon content, the exhaustion of interstitial atoms is not presented and they can pin the mobile dislocations (recent insights of DSA phenomenon in iron alloys supported by *in situ* observation has been presented by [196]. Remarkably, Liang *et al.* (2016) [197] and Zhou *et al.* (2016) [198] showed in a Fe-18Mn-1.7Al-0.75C steel, the contribution of different metallurgical factors on the flow stress and work hardening (stage II) during the true stress vs true strain curve. From Figure 24, can be observed that the major contribution is caused by the dislocation structure evolution and the twin formation is almost negligible. They explain that in low SFE in TRIP steels the dislocation cores are wider which reduce or suppress the dislocation annihilation by cross slip and thereby promote dislocation accumulation. This result is important because shows evidence of other factors on the work hardening that can be more important than the twinning effect. The discussion around these phenomena is still open.

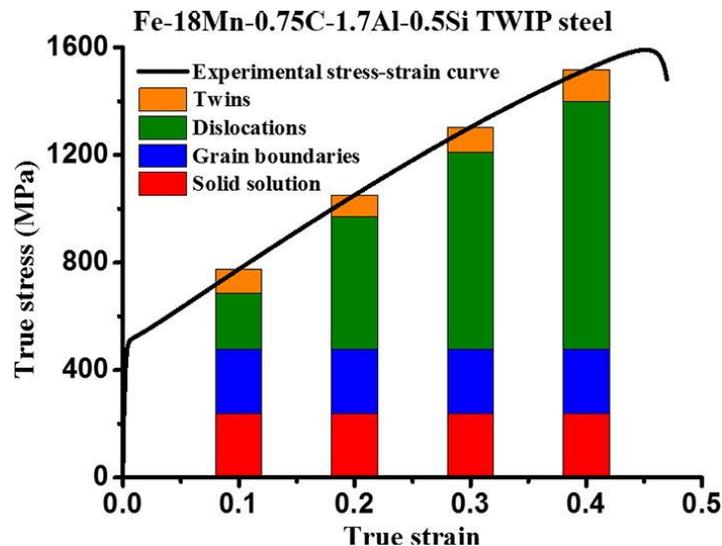

Figure 24. The respective contributions of twins, dislocations, grain boundaries and solid solution to the flow stress during plastic deformation, reproduced from [197] with permission of Elsevier[®]

In stage III, the dynamic recovery and the interaction between glide dislocations and stacking faults are dominant and are the main contribution to this plateau. This occurs despite the mechanical twin formation due to the low volume fraction 0.6%. In stage IV, the dislocation annihilation (i.e. recovery) rate is greater than the dislocation multiplication (i.e. hardening) rate induced by the mechanical twinning and/or stacking faults. This means a decrease in the work-hardening rate. On the other hand, in some cases or depending on the test temperature, and oscillation effect could be observed due to the DSA phenomenon [199]. Is important to have in mind, that other factors like the test temperature can change the work-hardening behavior (in general, a drastic decrease), avoiding some of the stages discussed previously [184]. On the other hand, most of the investigations about the strain-hardening rate has been focused on the first stages, however, during the last stages an interesting phenomenon occurs. That is, a sudden drop of the strain-hardening rate just before the necking. This phenomenon was recently addressed by Yang *et al.* (2017) [200], and it was concluded that macroscopic voids instead of dislocations annihilation or dislocations exhaustion were responsible for the prior-necking phenomenon.

Other factors could also influence the work-hardening rate in Fe-Mn-Al-C. For instance, Ding *et al.* (2014) [201] studied the effect of carbon on the mechanical properties of two alloys; i) Fe-18Mn-10Al-0.8C and ii) Fe-10Al-1.2C-18Mn. These alloys have high SFE, so the TWIP and TRIP reactions cannot be present. It is observed that despite no TRIP or TWIP effect was possible, a high plastic deformation occurs without fracture, as shown in Figure 25. Thus, the authors suggested that this behavior was due to the κ-carbides precipitates.

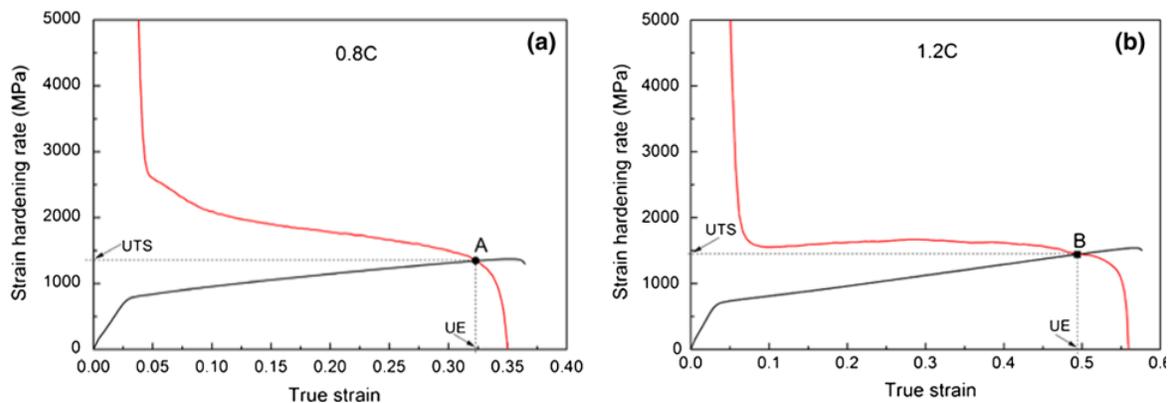

Figure 25. True stress-strain and strain hardening rate curves of the (a) 0.8C steel and (b) 1.2C steel annealed at 950 °C, reproduced from [201] with permission of Springer®

Moreover, this is one of the first research showing that κ-carbides improved the plasticity and the strain hardening coefficient as shown in Figure 26, where the steel with 1.2% carbon, with the high κ-carbides amount, showed the highest strain hardening at low and high plastic strains.

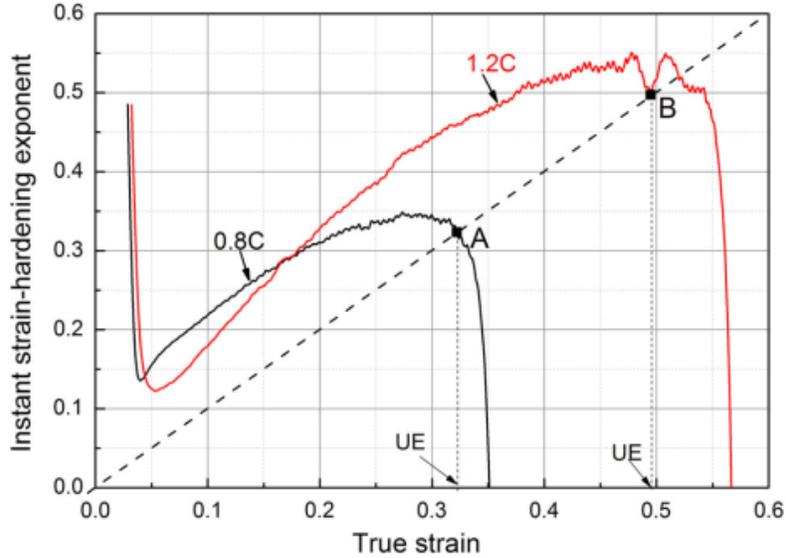

Figure 26. Instantaneous strain hardening coefficient (n) for (a) a Fe-18Mn-10Al-0.8C steel and (b) a Fe-18Mn-10Al-1.2C steel, both annealing at 950 °C, reproduced from [201] with permission of Springer®.

Besides of the carbon effect on the work-hardening. Hamada (2007) [202] showed the effect of aluminum additions in the work-hardening of the Fe-25Mn-xAl-xC alloys. In Figure 27 is evident that increasing the aluminum content from 0% to 8%, the ductility decreases drastically.

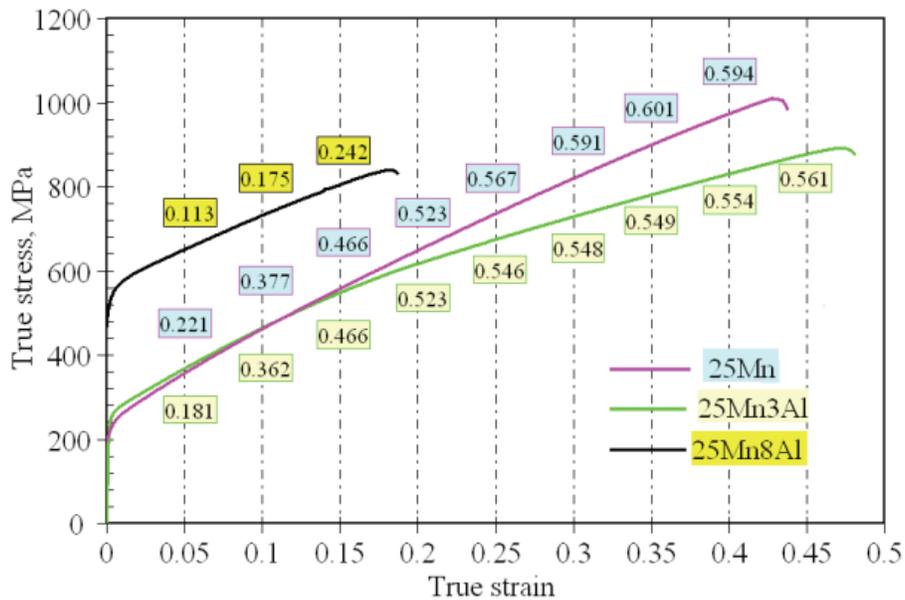

Figure 27. Strain-hardening exponents (n-values) from the Hollomon equation vs. true strain for three high-Mn TWIP steels at room temperature [202]

This trend of decreasing the strain hardening, is mainly due the aluminum decreases the C atoms diffusion and increases the SFE which restrict the deformation twinning. This behavior is consistent with the results of Lai *et al.* (1989) [203], Tian *et al.* (2013) [167], Hwang *et al.* (2011) [204] and Abbasi *et al.* (2010) [205]. However, discrepancies are still presented with this aspect. For instance, Canadinc *et al.* (2005) [206] showed that Hadfield steels with Al additions increases dramatically the strain hardening capacity due to the creation of dislocations walls of high density which prevent further slip [207]. Finally, it is important to note that these steels generally have strain hardening coefficients (n) between 0.4 to 0.9 [35]. However, coefficients as high as 1.4 could be obtained (Figure 28).

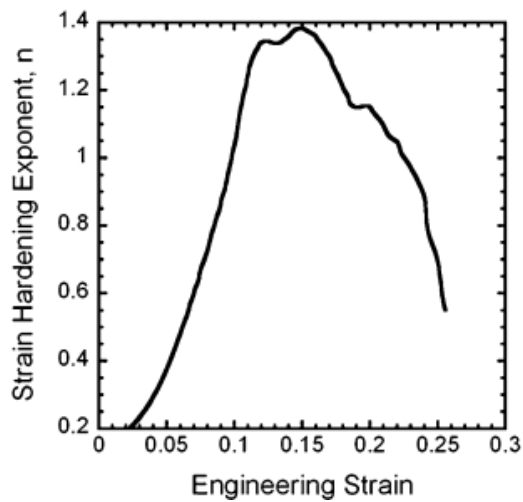

Figure 28. Strain-hardening exponent (n) for a Fe-15.3Mn-2.4Al-0.07C-2.85Si steel as a function of strain, reproduced from [208] with permission of Springer®.

Furthermore, as shown Hamada (2007) [202], the rate of hardening decreases with increasing the percentage of aluminum in the alloy, which depends on the prevailing mechanism of deformation, i.e. there is a direct dependency of the SFE. These results are consistent with the findings of Gutierrez-Urrutia *et al.* (2013) [53] and Jung *et al.* (2013) [153].

This year, Park *et al.* (2017) [209] showed in a Fe-15Mn-1.2Al-0.6C steel, the microstructural evolution under static and dynamic conditions (strain rates at 2000 s$^{-1}$) using tensile tests. They found that the dynamic tensile strength was higher than the quasi-static tensile strength. This was indirect inferred and supported by comparing the kernel average misorientation (KAM) in both microstructures. Basically, in a high KAM values (dynamic conditions) promote a higher tendency for slip and twinning than in the static conditions.

This phenomenon, i.e. the positive strain rate sensitivity, has been reported previously to operate in Hadfield steels [210] and low aluminum Fe-Mn-A-C steels [211] at lower strain rates range ($10^{-3}$ to $10^3$). In fact, the strain rate sensitivity has recently reported to change from negative to positive values as function of the test temperature [212] and Al content (SFE) [213]. Finally, I would like to suggest the interesting reviews performed by De Cooman *et al.* (2017) [214] and Bouaziz *et al.* (2011) [215], to the reader that wants to expands the concepts showed here about the TWIP phenomenon in high manganese steels and its role in different phenomena.

## *3.3 κ-carbide hardening*

It is well known that the κ-carbides increase the opposition or difficulty to the sliding of dislocations, which increase hardness, yield strength, and ultimate tensile stress. One of the most important works related to the effect of the κ-carbides in the Fe-Mn-Al-C system was performed by Springer *et al.* (2012) [63]. They performed aging treatments at different aging times and temperatures, with the aim of increasing the κ-carbides content in the austenite. The summary of the mechanical properties as a function of the Al content for each aging treatment is shown in Figure 29.

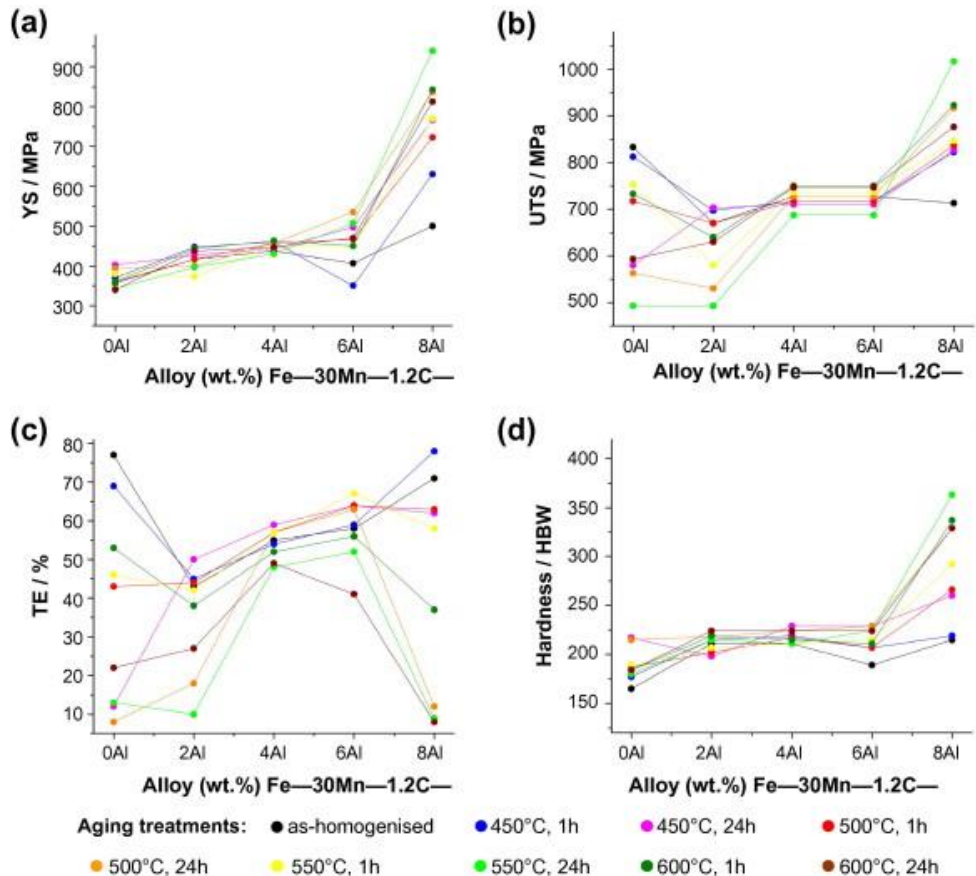

Figure 29. Overview of the mechanical properties obtained from the RAP experiments as a function of alloy composition and the applied aging treatment: (a) yield stress (YS); (b) ultimate tensile stress (UTS); (c) total elongation (TE); (d) hardness. RAP, rapid alloy prototyping. Reproduced from [63] with permission of Elsevier®

There is a smooth increase of the yield strength with the increase of the Al content until 4Al, which is caused by the solid solution strengthening. In this scenario, the effect of the aging treatment (κ-carbide precipitation) is negligible. However, once the Al content is above of 4Al, the yield strength increases drastically (except for the lowest aging temperature) due to the κ-carbide precipitation in the austenite. It is important to remember that the κ-carbide has a close chemical composition of $(Fe,Mn)_3AlC$ (*see Section* 2.3), and it is evident that these carbides require a high amount of Al and C atoms to be formed. This explains why before 4Al these precipitates did not form. On the other hand, there is a substantial increase of the yield strength due to the increase of aging time for a constant aging temperature.

An interesting debate rise during the last years related to κ-carbides was to reveal the exact deformation mechanism that operates in precipitate hardened Fe-Mn-Al-C steels. That is, how κ-carbides interact with the dislocations during plastic deformation. Initially, it was argued the κ-carbides were sheared by gliding dislocations [216, 217], later some observations and calculations suggested that the predominant deformation mechanism was the Orowan bypassing of κ-carbides [55]. Recently, it was shown that effectively the κ-carbide precipitates are primarily sheared by gliding dislocations as is shown in Figure 30, and also an interesting contribution is to show the fragmentation and dissolution phenomenon of κ-carbides at high strains [67].

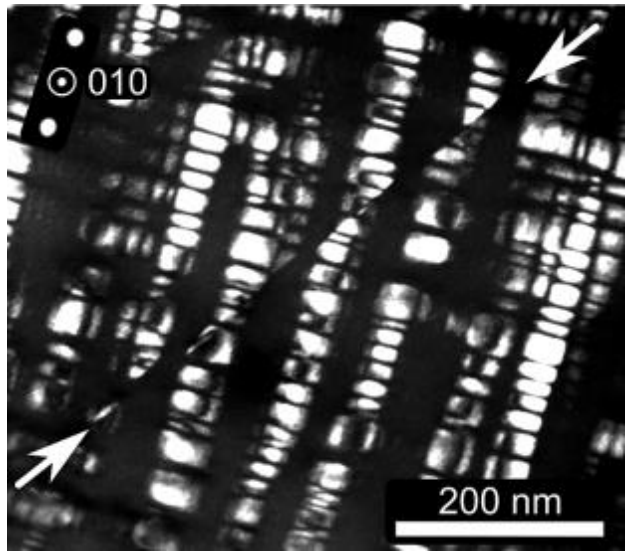

Figure 30. DF-TEM images showing sheared κ-carbides at a true strain of 0.02, reproduced from [67] with permission of Elsevier®

## 3.4 Microstructural evolution of Fe-Mn-Al-C steels during the σ vs ε curve

One of the main features of the Fe-Mn-Al-C steels is the possibility to get a great strain hardening as well as an increase of elongation simultaneously. This behavior is basically due to the special microstructural evolution of this kind of steels. An interesting systematic research was performed by Hua *et al.* (2013) [218], who studied the effect of the aluminum content ( an increase of the SFE) on the microstructural evolution for the Fe-26Mn-XAl-1C system. They showed that with the lowest aluminum content (SFE of 37 mJ/m$^2$, preponderant twinning regime) the microstructure at a deformation of 5% starts with a dislocation tangles arrangement, then the slip traces (white arrow) are formed at a deformation of 10% with the presence of dislocation tangles, and at a deformation strain of 30% the deformation twins can be appreciate (corroborated by the selected area diffraction pattern), finally, at the fracture of the sample, a high population of primary and secondary twins are found as is shown in Figure 31.

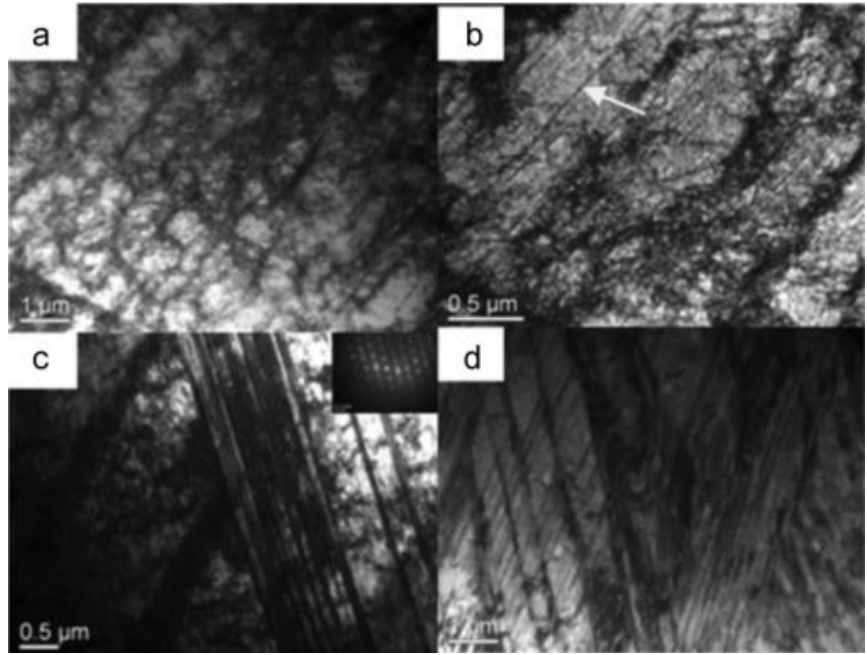

Figure 31. Microstructures (TEM) of a Fe-26Mn-3Al-1C steel after deformation (a) 5%, (b) 10%, (c) 30%, (d) fractured, reproduced from [218] with permission of John Wiley and Sons®

This results are consistent with the works performed by Hwang *et al.* (2015) [219] and Yang *et al.* (2013) [220] in a Fe-20Mn-1Al-0.6C (TWIP, 23.5 mJ/m$^2$) and a Fe-22Mn-1.5Al-0.6C (TWIP, 30 mJ/m$^2$) steel, respectively where there are two conditions; (i) a decrease of the twins inter-spacing as the deformation is increased how is shown in Figure 32 (a) and (ii) an increase of twinned grains fraction as the deformation is increase how is shown in Figure 32 (b).

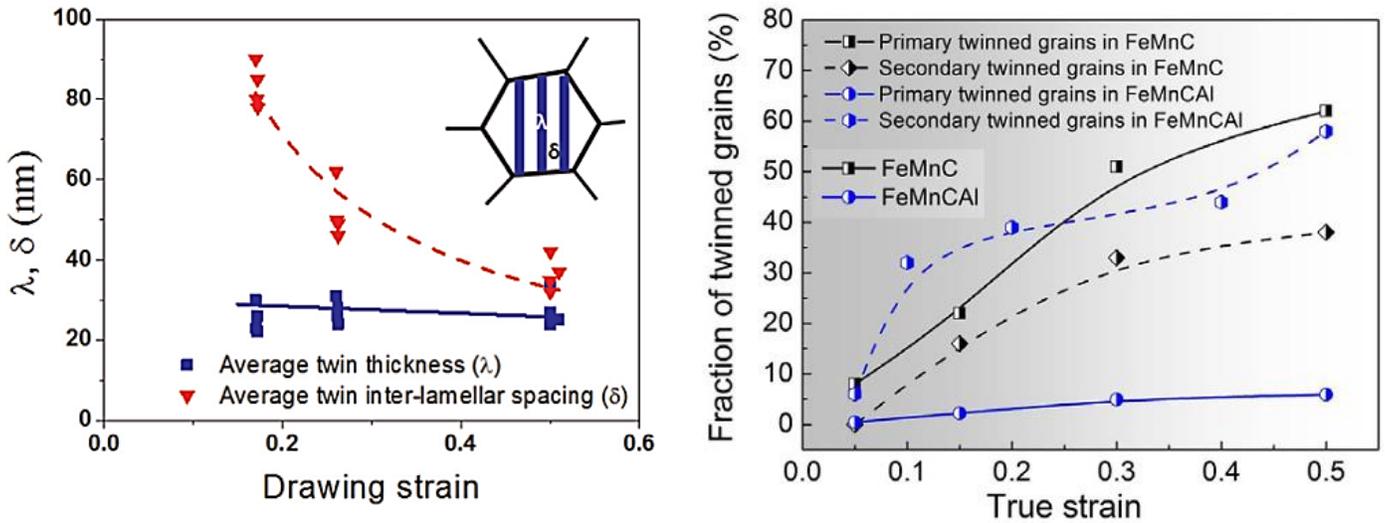

Figure 32. Variations in the (a) average twin thickness (λ) and twin inter-lamellar spacing (δ) in twin bundles withdrawing strain, reproduced from [219] with permission of Elsevier® and (b) fraction of twinned grains with the true strain, reproduced from [220] with permission of Elsevier®

Continue with the work of Hua *et al* (2013), for a 6% of aluminum content (SFE of 60 mJ/m$^2$, preponderant microband regime) the microstructure at a deformation of 5% starts with planar slip, then as the deformation increase the dislocations are arrays with similar spacing between them along the main directions. Besides, the dislocation densities increased without change the slip directions. At the fracture point, the sample develops a well-identified microbands as is shown in Figure 33.

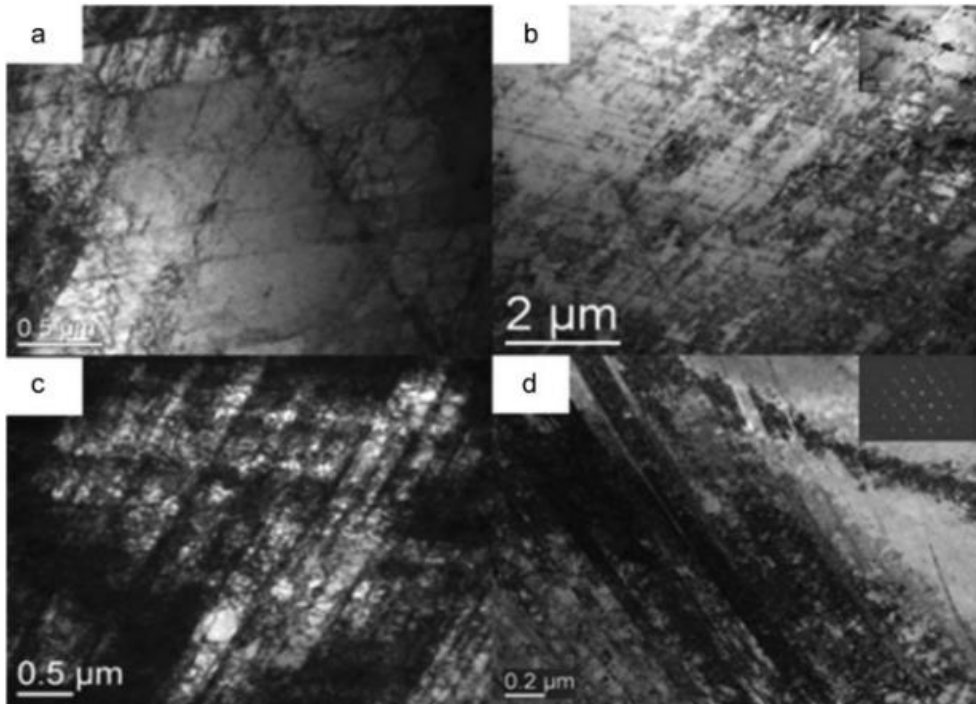

Figure 33. Microstructures (TEM) of a Fe-26Mn-6Al-1C after deformation (TEM) (a) 5%, (b) 10%, (c) 30%, (d) fractured, reproduced from [218] with permission of John Wiley and Sons®

For the highest aluminum content of 12% (SFE of 96 mJ/m$^2$, preponderant microband regime) the microstructure at a deformation of 5% starts with a uniformly arranged of a high number of slip bands, then the density of slip bands increase as the deformation increase until 10%. As deformation proceeded until 30% and fracture, the slip bands of high density are found and also intersecting each other, as is shown in Figure 34.

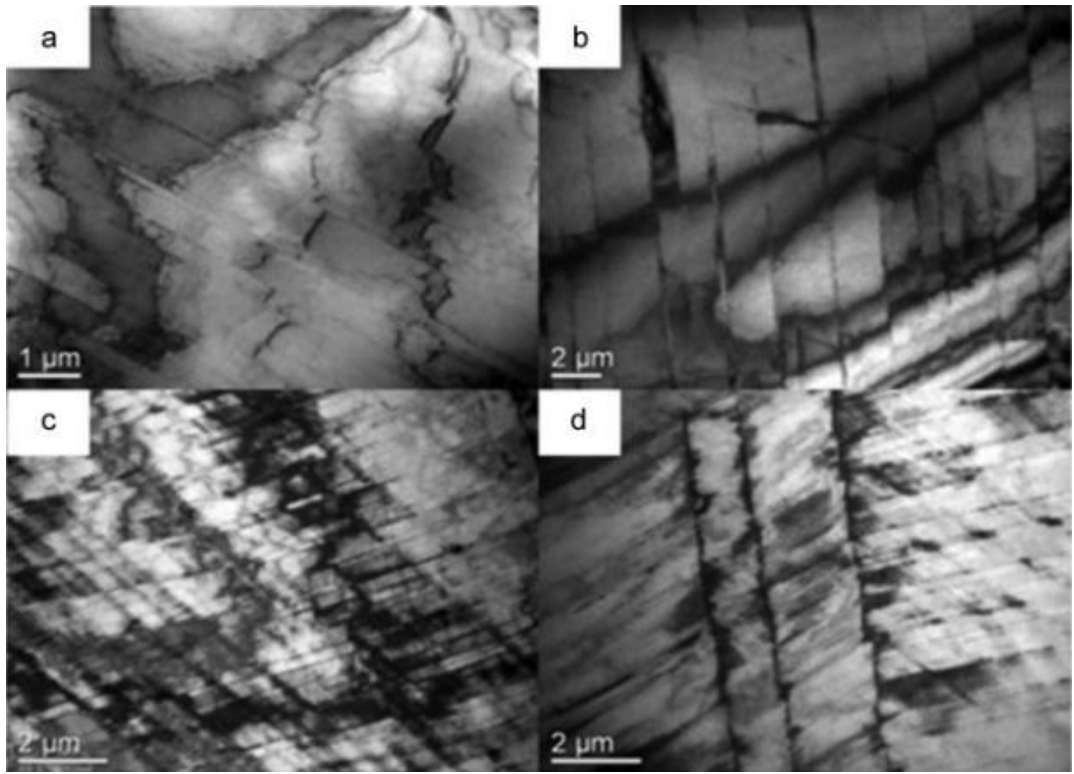

Figure 34. Microstructures of 12Al steel after deformation (TEM) (a) 5%, (b) 10%, (c), and (d) 30%, reproduced from [218] with permission of John Wiley and Sons®

Recently, Galindo-Nava *et al.* (2017) [221] propose a framework to predict the evolution of the ε, $\alpha'$ and twins with the strain and as a function of the SFE. To this end, it was formulated a mechanism for the ε and twin formation, based on the formation and overlapping of stacking faults to form micro-bands. Additionally, they modify the energy of embryo using the well-known Olson and Cohen formulation to include the SFE parameter. Interestingly, they formulation allows to determine the activity of ε, $\alpha'$ and twins not only in terms of the SFE but also in terms of the strain level as is shown in Figure 35. This is an important achievement because stablished an unified description for the evolution of ε and $\alpha'$ martensite, and as well as twinning in austenitic steels.

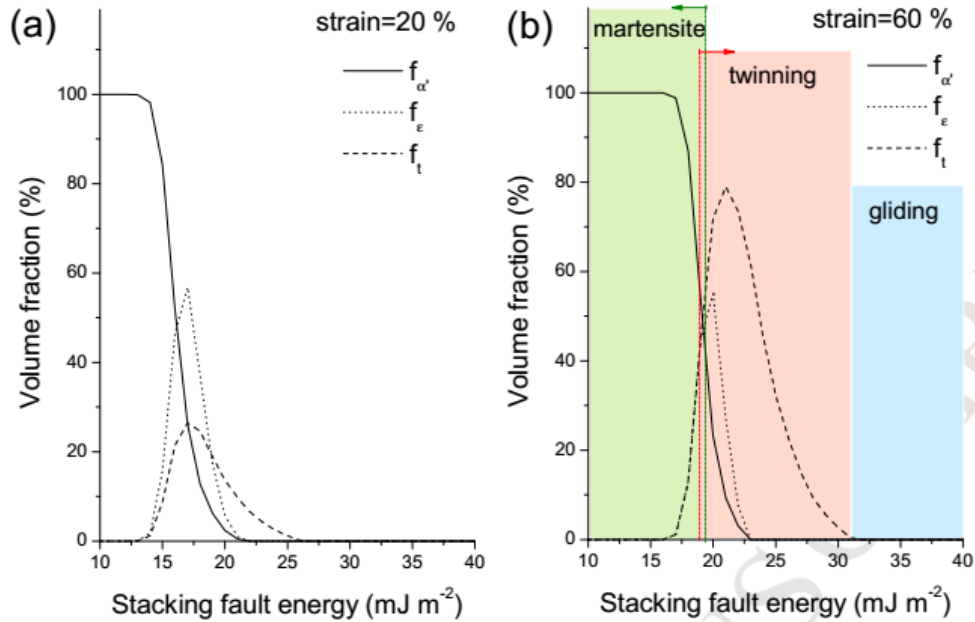

Figure 35. Variation in the volume fraction of the $\alpha'$, $\varepsilon$ and twinning for different SFE at (a) $\varepsilon=20\%$ and (b) $\varepsilon=60\%$, reproduced from [221] with permission of Elsevier®

The previous model was focused on austenitic steels in general, which can be, in principle, apply for austenitic Fe-Mn-Al-C steels. Some recent efforts have been made particularly in Fe-Mn-Al-C [222] steels, where it was obtained a representative kinematic model which takes into account the role of the SFE on the deformation twinning and martensitic transformation as well as the deformation temperature, and chemical compositions.

## 3.5 Fatigue behavior

One of the first systematic studies in the fatigue of Fe-Mn-Al-C were performed by Chang *et al.* (1989) [223], they studied the Fe-28Mn-9Al-XC system in fatigue bending tests. For this purpose, it was studied three alloys (A: γ, B: 10%α+90% γ, and C: 55%α+45%γ) and the results are depicted in Figure 36. The figure shows a small difference between the S-N curves between the A, B and C steels. Due to the usual statistical nature of the fatigue data, it is concluded that the three steels have approximate equal fatigue life under equal stress amplitude. Nevertheless, a slight improvement is observed for the steel B (10%α+90% γ), this improvement is attributed to the existence of the small grains of ferrite, which work as fatigue crack retarders. For the other two specimens, the fatigue cracks initiate in the slip bands inside the austenite

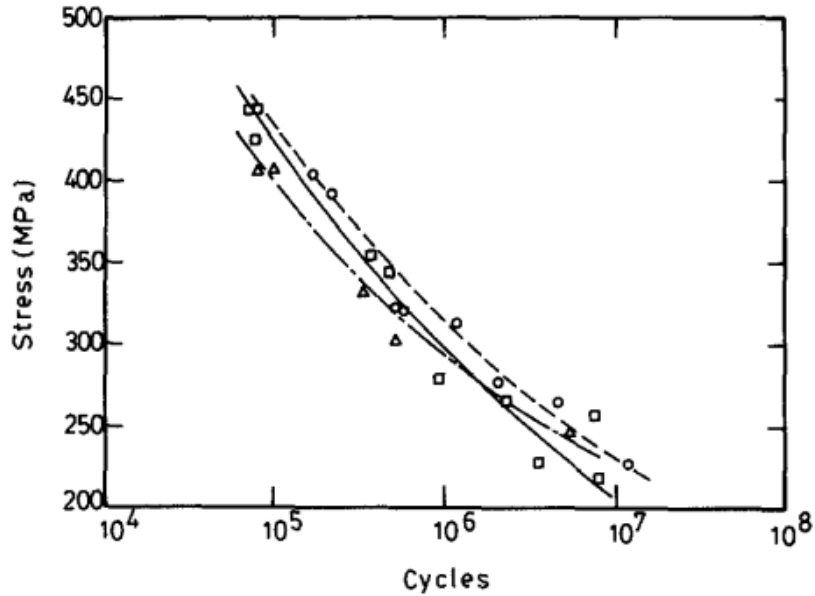

Figure 36. S-N curves of the Fe-Mn-Al-C specimens (A: □, B: ○, and C:Δ) in reverse bending fatigue test, reproduced from [223] with permission of Springer®

Hamada *et al.* (2009) [224] studied a high manganese TWIP steel in high cycle fatigue (HCF) conditions and studied crack nucleation and propagation processes. It was found that this alloy has a fatigue strength limit between 42% and 48% of ultimate strength, very similar to the fatigue behavior of austenitic stainless steels as is shown in Figure 37.

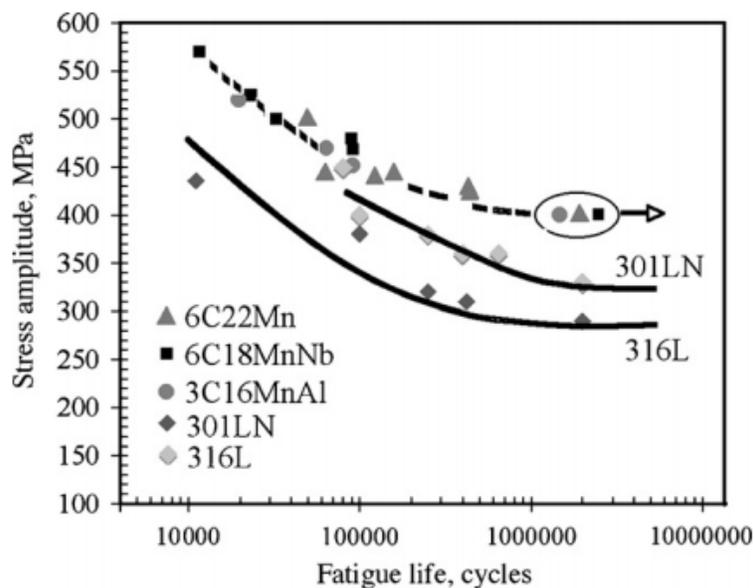

Figure 37. Fatigue behavior of some austenitic and Fe-Mn-Al-C steels, reproduced from [224] with permission of Elsevier®

In addition, an early appearance of cracks and a tendency to nucleation on grain and twin were identified, which impairs fatigue resistance. A microscopic analysis revealed that neither martensitic transformation nor twinning was produced, concluding that there is no contribution of these deformation mechanisms and therefore there is no evidence of an improvement in fatigue properties. Similar analysis and results were obtained in HCF using an austenitic TWIP steel by Hamada *et al.* (2010) [225], who found that the cracks propagate by the twins and grain boundaries, as well as slip bands, as is shown in Figure 38.

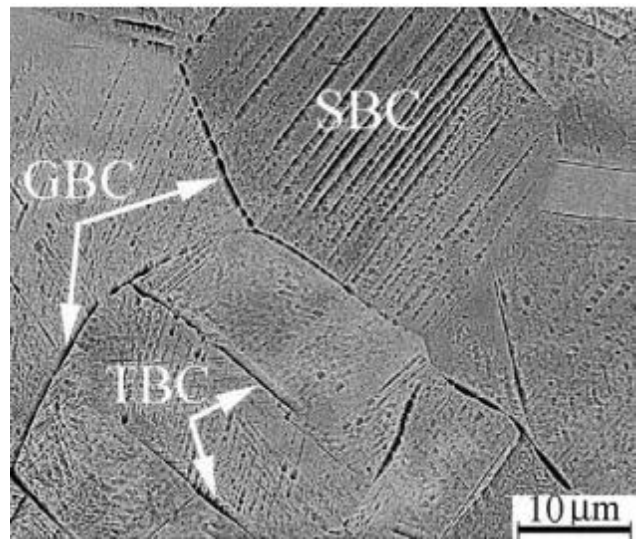

Figure 38. SEM photo of a twip steel after fatigue with a stress amplitude of 500 MPa for 50,000 cycles. Numerous slip band cracks (SBC), grain boundary cracks (GBC) and cracks along twin boundaries (TBC), reproduced from [225] with permission of Elsevier®

In contrast, the TRIP effect causes an excellent performance of the material subjected to cyclic plastic deformation [226]. This is mainly because the martensitic transformation delays the propagation of the crack by absorbing the energy delivered to the material in each cycle avoiding the accumulation of the damage, as was found by Cheng *et al.* (2008) [227]. It has also been found that this transformation is localized and only occurs within the monotonic deformation zone and depends on the rate of deformation and stability of the retained austenite [226, 228].

The fatigue strength of TRIP steels depends strongly on the rate of transformation, which is directly proportional to the SFE of the austenite. It has been shown that a more stable retained austenite produces a more gradual transformation, and the best properties are obtained under fatigue [229, 230]. Figure 39 presents the stress-life curves of a material that when subjected to two different thermal treatments has different levels austenite of stability (SFE). The values in parentheses indicate the percentage of transformed austenite, from which it can be concluded that the steel with higher SFE has the best fatigue behavior [231].

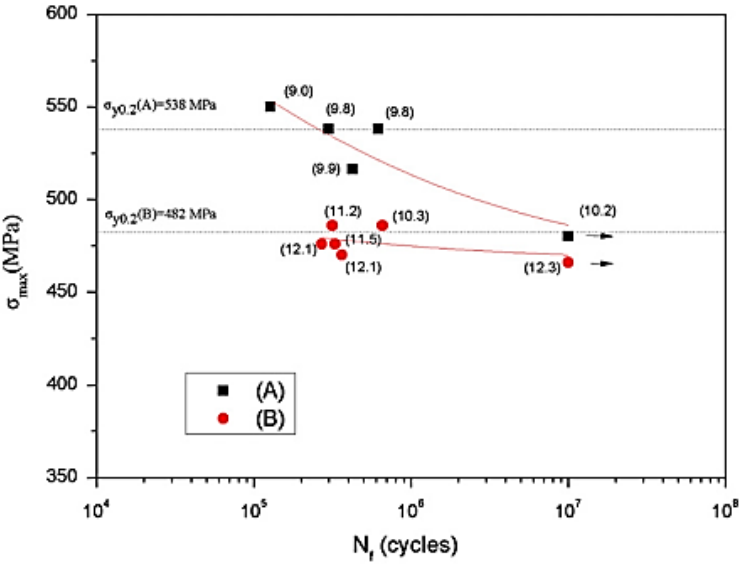

Figure 39. S–N curves of TRIP steel with A and B treatment. The volume fraction of retained austenite after the test is shown in parentheses, reproduced from [229] with permission of Elsevier®

The martensitic transformation occurs according to the level of deformation applied and the stability of the retained austenite. The research of Haidemenopoulos *et al.* (2013) [229] compared the stress-strain curves of a material with different degrees of austenite stability (Figure 40). The condition with less stable austenite exhibits a lower creep stress due to the softening that is carried out because of a very prominent transformation, and in turn, exhibits a higher ultimate stress due to the higher initial strain hardening rate. The relationship between the ultimate stress and the fatigue strength limit should be highlighted. In general, a steel with a higher ultimate stress than the other will also present a higher fatigue strength limit and will be approximately 50% of the ultimate stress.

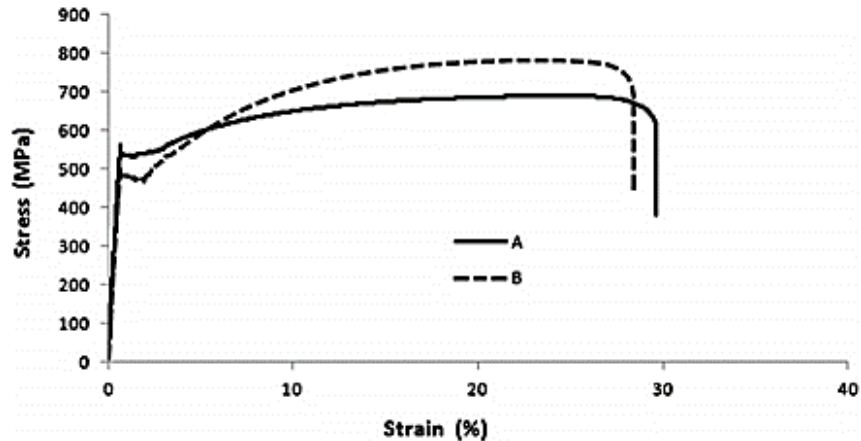

Figure 40. Engineering stress-strain curves in TRIP 700 steel with two treatments (a) and (b), reproduced from [229] with permission of Elsevier®

However, the condition that presented the lowest ultimate strength was the one that obtained a greater limit of resistance to the fatigue. This is due to the greater stability and more gradual transformation, highlighting the high value of the resistance limit to fatigue, which corresponds to 70% of the last material effort, which is much higher than the values reported in the literature for the conventional ferrous alloys.

There are few investigations on the fatigue strength of Fe-Mn-Al-C steels. Aspects like the effect of the deformation mechanisms in the crack propagation and fatigue strength are yet unclear. In this regard, some recent studies tried to elucidate this effect in Fe-Mn-C and Fe-Mn-Al-C steels. For instance, the work of Shao *et al.* (2016) [75] in LCF and ELCF (extremely low cycle fatigue) regimes were studied for Fe-30Mn-XC (X= 0, 0.3 and 0.9), obtaining the next conclusions; the C content decreases the capability of cyclic plastic deformation, causing the concentration of fatigue damage and induce faster crack propagation speed. Moreover, as the C content increases, the substructure evolution changes during low cycle fatigue (LCF) regime from twins and stacking faults to only dislocations bands without any evidence of stacking faults or twins. The absence of this features in the high carbon content samples promotes the fatigue damage and propagation.

Ju *et al.* (2016) [232] evaluate using *in-situ* observations during the LCF regime, the crack growth of three manganese steels (one of them a Fe-Mn-Al-C steel). Interestingly, it was found that the ε-martensitic transformation reduced the strain localization, and promote a zigzag crack propagation path, which promotes the roughness-induced crack closure. On the other hand, Yang *et al.* (2017) [233] showed in two austenitic Fe-Mn-Al-C steels (with and without Al additions), the effect of the SFE in the LCF regime. It was found that the Fe-22Mn-0.6C steel (SFE: 21.5 mJ/m$^2$) showed an increased resistance to fatigue crack growth i.e. the crack closure effect was highly pronounced due to the twinning phenomenon. In contrast, the Fe-22Mn-0.6C-3Al steel (SFE: 27 mJ/m$^2$) showed a less plasticity-induced crack closure effect. Similar results were obtained by Ma *et al.* (2015) [234] and recently by for two Fe-Mn-Al-C steels (with and without Al addition), where the Al additions significantly lower crack growth resistance and the fatigue lifetime, respectively.

Recently, Seo *et al.* (2017) [235] studied a TWIP steel in an HCF regime using two test temperatures; room and cryogenic temperatures. They found as is shown in Figure 41(a) that the fatigue resistance in the HCF regime increase as the temperature decreases. Since the strength of metal increases with decreasing temperature, this behavior is usually expected. But, the most important finding is that fatigue resistance in the HCF regime is directly correlated with the tensile strength (instead of yield strength) of the alloy, as is shown in Figure 41(b). They also found that the nucleation of fatigue cracks occurred in grain boundaries and slip bands.

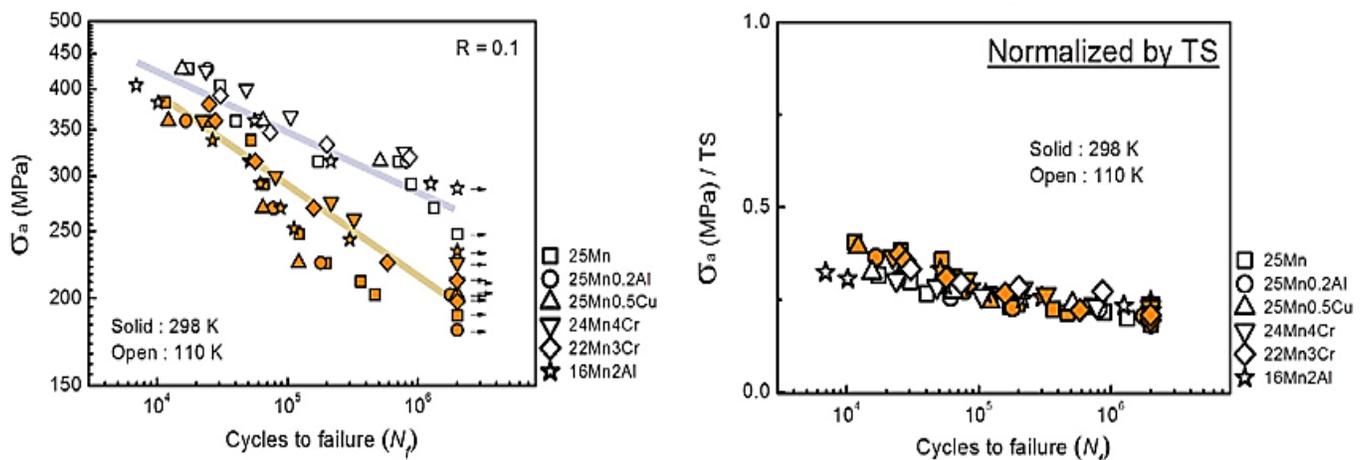

Figure 41. The S–N fatigue curves of high-Mn steels (a) at 298 and 110 K, and (b) date normalized with the tensile strength, reproduced from [235] with permission of Elsevier®

In summary, features like the high elongation at fracture of these materials and the deformation mechanisms allows a more controlled failure of the components instead of a brittle and catastrophic failure, making the Fe-Mn-Al-C steels alloys with a great potential in the industry. That is, bearing in mind that the economic growth of industry and companies claim in turn to impose new and more demanding conditions of process on materials.

Recently, Habib *et al.* (2017) [236] studied the role of the Mn-C pairs (a recent study in Fe-Mn-Al-C steels using *ab initio* methods showed that the Mn-C pairs are energetically favorable [237]) on the fatigue crack growth in austenitic steels. First of all, they performed tension test for two alloys (Fe-Cr-Ni and Fe-Mn-C steels), and they observed in the stress vs strain curve, the serration phenomenon in the Fe-Mn-C steel, this is well-known to occurs due to the DSA phenomenon. Then, they measured the increase of the crack length by the increase of fatigue cycles, as is shown in Figure 42. It was proposed that the fatigue limit was enhanced through DSA associated with the formation of Mn-C couples which caused a local hardening.

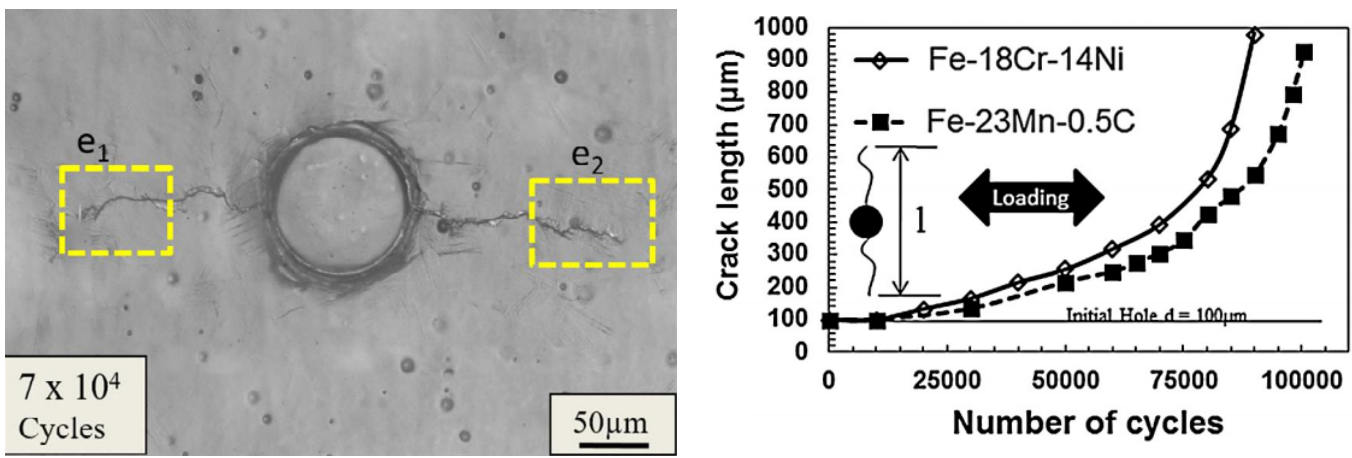

Figure 42. (a) Optical micrographs of the Fe-18Cr-14Ni steel during the fatigue test (b) crack length versus number of cycles for the two steels at the same ratio between stress amplitude and tensile strength, reproduced from [236] with permission of Elsevier®

To close, a recent work about the fatigue behavior of Fe-Mn-Al-C steels showed the synergistic effect of aluminum additions as well as the hydrogen embrittlement [238]. It was found that under cyclic loading at high strain amplitudes, the Al increases the LCF life. Conversely, at low strain amplitudes the Al decreased LCF life.

# 4 Wear

## 4.1 Abrasive wear

To date, there are very few tribological studies on Fe-Mn-Al-C alloys. First, some investigations carried out on Hadfield steels with additions of aluminum, and later Fe-Mn-Al-C steels will be discussed. One of the first tribological studies of these steels was performed by Zuidema *et al.* (1987) [239] who evaluated the effect of adding aluminum and carbon to a Hadfield steel on abrasive wear resistance to two and three bodies as observed in Figure 43 and Figure 44, respectively.

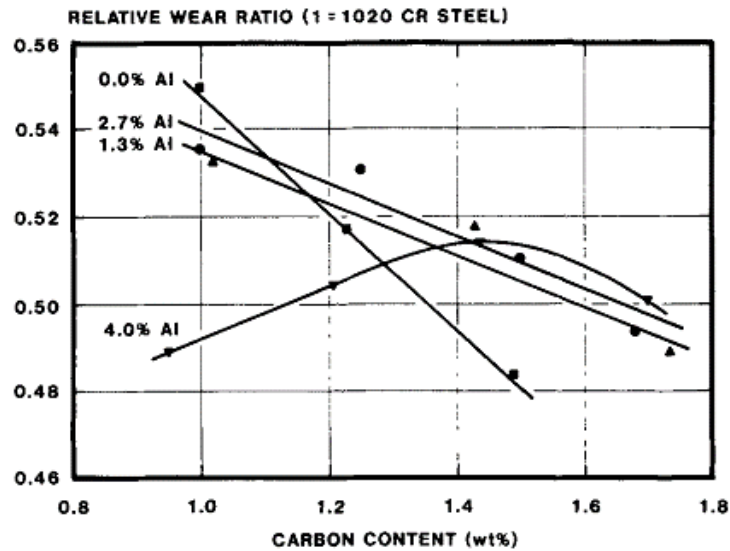

Figure 43. Effect of the carbon and aluminum content on the relative wear resistance in the high-stress pin-abrasion test, reproduced from [239] with permission of Springer®

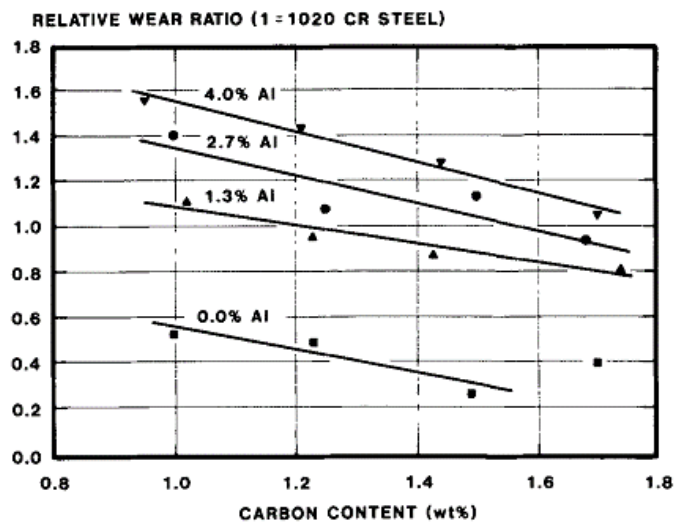

Figure 44. Effect of the carbon and aluminum content on the relative wear resistance in the rubber wheel abrasion test, reproduced from [239] with permission of Springer®

In high-stress wear (pin-abrasive paper) the samples with the lowest carbon content, the increase of Al content increases the wear resistance of the alloys. The authors suggest that this result is due to the improvement in carbon solubility in austenite due to aluminum content, which substantially increases the solid solution hardening of the austenite. For this reason, a higher C content, the wear resistance is evident increased. On the other hand, for low-stress wear (rubber-wheel) the increase of aluminum drastically reduces the wear resistance, mainly due to the formation of coarse carbides during aging. The authors recommend looking for other chemical compositions that guarantee a homogeneous precipitation and a high dissolution of the carbon in the matrix.

In the investigation of Acselrad *et al.* (2004) [240], the abrasive wear resistance was evaluated and compared by a micro-abrasion equipment of Fe-Mn-Al-C steels (Fe-28Mn-8,5Al-1C), M2 and AISI 304. This investigation determined that, in the solubilized state, Fe-Mn-Al-C steel was worn 25% faster than M2 tool steel. On the other hand, the aging treatment performed at 550 °C for 16 hours increased the hardness and strength of the alloy, however, the wear rate was 55% higher compared to M2 tool steel. However, when was used a controlled cooling treatment of 700 °C to 550 °C in the Fe-Mn-Al-C (L2CC) alloys, the abrasion resistance was comparable to that of AISI 304 stainless steel and M2 tool steel as is shown in Figure 45.

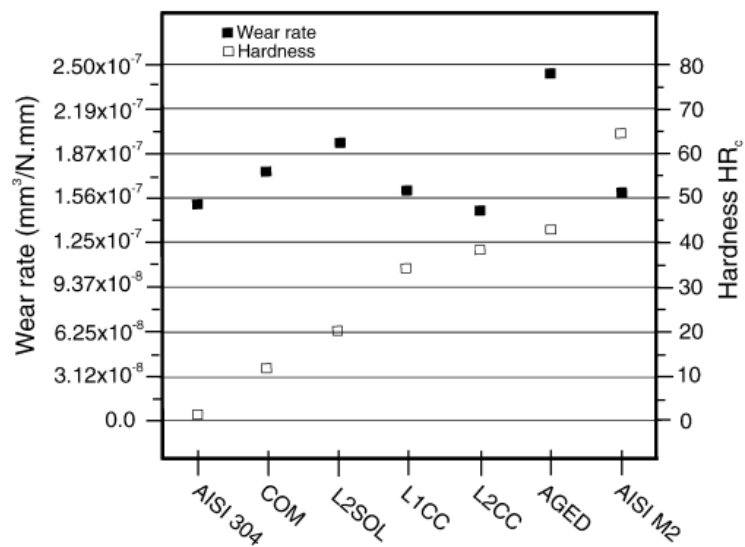

Figure 45. Wear rate and hardness for the different Fe-Mn-Al-C alloys and treatments, reproduced from [240] with permission of Elsevier®

A methodical study on the abrasion resistance of these steels was carried out by Van Aken et al (2013) [23], who determined the effect of carbon and aluminum on abrasive wear resistance in high manganese steels. For the wear analysis, the lost volume of the alloy was used and normalized with the lost volume of an AISI 1020 hardness steel 137 BHN. Several important results are concluded; (1) The solution condition (ST) offers better wear resistance results than the same alloys in the aged (solution treated and aged) condition, (2) the carbides-κ precipitated at the grain boundaries, generating excessive brittleness (3) increase of aluminum favors the increase of volume removed and (4) the increase of the carbon content favors the increase of the resistance to the wear. This is shown in Figure 46.

It is important to note that in this investigation the precipitation of the κ-carbides was produced by the conventional treatment of quenching + aged, where it was observed that the carbides precipitate along grain boundaries, thus causing brittleness.

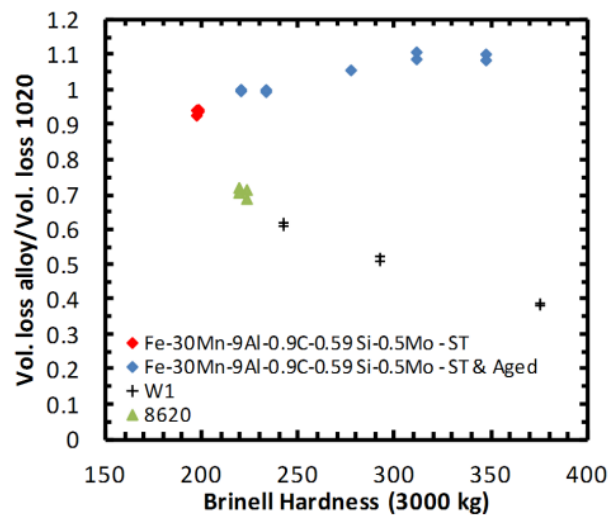

Figure 46. Wear behavior of Fe-Mn-Al-C steels varying the thermal treatment, and comparing with some commercial steels [241]

Mejía *et al.* (2013) [242] studied under dry sliding conditions (''pin-on-ring'') the wear behavior of a Fe-22Mn-1.5Si-1.5Al-0.4C TWIP steel with and without Nb additions. They found that the sliding velocity changes drastically the wear resistance for both steels, due to the formation of a protective iron oxide layer (mainly magnetite and hematite). On the other hand, the Nb additions seem not improve the wear resistance, mainly due to the presence of NbC precipitates, which prompt detachment of the oxide layer.

Another interesting work was performed by Ramos *et al.* (2015) [243] who studied the abrasive wear of a Fe-29.0Mn-6Al–0.9C-1.8Mo-1.6Si-0.4Cu steel in "as cast" condition. In this work, they found that the austenite was transformed into martensite during the wear process. They used X-ray diffraction and Mössbauer spectroscopy to quantify and identify the austenite and martensite content. This transformation was responsible for the increase in samples' hardness. This behavior has not been previously reported and special attention from researchers is suggested to this phenomenon.

In the recent work of Peng *et al.* (2016) [244], they studied the wear behavior of a Fe-25.1Mn-6.6Al-1.3C steel and the effect of aging treatments under impact abrasion conditions. The main results are shown in Figure 47.

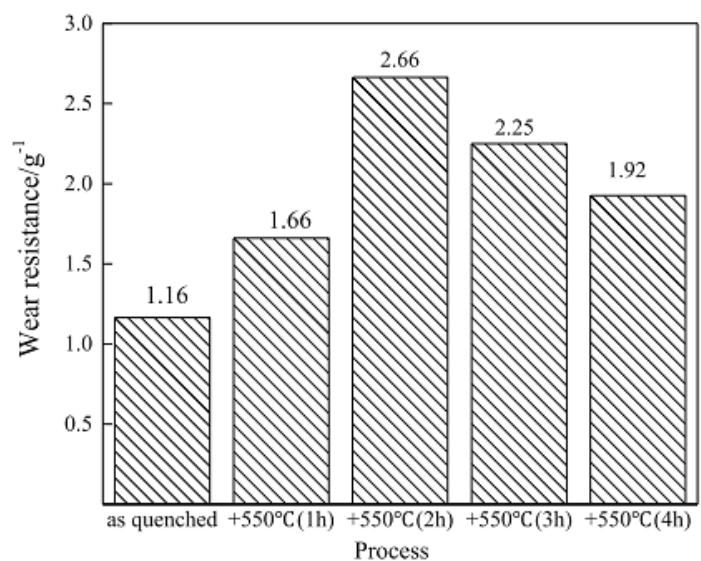

Figure 47. Wear resistance Fe–25.1Mn–6.6Al–1.3C steel with different aging treatments, reproduced from [244] with permission of Springer®

They explain that the (Fe, Mn)$_3$A1C carbides act as obstacles for the grooving abrasives and reduce the mass loss, leading to an increase of the wear resistance. However, when the aging time is longer than 2 h, the (Fe, Mn)$_3$A1C carbides get coarse and leads to its microcracking, and then this breaking carbides act as additional abrasive particles in the tribosystem, which decrease the overall wear resistance of the alloy. They validate that the nanoscale κ-carbide precipitates increase the critical stress required for dislocation slip and thus restrict slip band formation at the initial stages of plastic deformation in aged conditions. In this sense, they found two important scenarios depending on the volume content of κ-carbides in the aged samples; (i) when κ-carbides appear massively, there is a high chance for dislocations meeting with κ-carbides and shearing them. In such a case, the deformation will easily proceed via the formation of a small number of slip bands since the number of already sheared κ-carbide precipitates is large. (ii) when κ-carbide appear in fewer proportions and dimensions, a higher slip bands are required to accommodate the plastic strain in the samples. Such difference in the activity of slip band formation between the specimens subjected to different aging treatments can explain the decrease in the work-hardening rate in the Fe-25.1Mn-6.6Al-1.3C steel in aging condition.

One of the last published works related to the abrasion of these steels was carried out by Zambrano *et al.* (2016) [245], who studied the role of stacking fault energy (SFE) in the two-body abrasive wear of austenitic steels. Using a pin-abrasion test with 220 grit garnet paper as the counterbody, and three austenitic steels with different SFE were compared. The steels were: (i) FeMnAlC (medium SFE), (ii) Hadfield steel (lowest SFE), and (iii) AISI 316 L (highest SFE) steel. They showed that the FeMnAlC steel had a higher wear resistance than AISI 316 L steel but lower wear resistance than the Hadfield steel. However, at the highest test load, all three steels had similar wear resistance. The steel with the lowest SFE had the highest abrasive wear resistance and the steel with the highest SFE had the lowest abrasive wear resistance as is shown in Figure 48. In these conditions, they found that the main wear mechanisms were microcutting and microploughing, and a transition from microploughing to microcutting occurred as the normal load was increased. It is interesting to point out that these results were obtained without a clear correlation with the initial or worn hardness of the steels. So, the authors suggest that the SFE could be a more general parameter to characterize the wear resistance in austenitic steels. However, more research is necessary to validate this statement.

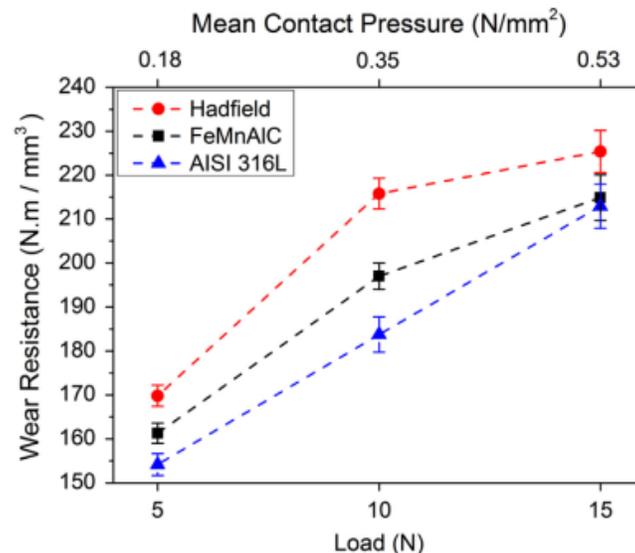

Figure 48. The effect of the normal load on the wear resistance for the FeMnAlC steel, Hadfield steel, and AISI 316L steel (the bars correspond to the interval of confidence at 95%), reproduced from [245] with permission of Elsevier®

To end this section about the abrasive wear of Fe-Mn-Al-C steels, it is important to name that some recent results showed a better wear resistance of Fe-Mn-Al-C alloys in comparison with conventional Hadfield steels [246]. Actually, this year was patented a series of Fe-Mn-Al-C alloys with improved wear resistance and impact toughness [247].

## *4.2 Cavitation wear*

In later years, Chang *et al.* (1995) [248] determined that for Fe-30Mn- (8-10%) Al- (0.2-1%) C alloys having the highest hardness after the treatment of solution, presented better resistance to cavitation in comparison to AISI 304 stainless steel (Figure 49). Nevertheless, the aging treatment did not improve the resistance to cavitation.

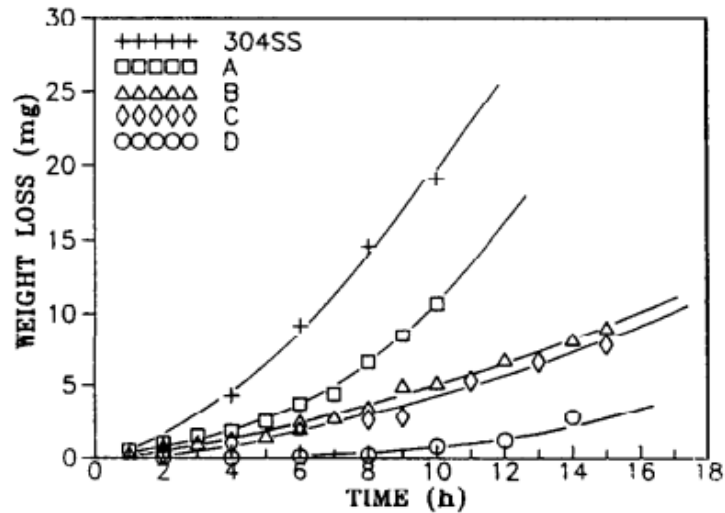

Figure 49. The accumulative weight loss of Fe-Mn-Al alloys (A-10%Al, B-6.72%Al, C-8.38%Al, D-8.35%Al+2.6%Cr) and 304 stainless steel in distilled water, reproduced from [248] with permission of Elsevier®

On the other hand, the cavitation resistance of the Fe-Mn-Al-C alloys showed higher mass losses in the more severe conditions (saline medium), almost independently of the thermal treatments used (Figure 50).

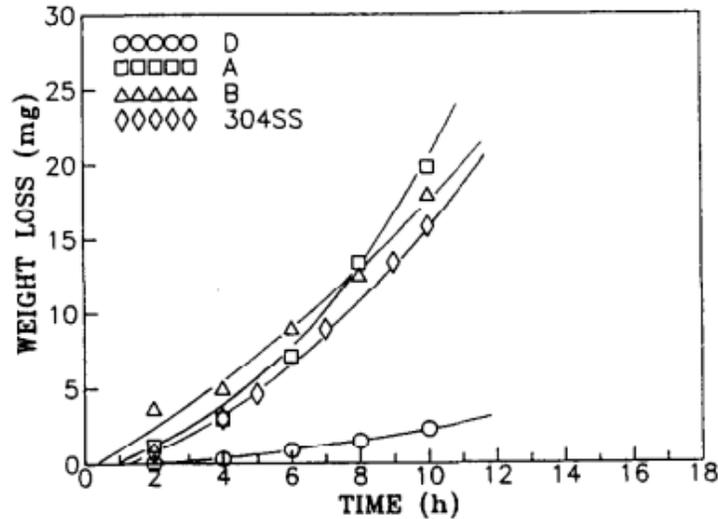

Figure 50. The accumulative weight loss of Fe-Mn-Al alloys (A-10%Al, B-6.72%Al, C-8.38%Al, D-8.35%Al+2.6%Cr) and 304 stainless steel in 3.5% NaCl solution, reproduced from [248] with permission of Elsevier®

However, steel with addition of Cr (Fe-Mn-Al-2.6% Cr) showed best results, even more than the AISI 304 steel. It is highlighted from this investigation that the alloy (without chromium content) with the higher carbon and aluminum content, showed the best resistance to cavitation. This result is very important since the cavitation or impact resistance of Fe-Mn-Al-C alloys is generally associated with their ability to absorb energy (implosion of bubbles and subsequent emission of elastic waves) through the mechanisms of plasticity such as TRIP, TWIP, MBIP or DSBR.

## 4.3 Erosion wear

The resistance to erosion and corrosion of a Fe-Mn-Al-C alloy (with the addition of chromium) was evaluated by a block device (alumina) on the cylinder, both in the air and in a solution of NaCl at 3.5 % by Huang *et al.* (2000) [249]. In the corrosion tests where no normal loading was used, it was determined that the samples treated with solution + quenching presented higher resistance to corrosion than the aged samples, which is due to the fact that in the aged samples the β-manganese, responsible for the destruction of the passive zone. However, in the tests where normal loads were used, the aged samples showed a better performance than the solution + quenching samples, although these results appear not to be statistically significant. From the above, it was concluded that in the solution treated samples (in which the ferrite and austenite were present) a preferential attack occurred in the ferrite phase whereas, in the aged samples, a preferential attack was presented in the austenite where it was present the β-manganese phase. On the other hand, in wear tests performed in air with a load of 11.76 N, showed that the solubilized sample was worn twice as much as the aged samples, which is mainly due to the increase of hardness and the presence of the phase β-manganese in austenite.

Aperador *et al.* (2012) [250] studied the erosion wear of a series of Fe-Mn-Al-C steels varying the Al and C content. They show in Figure 51 the wear resistance of the alloys evaluated. Basically, the stainless steel showed a superior wear resistance than the Fe-Mn-Al-C steels. But, the Fe-Mn-Al-C steel with the higher manganese and aluminum content showed a close behavior with the AISI 316, due to the high solid solution strengthening. However, in their study the phenomena involved and its relation with the wear response is still unclear.

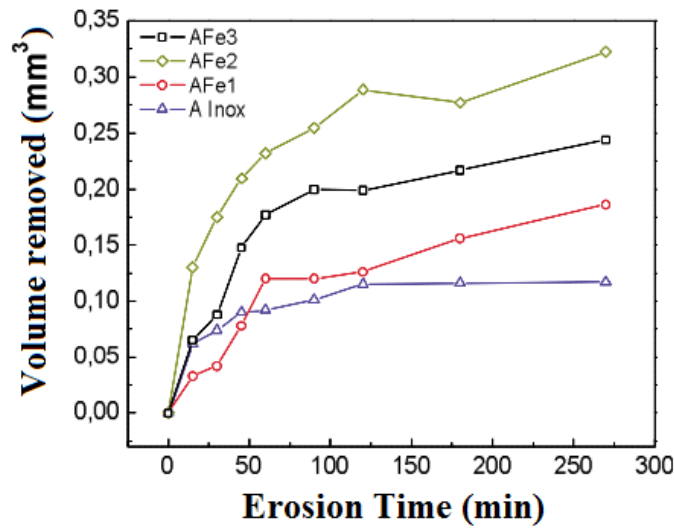

Figure 51. Volume removed as a function of erosion time for a stainless steel and three Fe-Mn-Al-C steels (AFe1: high Al and Mn, AF3: low Al, Mn and C, AF2: Intermediate values), adapted from [250]

## 5  Processing

The adequate processing of Fe-Mn-Al-C alloys is a challenging task due to the presence of some phases at the grain boundaries which produces embrittlement such us the κ-carbides [251, 252]. In fact, some patents have been made reporting to eliminating this problem at the grain boundaries with Nb, V, Ti, Zr, and some rare earth elements additions [253-255], controlling the reduction passes [256], the chemical composition [257], the phases ratio in the microstructure [258] or other process parameters [259]. Additionally, the knowledge about the physical metallurgy of these steels is still incomplete and under investigation (*see Section 0*), which add an extra difficulty. In addition, the effect of variables such us the strain rate, temperature and chemical composition on the mechanical properties and oxidation resistance have not been fully understood. In this regard, one of the most important contributions that have been made towards the industrial up-scaling of Fe-Mn-Al-C steels was described by Frommeyer *et al.* (2013) [260]. In this final report, it is described a several number of drawbacks during the casting and processing, and how the problems were solved, and the reader is referred to this work. For the extension of this manuscript, it will be describing briefly the progress made by different author from the processing perspective.

## 5.1 Thermo-mechanical treatments (TMT)

The thermo-mechanical treatments, in general, allow to reduce, to diminish or to remove a great variety of defects that present either in the "as cast" or in the solid solution state. Defects such as chemical micro-segregation, micro-porosity and excessive grain size growth (phenomena occurring during casting, solidification and solution treatments) promote poor mechanical properties, which adversely affect performance and durability of steel.

For example, the use of thermal solubilization treatments to destroy the "as cast" structure is adequate to avoid chemical micro-segregation (at the interdendritic level) and prompt a homogeneous microstructure. However, the chemical macro-segregation (distances from 1 cm to 100 cm) and the microporosities in the alloy cannot be eliminated by simple solubilization [261, 262]. In addition, solution treatments have the drawback that as the solution temperature increases, the austenitic grain size also grows rapidly as is shown in Figure 52 (these results are also consistent with the works of Yuan *et al.* (2015) [158], Yoo *et al.* (2009) [92] y Jun *et al.* (1998) [172]). This produces a detriment of the mechanical properties of the alloy. It is here that thermo-mechanical treatments (for instance hot rolling) play a major role in reducing these negative effects [263-266]. Following, some work on Fe-Mn-Al-C steels related to their formability will be discussed.

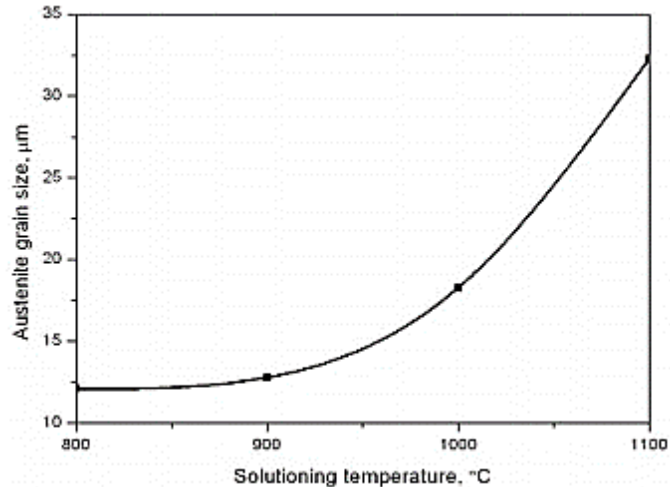

Figure 52. Austenitic grain size as a function of the solution treatment, reproduced from [267] with permission of Elsevier®

For example, Han et al. (1988) [48] studied in a Fe-32Mn-8Al-0.9C-0.2Mo alloy, the effect of thermo-mechanical treatment on the resulting microstructure and hardness. For this purpose, they performed the following initial procedure: homogenization, hot forging at 1200°C, followed by rolling at 1150 °C. Subsequently, cold rolling was used without intermediate annealing and after that, 3 different treatments were performed to evaluate:

I. Solution treatment (1027ºC, 10 min, quenching in water) + 550ºC aging.

II. Aging at 550 ° C without solution treatment (freshly deformed samples).

III. Solution treatment (1027 ° C, 10 min) + pre-aging 550 ° C (40h) + deformation in warm (550°C, 15% reduction) + final aging at 550 ° C

The results of hardness obtained are shown in Figure 53, where it is observed that: the increase of the microhardness is almost equal for treatment II and III (the highest). However, after $10^5$ s, the treatment II reports higher hardness values. On the other hand, the eutectoid decomposition of austenite in α and β-manganese accelerates with the treatment II, for this reason, the authors do not recommend using cold deformation before the treatment of aging. Thus, it is clear that these treatments can improve the hardness and strength of the alloy since they do not modify the mechanism of aging and in the cases described to avoid the precipitation of brittle phases.

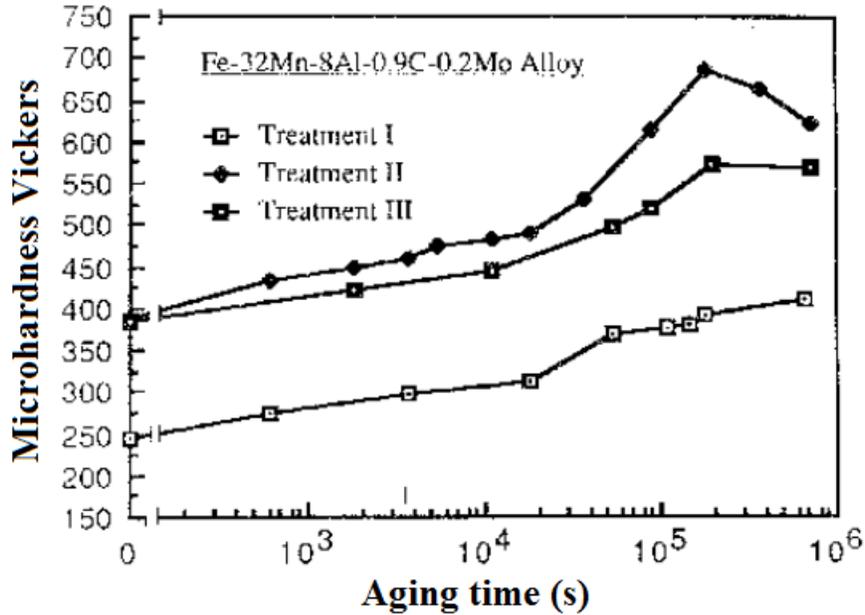

Figure 53. Aging curves at 550ºC with different pretreatment treatments: treatment I (none), treatment II (cold deformation) and treatment III (pre-aging and warm deformation at 550ºC), adapted from [268]

The conformability of high-manganese steels (Fe-Mn-Al-C) at high temperatures presents great challenges due to the ease of the elements that make up the alloy in producing oxidation reactions, low melting point intermetallic compounds, Intergranular precipitation of fragile phases, sensitivity to the rate of deformation, among other phenomena that are associated with cracking during processing. According to the above, some investigations such as that advanced by Kim *et al.* (1997) [253] have sought to avoid or reduce cracking during the hot rolling of steels in the range of Fe- (15-35) Mn- (1-6) Al - (<1,5) C. To this end, they suggest two solutions: i) use additions of Nb, Ti, Zr, Ce and/or ii) a careful control of the reduction steps and the deformation rates. In the first method, Ti and Nb: (1) favor thin dendritic structures (2) avoid the formation of coarse aluminum nitrides or conglomerates (3) increase the melting temperature of intermetallic compounds and (4) spheroidize inclusions. With the second method, it was determined that the procedure to obtain an optimal formability was: (1) to heat the ingot between 1150 ° and 1300 ° C (to destroy the casting structure), (2) then a hot rolling (reduction by maximum step of 7%, until reaching a total reduction of 40%), maintaining a deformation rate of less than 2 $s^{-1}$ during soft rolling (3) after reaching the total reduction of 40% (20% reduction per step and speeds as high as $10s^{-1}$). In this way no cracks occur in the ingots during rolling, however, it should be clarified that this procedure already takes into account the addition of Nb and Ti. In this sense, Han (2000) [269] reported that the use of a controlled thermomechanical treatment (Figure 54) in an Fe-Mn-Al-C alloy with additions of Nb, W and Mo, favors the generation of sites of nucleation for the precipitation of niobium carbides and niobium nitrides in the matrix (deformation induced precipitation), where the modulated structure is added, which generates a greater increase in the mechanical properties (this effect that does not occur in the microstructures Of fast cooling). The author pointed out several important facts; (1) the mechanical properties obtained with the thermo-mechanical treatment are even greater than alloys of similar chemical composition after a solubilization and aging treatment, (2) decomposition of the phase started during air cooling after final rolling , (3) the precipitation of the β-manganese phase in the grain boundaries was not evidenced by the thermo-mechanical treatment used and (4) in non-deformed alloys the precipitation occurs at the grain boundaries.

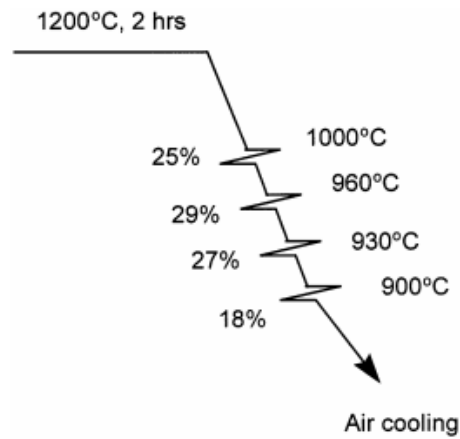

Figure 54. Schematic of controlled rolling schedule, reproduced from [269] with permission of Elsevier®

Another important research was performed by Grajcar *et al.* (2008) [270] who studied the effect of TMT on recrystallization in two Fe-Mn-Al-Si steels (with additions of Nb and Ti). For this purpose, they evaluated 3 reduction percentages (20%, 40%, and 60%) at temperatures in the range of 900 ° C to 1100 ° C. Determining that at 20% reduction, no grain refinement was produced by dynamic recrystallization. However, a 40% reduction resulted in a grain refinement and with 60% reduction a drastic grain refinement occurred. It is also important to mention that the rate of deformation and the temperature at which the deformation is carried out directly affect the possibility of producing or not, dynamic recrystallization, which can produce a grain refinement.

Another interesting research that relates the thermo-mechanical treatment with the mechanical properties and the final microstructure was performed by Zamani *et al.* (2011)

[271] for a Fe-31Mn-4Si steel and a Fe-31Mn-4Si-2Al-0.06Nb-0.09Ti. Where the alloys homogenized at 1100 °C for 1 hour under an argon atmosphere, they were subsequently hot rolled between 900 °C and 1100 °C with a 50% reduction percentage. Cold rolling with reduction percentages of 50, 65 and 80% and subsequent recrystallization for 30 minutes at 650 °C, 700 ° C and 750 °C and air cooling were performed. Finally, it was determined that the alloy with additions of Nb and Ti produces a grain refinement during the thermo-mechanical treatment and also favored the precipitation of additional carbides, which increased the mechanical properties. Further studies such as Hamada *et al.* (2007) [272] and Hamada *et al.* (2011) [273] on TWIP steels showed that the use of certain thermo-mechanical treatments generates a grain size refinement due to dynamic recrystallization or to the type of cooling after the treatments. This last phenomenon has been investigated by Dobrzański *et al.* (2012) [274] in TRIP and TWIP steels. In this study, samples quenched in water after finished the thermo-mechanical treatment, showed the lowest grain sizes of austenite (8-10μm) followed by cooling in air (12-15μm) and finally by the holding treatment to the final deformation temperature at 30 s and then followed by quenching in water (20 μm). It was also determined that the stability of the austenite is not affected by the type of cooling employed.

More recently, Khosravifard *et al.* (2013) [275] determined the effect of temperature and strain rate on behavior at high temperatures for two TWIP steels. They determined that the rate of deformation directly affects the reduction of the grain size of the alloy and this relationship is not linear, since at low rates of deformation there is no reduction of grain size by dynamic recrystallization, at intermediate deformation rates there is a drastic reduction in grain size, however, employing higher deformation rates does not lead to a larger reduction in grain size. Also, they determined that a critical temperature (about 1100 ° C) is required to facilitate the dynamic recrystallization process and facilitate the decrease in grain size.

On the other hand, it is important to highlight the research done by Liu (1990) [276] on Fe-Mn-Al-C steels for a chemical composition range of Fe- (15-35) Mn- (1-6) Al- ( <1.5) C, who evaluated the effect of alloying elements such as Ti and Nb on mechanical properties (Figure 55), microstructure and formability during hot rolling. For alloys with no additional elements, it was determined that by laminating the specimens in hot and followed by quenching in water, no κ-carbides were precipitated into the matrix. On the other hand, in the hot rolling and cooling in the air, there is a consistent precipitation of the carbides-κ within the matrix, with carbide size from 360 nm to 3200 nm in length and from 52 nm to 220 nm in width. Under these conditions, the alloy did not reach satisfactory strength values. However, when using additions of Ti, Nb, and V, it was shown that after hot rolling and subsequent quenching in water from the final rolling temperature, κ-carbides were precipitated coherently in the matrix. Similarly, when hot rolling and cooling was used in air, the precipitation of carbides-κ with sizes from 10 nm to 30 nm within the matrix was observed, thus favoring satisfactory properties.

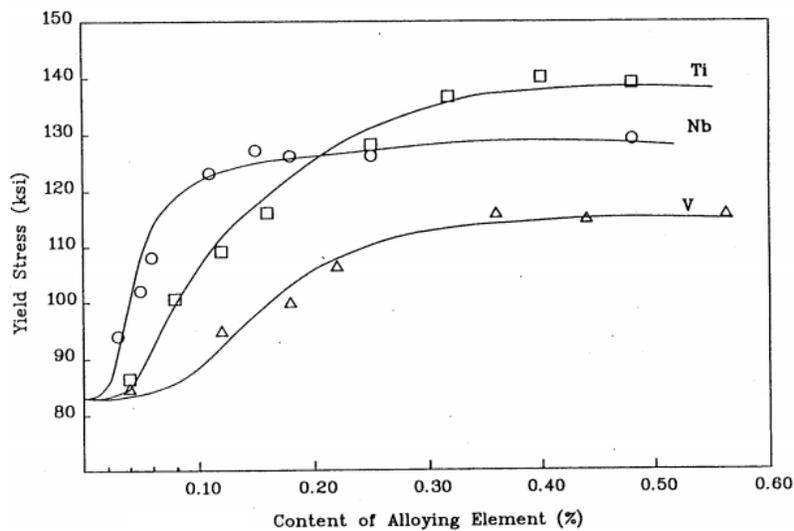

Figure 55. Influence of the Ti, Nb, and V content on the yield stress [276]

To date, few studies have evaluated the microstructural evolution during hot deformation, as well as the effect of temperature or strain rate on dynamic recrystallization in Fe-Mn-Al-C steels. In addition, metallurgical factors such as SFE, which will determine the recrystallization mechanisms that operate at high temperatures, have not been properly discussed to date in Fe-Mn-Al-C steels. These concerns have been recently [277] in a high aluminum and high carbon Fe-Mn-Al-C steel. In this work, it was studied the effect of test temperatures between 900 °C and 1150 °C, and the strain rates varying from 0.01 s$^{-1}$ to 1 s$^{-1}$ on dynamic recrystallization behavior. They found even with a high SFE of 92 mJ/m$^2$ the recrystallization mechanism was, instead of dynamic recrystallization, the operation of dDRX as is shown in Figure 56.

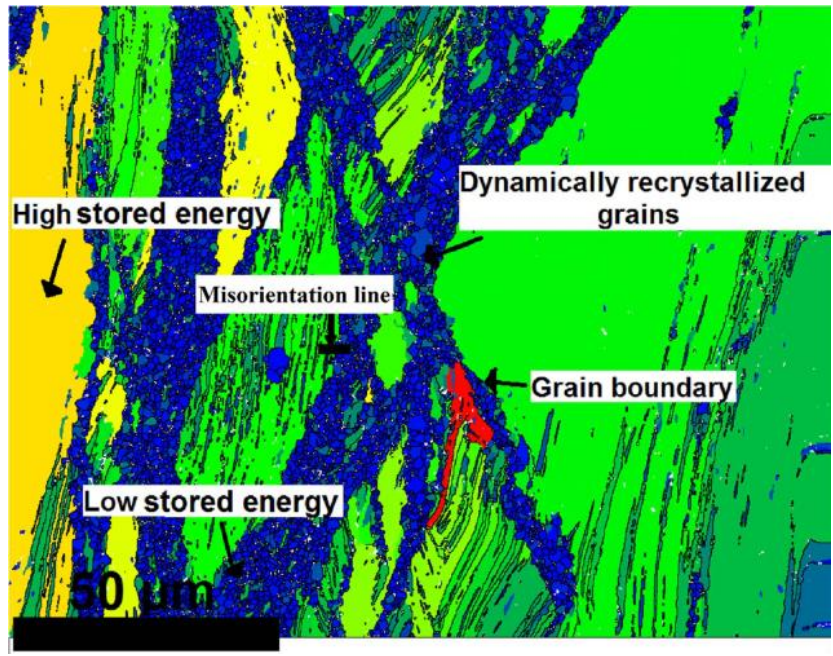

Figure 56. Grain orientation spread (GOS, misorientation) of the sample deformed to a true strain of 0.7 at 900 °C and a strain rate of 0.01 s-1 with initial austenitic grain size of ~100 μm, reproduced from [277] with permission of Elsevier®

Interestingly, they also found a correlation between the new DRXed grains and annealing twins as is shown in Figure 57.

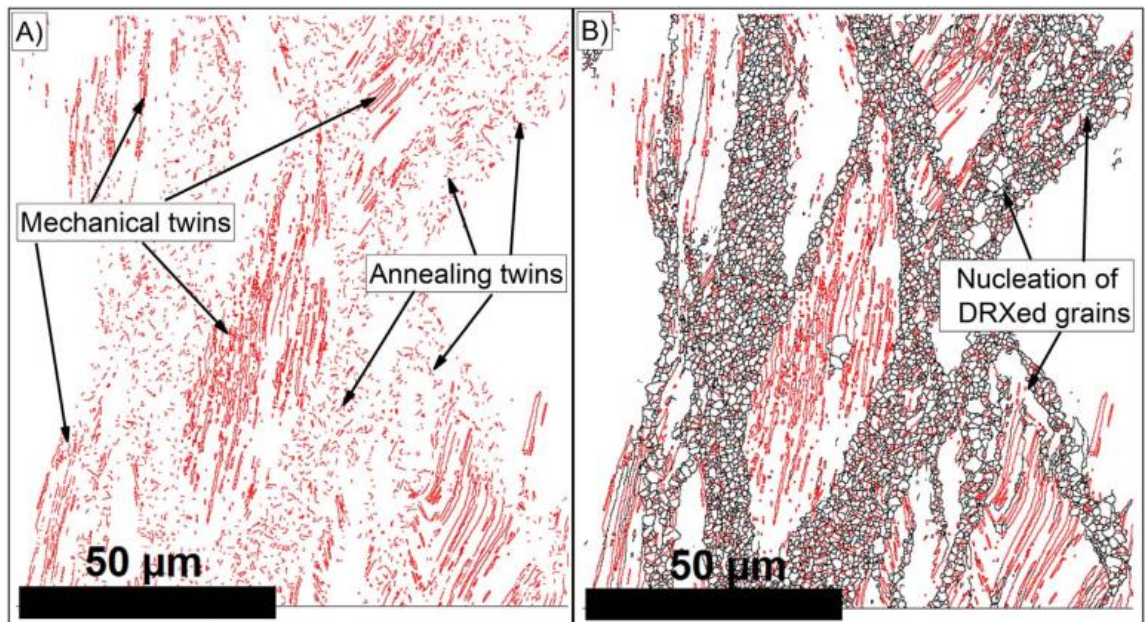

Figure 57. OIM maps of the sample deformed to a true strain of 0.7 at 900 °C and a strain rate of 0.01 s-1 (a) Σ3 boundaries and (b) high angle boundaries ( > 15°) with Σ3 boundaries, reproduced from [277] with permission of Elsevier®

Finally, due to the high carbon and aluminum content of the alloys, as well as the conditions imposed, the steel precipitate dynamically κ-carbides as is shown in Figure 58. This occurs in few seconds instead of the conventional aging treatment which takes hours. This produced an important enhanced in the hardness. However, the mechanism of this dynamic precipitation is still hidden or unrevealed.

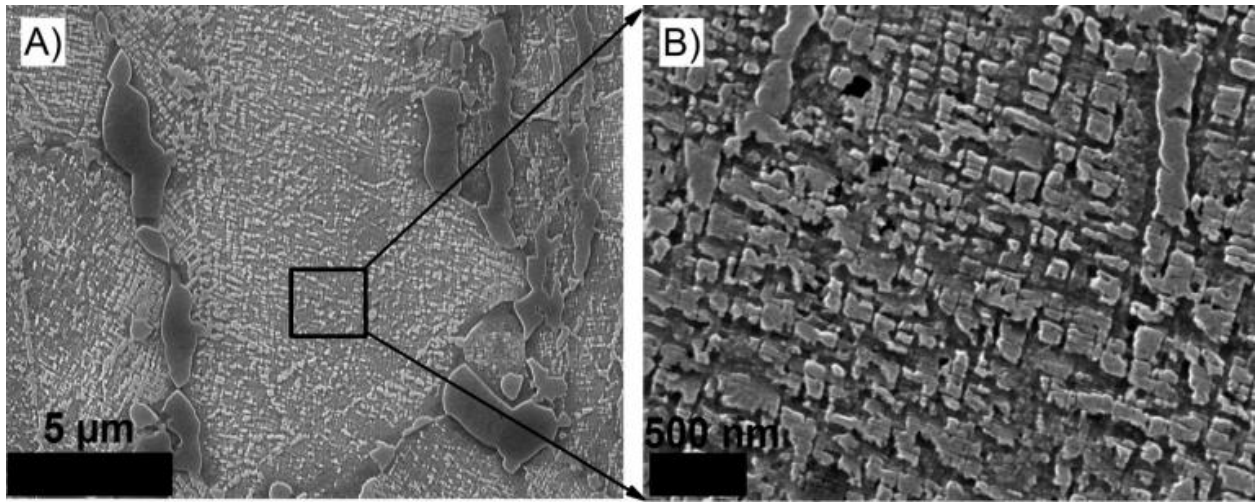

Figure 58. SEM images of the sample after TMT at 1150 °C–0.1 s-1 (a) 5 µm and (b) 500 nm-modulated precipitates in austenite matrix, reproduced from [277] with permission of Elsevier®

## *5.2 Welding*

There are only a few investigations related with the weldability of Fe-Mn-Al-C steels. One of the main investigations in this aspect was conducted by Chou *et al.* (1990) [278], who studied the effect of the carbon content and welding conditions in Fe-30Mn-10Al-xC steels with autogenous gas tungsten arc (GTA). They found that all of the compositions studied presented satisfactory properties (Figure 59), but lowest properties were achieved by the composition with the highest contents of carbon.

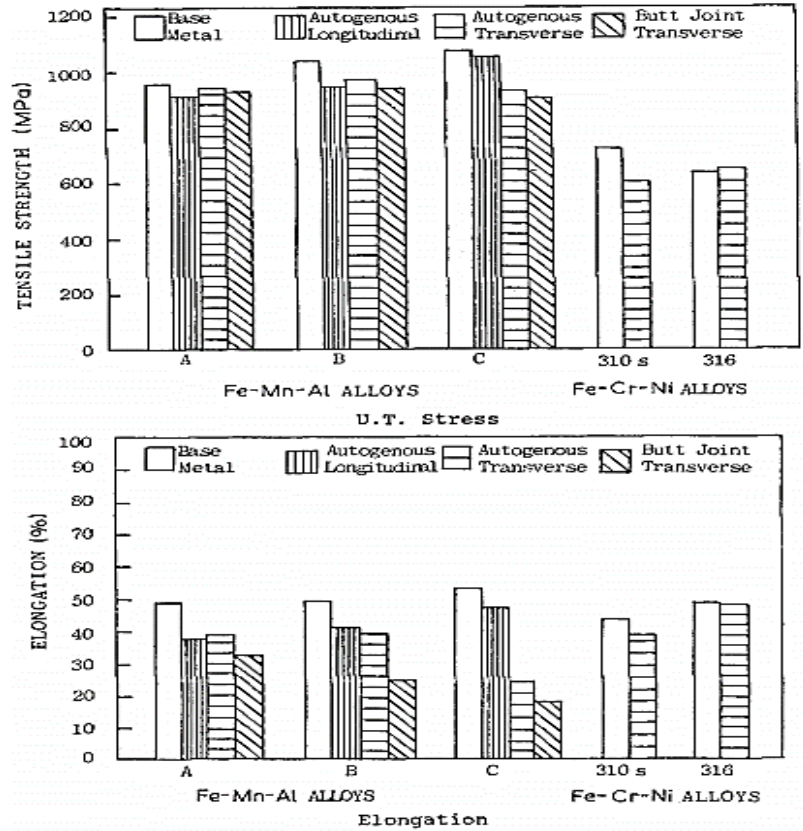

Figure 59. Tensile test results of Fe-Mn-Al-C weld metals A (low C), B (medium C) and C (high C), reproduced from [278] with permission of Springer®

These results are in somehow expected because the increase of carbon content decreases the ferrite content, which is known to reduce thermal contractions and expansions (hot cracking) and also the help to reduce the hydrogen solubility. It is also inferred that the highest carbon content can increase the SFE which can reduce the facility of twinning or martensitic transformation induced.

# 6 Corrosion

Since these steels were initially developed to replace the stainless steels (Section 1.1), the corrosion behavior of Fe-Mn-Al-C steels is one of the most striking features. So, in this section, we are going to revise the most important results pointing in this direction, which forks in three branches; the electrochemical behavior, the stress corrosion cracking phenomenon, and oxidation resistance.

## *6.1 Electrochemical corrosión*

Cavallini *et al.* (1982) [279] evaluated the effect of the Mn content on the electrochemical corrosion behavior of the Fe-(19.9-32.5%)Mn-(7.1-10.2%)Al-(0.76-0.99%)C system in NaOH (basic), H$_2$SO$_4$ (acid), and neutral solutions. They showed that the increase of the Mn content increase the $I_{corr}$ and decrease the $E_{corr}$. studied with anodic polarization technique the aqueous corrosion behavior of a Fe-8.7A1-29.7Mn-1.04C alloy with and without aging treatment, and the results were compared with a traditional 316 SS. The results showed that the $I_{corr}$ and $E_{corr}$ for the Fe-Mn-Al-C steel is very high and low, respectively in comparison of conventional 316 stainless steel. Besides, the aging treatments showed a poor corrosion resistance in comparison with the solution treatment. During the same year, Altstetter *et al.* (1986) [280] also evaluate the electrochemical corrosion behavior for several Fe-Mn-Al-C steels in a specific saline solution and H$_2$SO$_4$, respectively. They found that the Fe-Mn-Al-C alloys had a lower electrochemical resistance in comparison with the conventional stainless steel. Then, Gau *et al.* (1992) [281] studied the galvanic corrosion behavior of Fe-Mn-Al alloys in sea water and were compared with a conventional 304 SS. They showed that the Fe-Mn-A-C alloys had a $I_{corr}$ ten times higher than the 304 SS. But, 5 times less than a simple carbon steel. Additionally, it was found that the austenitic Fe-Mn-Al-C alloy showed a better corrosion resistance than the duplex Fe-Mn-Al-C alloy. Similar results were obtained in the evaluation of pitting corrosion in artificial sea water in Fe-Mn-Al-C steels by Rfiscak *et al.* (1993) [282] and Shih *et al.* (1993) [283], where pitting occurred preferentially in the ferrite grains and in the interface between the austenite/ferrite. Additionally, it was found that the passivation region depended on the scanning rate, where at the lowest scanning rate the passivation had more time to occurred.

In Zhu *et al.* (1998) [284] investigated the electrochemical corrosion behavior and passive films for the Fe-Mn, Fe-Mn-Al, and Fe-Mn-Al-Cr systems in aqueous solutions. It was found that in 3.5 wt% NaCl solution, the three systems did not show passivation phenomenon. The increase of the Al content produces that the $I_{corr}$ decreases and the $E_{corr}$ increases (more noble behavior), and with the increase of manganese, the opposite behavior was observed. Similar results about the Al effect were obtained by [285] as is shown in Figure 60. One of the same authors reported [286] that a passive film was formed in sodium sulphate solution for a Fe-30Mn-9Al alloy, however, the presence of instable manganese oxides in the passive film avoid the substitution of the Fe-Mn-Al alloys for the conventional stainless steels. In summary, the corrosion resistance of the Fe-Mn-Al-C alloys were inferior to the conventional stainless steels. To this same conclusion was also reached by Abuzriba *et al.* (2015) [287].

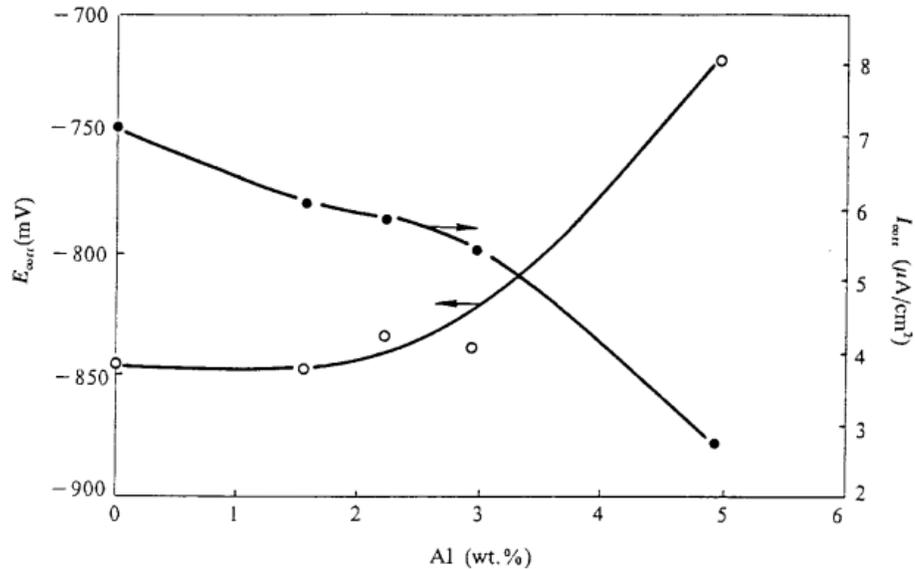

Figure 60. Effect of Al content on the $I_{corr}$ and $E_{corr}$ of Fe–Mn–Al–C steels in Na$_2$SO$_4$ solution, reproduced from [285] with permission of Elsevier®

Aiming to have most of the electrochemical responses of the Fe-Mn-Al-C alloys (corrosion rate and $E_{corr}$) reported in the literature, Table 3 was constructed. The results so far, showed an uncomfortable truth; the Fe-Mn-Al-C steels *per se* cannot surpass the conventional stainless steels (SS) corrosion resistance. Despite that some of the Fe-Mn-Al-C steels showed a close response as SS. With this concern, during the last years, several researches have been using coatings, thermal treatments, and alloy additions to increase corrosion resistance of the Fe-Mn-Al-C steels [250, 288-298], or apply anodic passivation currents to promote thick, protective and stable passive films [299]. These new strategies have been shown promising results.

| Alloy | $E_{corr}$ (V(SCE)) | Corrosion rate (μm year$^{-1}$) Tafel extrapolation | Reference |
|---|---|---|---|
| AISI 1020 (specific saline solution) | -0.71 | 150 | [280] |
| Mild steel | -0.72 | 185.4 | [281] |
| AISI 304 (specific saline solution) | -0.26 | 2 | [280] |
| AISI 304 (5% HNO$_3$) | -0.14 | 0.34 | [293] |
| AISI 304 (3.5%NaCl) | -0.36 | 3.1 | [281] |
| AISI 316 | -0.33 | 4.5 | [300] |
| Fe-25Mn-5Al-0.15C (10% Na$_2$SO$_4$) | -0.18 | 212050 | [285] |
| Fe-25Mn-5Al-0.15C (30% NaOH) | -1.15 | 88.1 | [285] |
| Fe-25Mn-5Al-0.15C (10% HCl) | -0.55 | 74160 | [285] |
| Fe-25Mn-5Al-0.15C (30% HNO$_3$) | -0.47 | 345307 | [285] |
| Fe-25Mn-5Al-0.15C (3.5%NaCl) | -0.72 | 133.3 | [285] |
| Fe-30Mn-9Al-1Si-1C (1 N H$_2$SO$_4$) | -0.61 | 195000 | [280] |
| Fe-32.7Mn-6.59Al-1.26Si-0.25C (1 N H$_2$SO$_4$) | -0.61 | 98493 | [301] |
| Fe-32.3Mn-8.54Al-1.31Si-0.54C (1 N H$_2$SO$_4$) | -0.53 | 278099 | [301] |
| Fe-28Mn–9Al (specific saline solution) | -0.74 | 15 | [280] |
| Fe-30Mn-14Al (specific saline solution) | -0.59 | 190 | [280] |
| Fe-30Mn–9Al-1.8C (3.5%NaCl) | -0.75 | N/A | [302] |
| Fe-18Mn-2Al-2Si -0.07C (Coarse-grained in 3.5%NaCl) | -0.76 | 23.2 | [303] |
| Fe-18Mn-2Al-2Si -0.07C (Fine-grained in 3.5%NaCl) | -0.71 | 46.3 | [303] |
| Fe-24.8Mn-7.3Al-0.90C (3.5%NaCl) | -0.57 | 33.2 | [281] |
| Fe-24.4Mn-9.2Al-0.40C (3.5%NaCl) | -0.59 | 48.3 | [281] |
| Fe-24.4-Mn-9.96Al-0.40C (3.5%NaCl) | -0.84 | 580 | [283] |
| Fe-30.7Mn-13.03Al-0.44C (3.5%NaCl) | -0.85 | 570 | [283] |
| Fe-26.6Mn-9.29Al-0.43C (3.5%NaCl) | -0.85 | 200 | [283] |
| Fe-30Mn-3Al-1.5Si-0.06C (3.5%NaCl) | -0.79 | 263 | [304] |
| Fe-17.3Mn-3.10Al-0.38Si-0.24C (3.5%NaCl) | -0.69 | 366 | [305] |
| Fe-20.6Mn-3.51Al-2.92Si-0.29C (3.5%NaCl) | -0.77 | 46.3 | [305] |
| Fe-30Mn-3Al-1.5Si-0.06C (0.1 M NaOH) | -0.16 | 22.9 | [304] |
| Fe-30Mn-3Al-1.5Si-0.06C (0.1 M H$_2$SO$_4$) | -0.64 | 25921 | [304] |
| Fe-25Mn-5Al-0.15C (10% HCl) | - | 74160 | [291] |
| Fe-30Mn-8Al-1C (10% HCl) | - | 126304 | [291] |
| Fe-25Mn-5Al-022C (5% HNO$_3$) | -0.41 | 1089 | [293] |
| Fe-25Mn-8Al-020C (5% HNO$_3$) | -0.39 | 869 | [293] |
| Fe–24Mn–8.3Al–5Cr-0.38Si-0.34Mo–0.45C (3.5%NaCl) | -0.58 | 89.5 | [290] |
| Fe-30Mn-10Al-1Si-1C (specific saline solution) | -0.83 | 20 | [280] |
| Fe-30.5Mn-7.5Al-1.5Si-1C (specific saline solution) | -0.54 | 50 | [280] |
| Fe-30.5Mn-7.5Al-1Si-1C (specific saline solution) | -0.86 | 350 | [280] |
| Fe-30.5Mn-7.5Al-0.5Si-1C (specific saline solution) | -0.54 | 230 | [280] |
| Fe-30.5Mn-8.68Al-1.8 C | -0.79 | 912 | [300] |
| Fe-26.4Mn-2.74Al-0.32C-1.13Cr (3.5%NaCl) | -0.72 | 783.3 | [298] |

Table 3. Summary of electrochemical behavior of several Fe-Mn-Al-C alloys ($corrosion\ rate$ and $E_{corr}$) reported in the literature as well as other conventional alloys for comparison purposes.

## 6.2 Stress corrosion cracking (SCC)

The evaluation of the stress corrosion cracking (SSC) behavior in Fe-Mn-Al-C steels is an important requirement for the industry. A good corrosion resistance in aqueous solution is not the only requirement to accomplish the life expectations of a component. For instance, some components are subjected to stresses and simultaneously exposed to corrosive environments. This leads to unexpected and premature failures of generally ductile materials. For this reason, the SCC will be briefly reviewed with the advances about this topic in Fe-Mn-Al-C steels.

Possibly, the first evaluation of the SCC behavior of an austenitic Fe-29.7Mn-8.7A1-1.04C steel was performed by Tjong (1986) [306], he used a 20% NaCl solution at 100 °C using slow strain rates. The results showed that at high strain rates, the specimen not showed SCC damage due to the mechanical failure which occurred prior initiation of SCC. On the other hand, at an extremely low strain rate, SCC does not occur because the surface of the alloy is repassivated. However, at intermediate strain rates, transgranular SCC occurs due to the repetitive rupture of the passive films. The same author [307] evaluate the effect of aging treatments at different times to understand the effect of the kappa carbide precipitation in SCC behavior. Prolonged heat treatment leads to the decomposition of austenite into ferrite and β-Mn phases at the austenite grain boundaries. The hydrogen-induced cracking was found to dominate at the grain boundary phases.

Shih *et al.* (1993) [308] studied the environmentally assisted cracking occurrence on two duplex Fe-Mn-Al-C alloys in chloride solutions. It was found that all of the alloys were quite susceptible to environmentally assisted cracking. Besides, the effect of the environment seems to start to occur at or after the beginning of plastic deformation. The ferritic phase was less resistant to the environmentally assisted cracking than the austenitic phase.

Chang *et al.* (1995) [309] evaluate the SCC phenomenon in 3.5% NaCl solution at room temperature and 165°C. It was found that the all austenitic Fe-Mn-Al alloys were susceptible to SCC at RT and 160°C NaCl solutions. The crack path was transgranular, and cleavage-like facets were observed on the fracture surface when relatively high potentials were used (cathodic potentials). However, when the applied potential was more negative (anodic potentials), it was observed a brittle behavior. So, intergranular fractures occurred in Fe-Mn-Al alloys. Both types of fractures are shown in Figure 61. The authors showed the main cause was due to the hydrogen embrittlement resulted from the water reduction. This phenomenon is almost always presented in SCC conditions, and in fact, due to the importance of the hydrogen embrittlement problems in Fe-Mn-C and Fe-Mn-Al-C alloys, it should deserve its own section and discussion in this manuscript. However, the reader is referred to the works of Motomichi Koyama *et al* [68, 310-315] who studied systematically several variables on the hydrogen embrittlement in these steels. Further interesting reading on this topic can be found in [238, 316-325].

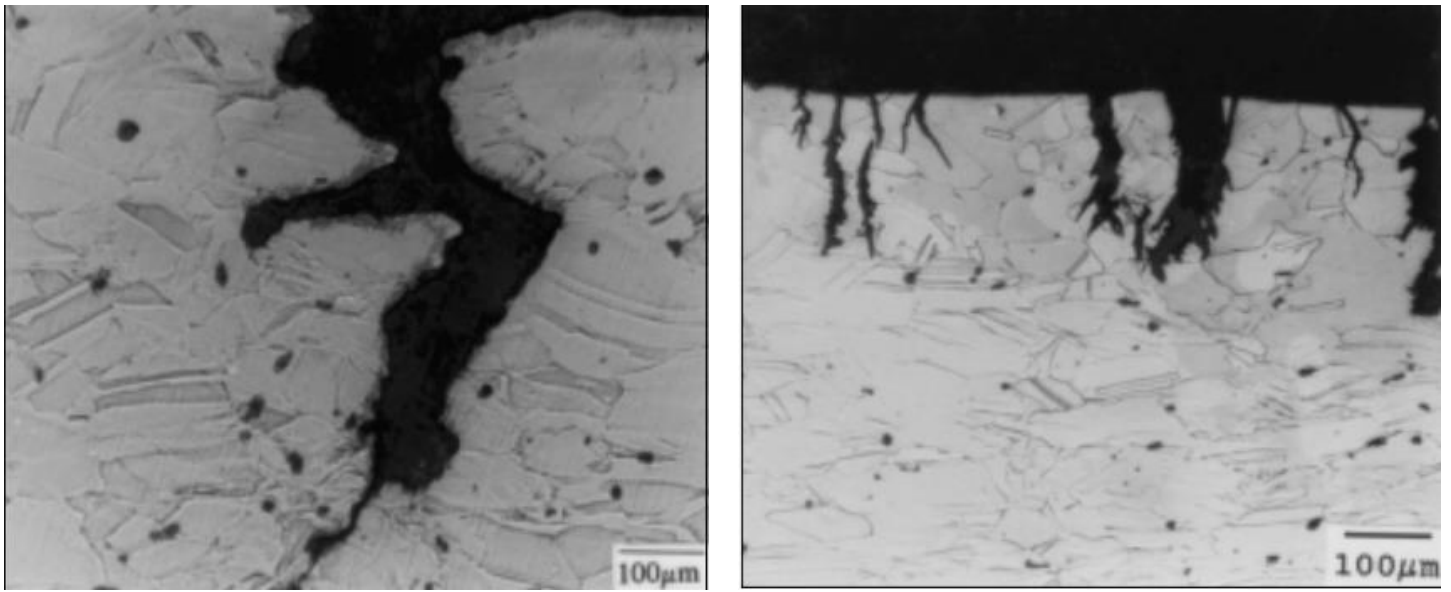

Figure 61. (a) Intergranular crack observed in a Fe-32.16Mn-9.41Al-0.93C steel tested at RT 3.5% NaCl solution with an applied potential of –1200 mVSCE, and (b) transgranular cracks in a Fe-32.16Mn-9.41Al-0.93C steel at 160°C using a load equal to 85% of the tensile strength [309]

## 6.3 Oxidation

One of the first works, which laid the foundations to understand the oxidation resistance of Fe-Mn-Al-C steels was performed by Boggs (1971) [326], who studied the effect of the Al content (up to 8% Al) in the iron-aluminum-carbon system (with up to 0.1% carbon) over the temperature range between 450°C and 900°C in water vapor, and in wet and dry oxygen. He showed that the oxidation rate decreases with increasing aluminum content, and observed the formation of $FeAl_2O_4$, and $Al_2O_3$ above 7% of Al (results that are fully consistent with those reported later by for the binary Fe-Al system), which imparted oxidation resistance to the alloy. However, the conventional iron oxides ($FeO$, $Fe_2O_3$, $Fe_3O_4$) were also observed. Almost 10 years later, Sauer *et al.* (1982) [327] published the earliest study about the oxidation resistance of Fe-Mn-Al-C alloys, properly speaking. They evaluate several austenitic Fe-Mn-Al-C steels containing C, Si, and Zr in an air atmosphere at 850°C and 1000°C. At 850°C, the alloy without additions showed discontinuous $Al_2O_3$, scale interspersed with a less protective $MnAl_2O_4$ spinel, and above 1000°C the alloys did not show suitable oxidation resistance due to the internal oxidation of Al, and formation of less protective iron oxides such as FeO. They also reported that the Mn provided oxidation protection in Fe-Mn-Al-C steels in an analogous form that by which Cr provides protection in Fe-Cr-Al alloys, i.e. through the MnO oxide, which has a thermodynamic stability intermediate to those for wustite (FeO) and alumina ($Al_2O_3$). Finally, they found satisfactory oxidation resistance when the formation of a continuous protective alumina layer was formed, which is promoted by the presence of Si and by a fine grain size. Similar results were obtained by Wang *et al.* (1984) [71] who showed a good oxidation resistance of the Fe-Mn-Al-C alloy in comparison with a 304 SS at 1100°C as is shown in Figure 62 (a). The reason is due to the formation of a protective $Al_2O_3$ scale. The $Al_2O_3$ scale growth on a Fe-30Mn-10Al-Si was characterized using *In-situ* SEM at 1100°C as is shown in Figure 62 (b). They showed a nodule nucleation mechanism, which then growth and then coalescence with others nodules to form a compact and protective oxide layer. However, they also point out that alloys with a low Al content, the internal oxidation of aluminum prevents the arrival of aluminum to the nodule/alloy interface, and the protective $Al_2O_3$ film cannot take place.

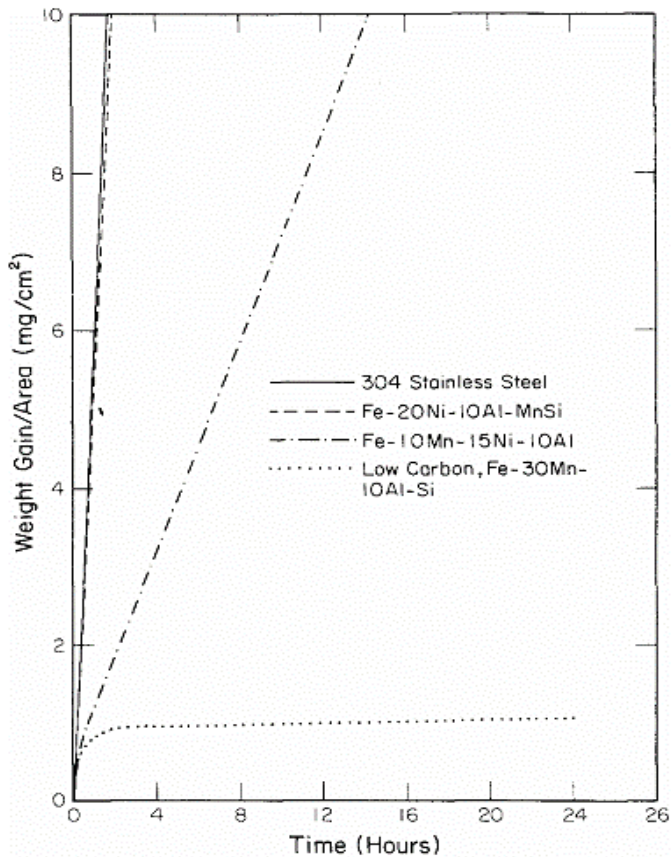
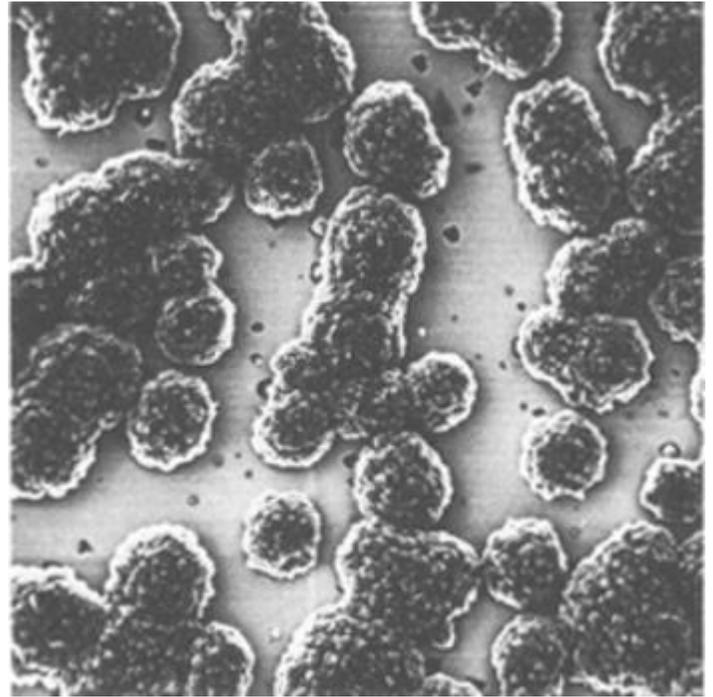

Figure 62. (a) oxidation kinetics at 1100°C for three alloys and type 304 stainless steel, reproduced from [71] with permission of Springer®, and formation and subsequent growth of nodules on a Fe-30Mn-10Al-Si alloy at 1100°C at 148 min, reproduced from [328] with permission of Springer®

Another systematic study about the oxidation resistance of the Fe-Mn-Al-C system was performed by Jackson *et al.* (1984) [329] at temperatures between 600°C and 1000°C in pure oxygen. They proposed oxidation maps or regions where established the limits of different oxides formation and its relation to oxidation resistance. For instance, the alloys in the region I was those which contained insufficient aluminum to form continuous $Al_2O_3$ scale. In region III, for example, alloys were characterized by very adherent continuous protective scales, which were not penetrated by nodules or other bulky oxides. They also found that the formation of austenite must be avoided due to the austenite facilitates the breakdown of preexisting $Al_2O_3$ scale.

Benz *et al.* (2012) [330] they showed a two important results: i) if the samples were cooled rapidly, the oxide formed on the surfaces were not protective and a linear rate of oxidation was obtained, but ii) when the samples were cooled sufficiently slowly, the samples formed a protective oxide layer on the surface, promoting a parabolic oxidation rate. Suggesting that the Fe-Mn-Al-C alloys can form protective oxides as long as oxides do not flake off by the thermal shock. In this direction, Kao *et al.* (1988) [72] evaluated from 600°C to 1000°C the oxidation of two Fe-Mn-Al-C alloys, one with low manganese and the other one with high manganese content. The results showed that increasing the manganese content produce an increase in the oxidation resistance, and also that different parabolic rates occurred depending on test temperature.  showed that the carbon addition in the Fe- Mn-A1 system decrease the oxidation resistance of the alloy, due to the formation of a porous oxide layer i.e. a carbon-induced oxidation phenomenon was observed which tends to perturb the compact integrity of the initial oxide layer and leaves cavities behind. The same role of the C in the oxidation resistance of Fe-Mn-Al-C steels was confirmed later by Duh *et al.* (1990) [331]. They also showed the beneficial role of Cr additions to retard the oxidation kinetics in Fe-Mn-Al-C steels.

Tjong (1991) [332] studied the oxidation resistance of Fe-(29-30)Mn-(8-l0)Al-(0-1)C in $SO_2$-$O_2$, gas mixtures between 800°C and 1000°C. This study revealed an important paradox (at least in oxidizing-sulfurized environments): the addition of C and Cr decrease the corrosion resistance owing to the formation of intergranular oxide, and fibrous sulphide structure. Nonetheless, without C and Cr the alloy has poor mechanical and oxidation properties and it is not usable for industrial applications. They also explain that the b.c.c (ferrite) has a higher diffusion coefficient of impurities than the fcc (austenite). Thus, Al can diffuse quicker from the matrix of the predominantly bcc Fe-30Mn-8Al alloy to the alloy/oxide interface, thereby enabling the $Al_2O_3$ scale, to heal more easily when damaged. In contrast to an austenitic microstructure.

Dias *et al.* (1998) [333] studied the oxides scales of a Fe-34Mn-7.3Al-1.4Si alloy at different temperatures (from 600°C to 1200°C) using X-ray diffraction. It was found that at each temperature, specific types of oxides were formed. The regimes were observed as a function of the oxides; $Mn_2O_3$, and $(Mn,Fe)_2O_3$ oxides appear up to 800 'C. Then, between 800 and 1000 C the oxides observed were $\gamma$-$Fe_2O_3$, $\gamma$-$Mn_2O_3$, $Fe_3O_4$, FeO, and $Mn_3O_4$. Finally, above 1100 °C, aluminum spinels were observed ($MnAl_2O_4$, $FeAl_2O_4$). The steel studied showed good oxidation resistance from 600°C to 900°C and poor oxidation resistance at temperatures above 1000°C. Similar results were obtained by Perez *et al.* (2002) [334] in an austenitic Fe-30Mn-5Al-0.5C steel. A good oxidation resistance from 600°C to 800°C, but at 900°C catastrophic damage was observed. The oxidation resistance of Fe-Mn-Al-C steels have not been only evaluating at high temperatures, but also in other interesting media such as a polluted atmosphere ($SO_2$ environments). The first approach in this direction was made by Agudelo *et al.* (2002) [335] in Fe–Mn–Al–C alloys. They found that the low corrosion rate of these alloys in this condition is the precipitation of a large amount of $Mn^{2+}$ species, in the form of $MnSO_4 \cdot H_2O$, and $MnSO_3 \cdot 3H_2O$, which accumulates in the outer part of the corrosion layer.

To end this short section, I would like also to name that in 1989 [336] and later confirmed in 2004 [337, 338], it was reported the formation of needle-like AlN (aluminum nitride) in Fe-Mn-Al-C steels when are subjected at 1000°C in an N2 rich atmosphere, as is shown in Figure 63. This phenomenon should be taken into account when prolonged heat treatments or thermo-mechanical treatments are planning to be done in Fe-Mn-Al-C steels.

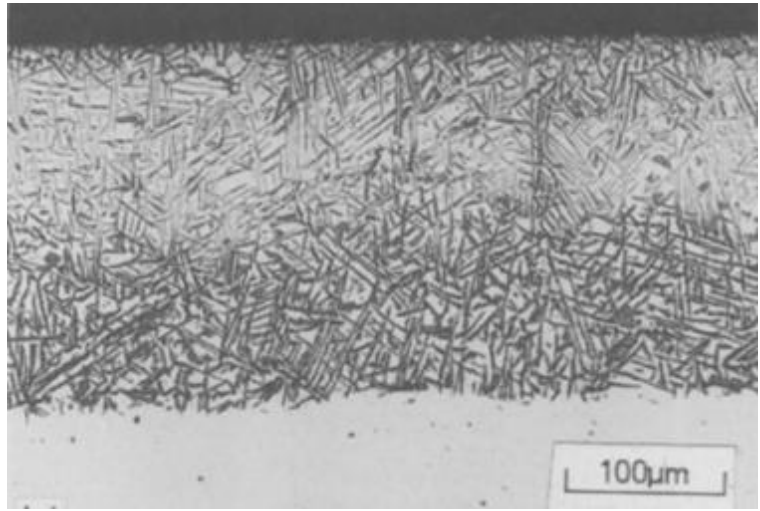

Figure 63. Metallographic cross-section of AlN needle-like structures on a Fe-Mn-Al-C alloy used after 12 h in nitrogen at 1000 °C, reproduced from [336] with permission of Springer®

# 7 Conclusions and future directions

In this review, the author has made an effort to tries to show the main properties of Fe-Mn-Al-C steels from different points of view as well as discuss the recent advances. During the last decade, research efforts have been made with the aim of elucidating and explain the mechanisms involved in the astonishing behavior of these steels. In addition, new research directions around the Fe-Mn-Al-C alloys has emerged. From the author's point of view, the topics that will surely be studied in greater depth during the next few years will be:

- ➢ The first works on High Entropy Alloys (HEA) were published in 2004 [339], since then, an extraordinary attention has been paid to develop compositions far from the classically equiatomically proportioned components, this with the aim of expanding the range of chemical space available to obtain a stable single phase. In less than 3 years, it was published one of the first attempts to use the initial concepts of HEA into steels, that is, use a non-equiatomic compositions and include the effect of interstitial atoms like C, N, etc (in the configurational entropy) to obtain a one single phase, with the inclusion of the TRIP/TWIP deformation mechanisms (a metastable phase). This leads to the first HEA (or more properly, High Entropy Steels) base on the Fe-Mn-Al-C-Si and Fe-Mn-Al-C systems [340]. The results in tension tests showed that the mechanical properties of the high entropy steels are superior compared to many conventional austenitic Fe-Cr-Ni steels. Besides, at cryogenic temperatures (for instance -50°C), the elongation to fracture was above 60%, which are very high values. The results obtained are promising and show a possible route of investigations during the following years in Fe-Mn-Al-C alloys.
- ➢ During this year [341], it was published the idea of obtaining a metastable hierarchical microstructure in a Fe-9Mn-3Ni-1.4Al steel that has a comparable fatigue crack resistance of bones. They demonstrated the effectiveness of a metastable multiphase nanolaminate microstructure concept for creating materials with exceptional fatigue resistance. Since the Fe-Mn-Al-C steels have the same capability of tuning these features, this motivates the systematic study of the austenite stability and different microstructures in Fe-Mn-Al-C steels on the fatigue resistance i.e. the role of the SFE and a hierarchical microstructure.

- Since in 2008 was published the first paper that suggested the use of Fe-Mn alloys as biomaterials [342], the biocompatibility behavior of Fe-Mn and Fe-Mn-C alloys, with Cr, Co, Pd, Al, and S additions [343-352] has caught the attention of different research groups. The preliminary results about the possible biocompatibility show favorable results. Obviously, a great effort has to be made to understand the biocompatibility potential of Fe-Mn alloy, particularly to the Fe-Mn-Al-C steels through *in-vitro* and *in-vivo* tests before thinking in any real biomedical application.

- Molecular dynamics simulations can be very useful to understand the interactions between stacking faults and twins, or dislocations and twins. Recent results elucidate these issues in Fe-Mn-C steels [353]. For this reason, this tool can be used to study and gain a depth insight about the interactions between martensite and dislocations, dislocations and κ-carbides, and to study the effect of variables such as the strain rate, the chemical composition and the test temperature on phase transformations in Fe-Mn-Al-C steels from an atomistic point of view. Also, it could be possible to evaluate the interface energy between $\gamma/\varepsilon$ phases ($\sigma^{\gamma/\varepsilon}$), and improve the current thermodynamic models to determinate the SFE.

- The superplasticity phenomenon is known several decades ago, however, to produce a drastic increase of the ductility without cracking, usually is requiring high material cost, high deformation temperature, and special fabrication process. This year, it was reported the superplasticity phenomenon in Fe-Mn-Al-C alloys [354] at test temperatures as low as 650°C, achieving ductility beyond 400%, and beyond 1300% at 850 °C. These results will bring the attention of the scientific community around this phenomenon in Fe-Mn-Al-C alloys.

- Recent works are trying to elucidate the austenite stability and deformation mechanisms through nanoindentation test [277, 355-357] in Fe-Mn-C and Fe-Mn-Al-C steels, and it is expected an accelerate progress in this direction for years to come.

- The conformability and the adequate parameters of processing are probably one, if not the most, of the main goals of the next decade. Only the control and knowledge of the processing of the Fe-Mn-Al-C alloys will permit to use and evaluate their properties in different industrial fields as well as imagine new applications so far unthinkable.

# Acknowledgment

The author gratefully acknowledges the support of the Universidad del Valle through the Project No. CI 71108, COLCIENCIAS (Colombia) for the funds required in pursuit of the doctorate degree, and FPIT (Banco de la República) convenio 201712-4.067 .

306. Tjong, S.C., *Stress corrosion cracking of the austenitic Fe-Al-Mn alloy in chloride environment.* Materials and Corrosion/Werkstoffe und Korrosion, 1986. **37**(8): p. 444-447.
307. Tjong, S.C. and C.S. Wu, *The microstructure and stress corrosion cracking behaviour of precipitation-hardened Fe·8.7Al·29.7Mn·1.04C alloy in 20% NaCl solution.* Materials Science and Engineering, 1986. **80**(2): p. 203-211.
308. Shih, S.T., I.F. Tsu, and T.P. Perng, *Environmentally assisted cracking of two-phase Fe-Mn-Al alloys in NaCl solution.* Metallurgical Transactions A, 1993. **24**(2): p. 459-465.
309. Chang, S.C., J.Y. Liu, and H.K. Juang, *Environment-Assisted Cracking of Fe-32% Mn-9% Al Alloys in 3.5% Sodium Chloride Solution.* Corrosion, 1995. **51**(5): p. 399-406.
310. Koyama, M., E. Akiyama, and K. Tsuzaki, *Effects of Static and Dynamic Strain Aging on Hydrogen Embrittlement in TWIP Steels Containing Al.* ISIJ International, 2013. **53**(7): p. 1268-1274.
311. Koyama, M., H. Springer, S.V. Merzlikin, K. Tsuzaki, E. Akiyama, and D. Raabe, *Hydrogen embrittlement associated with strain localization in a precipitation-hardened Fe–Mn–Al–C light weight austenitic steel.* International Journal of Hydrogen Energy, 2014. **39**(9): p. 4634-4646.
312. Koyama, M., E. Akiyama, and K. Tsuzaki, *Hydrogen embrittlement in a Fe–Mn–C ternary twinning-induced plasticity steel.* Corrosion Science, 2012. **54**: p. 1-4.
313. Koyama, M., E. Akiyama, K. Tsuzaki, and D. Raabe, *Hydrogen-assisted failure in a twinning-induced plasticity steel studied under in situ hydrogen charging by electron channeling contrast imaging.* Acta Materialia, 2013. **61**(12): p. 4607-4618.
314. Koyama, M., E. Akiyama, T. Sawaguchi, D. Raabe, and K. Tsuzaki, *Hydrogen-induced cracking at grain and twin boundaries in an Fe–Mn–C austenitic steel.* Scripta Materialia, 2012. **66**(7): p. 459-462.
315. Koyama, M., A. Bashir, M. Rohwerder, S.V. Merzlikin, E. Akiyama, K. Tsuzaki, and D. Raabe, *Spatially and kinetically resolved mapping of hydrogen in a twinning-induced plasticity steel by use of scanning Kelvin probe force microscopy.* Journal of The Electrochemical Society, 2015. **162**(12): p. C638-C647.
316. Ryu, J.H., S.K. Kim, C.S. Lee, D.-W. Suh, and H.K.D.H. Bhadeshia, *Effect of aluminium on hydrogen-induced fracture behaviour in austenitic Fe–Mn–C steel.* Proceedings of the Royal Society A: Mathematical, Physical and Engineering Science, 2012. **469**(2149).
317. Park, I.-J., S.-m. Lee, H.-h. Jeon, and Y.-K. Lee, *The advantage of grain refinement in the hydrogen embrittlement of Fe–18Mn–0.6C twinning-induced plasticity steel.* Corrosion Science, 2015. **93**(Supplement C): p. 63-69.
318. Timmerscheidt, T., P. Dey, D. Bogdanovski, J. von Appen, T. Hickel, J. Neugebauer, and R. Dronskowski, *The Role of κ-Carbides as Hydrogen Traps in High-Mn Steels.* Metals, 2017. **7**(7): p. 264.
319. Park, I.-J., S.Y. Jo, M. Kang, S.-M. Lee, and Y.-K. Lee, *The effect of Ti precipitates on hydrogen embrittlement of Fe–18Mn–0.6C–2Al–xTi twinning-induced plasticity steel.* Corrosion Science, 2014. **89**(Supplement C): p. 38-45.
97